\documentclass[11pt]{article}

\usepackage{bbm}
\usepackage{mathrsfs}
\usepackage{amsfonts}
\usepackage{amssymb}
\usepackage{amsfonts,amssymb, mathrsfs, amsmath,   amssymb,  theorem,  float}
\usepackage{color}

\newtheorem{thm}{Theorem}[section]

\newtheorem{cor}[thm]{Corollary}
\newtheorem{lem}[thm]{Lemma}
\newtheorem{defn}[thm]{Definition}
\newtheorem{rem}[thm]{Remark}
\newtheorem{exmp}[thm]{Example}
\newtheorem{prob}[thm]{Problem}

\floatstyle{ruled}
\newfloat{algorithm}{t}{loa}
\floatname{algorithm}{Algorithm}
\newcommand{\Inp}[1]
  {\noindent\begin{tabular}{@{}p{1.8cm}@{}p{13.2cm}@{}}
   {\bf Input: }&#1 \end{tabular}}
\newcommand{\Outp}[1]
  {\noindent\begin{tabular}{@{}p{1.8cm}@{}p{13.2cm}@{}}
   {\bf Output: }&#1 \end{tabular}}

\def \mathbbm{\mathbf}

\def\qed{{\vrule height5pt width2pt depth2pt}}

\def\and{\cap}

\def\bref#1{(\ref{#1})}
\def\proof{{\noindent\em Proof:} }

\DeclareFontFamily{U}{fsy}{} \DeclareFontShape{U}{fsy}{m}{n}{<->s*[.
9]psyr}{} \DeclareSymbolFont{der@m}{U}{fsy}{m}{n}
\DeclareMathSymbol{\diff}{\mathord}{der@m}{182}

\newcommand{\SPC}{\hspace*{15pt}}

\floatstyle{ruled}
\newfloat{algorithm}{t}{loa}
\floatname{algorithm}{Algorithm}

\def\X{{\mathbb{X}}}
\def\Y{{\mathbb{Y}}}
\def\U{{\mathbb{U}}}
\def\V{{\mathbb{V}}}

\def\A{{\mathcal A}}

\def\C{{\mathcal C}}
\def\D{{\mathcal D}}

\def\T{{\mathcal{T}}}

\def\SC{{\mathcal{S}}}
\def\PC{{\mathcal{P}}}

\def\PS{{\mathbb P}}

\def\D{{\mathbb D}}

\def\P{{\mathbb P}}
\def\Q{{\mathbb Q}}

\def\bu{{\mathbf{u}}}
\def\bv{{\mathbf{v}}}
\def\bc{{\mathbf{c}}}

\def\sat{\hbox{\rm{sat}}}
\def\asat{\hbox{\rm{asat}}}

\def\max{\hbox{\rm{max}}}

\def\lord{\hbox{\rm{Lord}}}

\def\deg{\hbox{\rm{deg}}}

\def\init{\hbox{\rm{I}}}
\def\ord{\hbox{\rm{ord}}}
\def\Eord{\hbox{\rm{Eord}}}

\def\lead{\hbox{\rm{ld}}}

\def\dim{\hbox{\rm{dim}}}

\def\lv{\hbox{\rm{lvar}}}
\def\mod{\hbox{\rm{mod}}}

\def\ord{\hbox{\rm{ord}}}
\def\rk{\hbox{\rm{rk}}}

\def\coeff{\hbox{\rm{coeff}}}
\def\norm{\hbox{\rm{N}}}
\def\I{\mathcal{I}}
\def\ff{{\mathcal F}}
\def\CI{{\mathcal{I}}}

\def\ee{{\mathcal E}}
\def\BF{{\mathbb{F}}}

\def\vol{\hbox{\rm{vol}}}

\def\Q{{\mathbb Q}}
\def\qq{{\mathbb Q}}

\def\F{{\mathcal {F}}}

\def\trdeg{\hbox{\rm{tr.deg}}}

\def\dtrdeg{\hbox{$\sigma$\rm{.tr.deg}}}
\def\Jac{\hbox{\rm{Jac}}}

\def\CQ{\mathcal{Q}}
\def\PK{\P^{[\overrightarrow{K}]}}

\def\YK{\Y^{[\overrightarrow{K}]}}

\def\buk{\bu^{[\overrightarrow{K}]}}

\def\codim{{\rm Codim}}

\def\and{\cap}

\newcounter{bean}
\def\bl{\begin{list}{Step \arabic{bean}}{\usecounter{bean}}\labelwidth=34pt}
\def\el{\end{list}}

\def\deg{{\rm deg}}

\def\init{{\rm I}}

\def\normalization1{{\rm normalization1}}
\def\normalization{{\rm normalization}}

\def\irrfactor1{{\rm irrfactor1}}
\def\irrfactor{{\rm irrfactor}}

\def\Res{\hbox{\rm Res}}

\def\sat{{\rm sat}}

\def\rank{{\rm rk}}

\def\SR{{\mathbf{R}}}

\def\TT{{\mathbb{T}}}
\def\JJ{{\mathbb{J}}}
\def\card{\rm card}

\def\m{\mathfrak{m}}

\def\conv{\hbox{\rm{conv}}}

\def\vf{\vskip5pt}

\begin{document}
\title{Sparse Difference Resultant\thanks{\quad Partially
       supported by a National Key Basic Research Project of China (2011CB302400) and  by grants from NSFC (60821002, 11101411).}}
\author{Wei Li, Chun-Ming Yuan, Xiao-Shan Gao\\
 KLMM,  Academy of Mathematics and Systems Science\\
 Chinese Academy of Sciences, Beijing 100190, China\\
 \{liwei, cmyuan, xgao\}@mmrc.iss.ac.cn }
\date{}
\maketitle

\begin{abstract}
In this paper, the concept of sparse difference resultant for a
Laurent transformally essential system of difference polynomials is
introduced and a simple criterion for the existence of sparse
difference resultant is given. The concept of transformally
homogenous polynomial is introduced and the sparse difference
resultant is shown to be transformally homogenous. It is shown that
the vanishing of the sparse difference resultant gives a necessary
condition for the corresponding difference polynomial system to have
non-zero solutions. The order and degree bounds for sparse
difference resultant are given. Based on these bounds, an algorithm
to compute the sparse difference resultant is proposed, which is
single exponential in terms of  the number of variables, the Jacobi
number, and the size of the Laurent transformally essential system.
Furthermore, the precise order and degree, a determinant
representation, and a Poisson-type product formula for the
difference resultant are given.

\end{abstract}

\section{Introduction}

The resultant, which gives conditions for an over-determined system
of polynomial equations to have common solutions, is a basic concept
in algebraic geometry and a powerful tool in elimination theory
\cite{canny1,cox,eisenbud,hodge1,joun1,sturmfels}. 
The concept of sparse resultant originated from the work of Gelfand,
Kapranov, and Zelevinsky on generalized hypergeometric functions,
where the central concept of $\mathcal {A}$-discriminant is studied
\cite{gelfand}.
Kapranov, Sturmfels, and Zelevinsky introduced the concept of
$\mathcal {A}$-resultant \cite{Kapranov1}.
Sturmfels further introduced the general mixed sparse resultant and
gave a single exponential algorithm to compute the sparse resultant
\cite{sturmfels,sturmfels2}.
Canny and Emiris showed that the sparse resultant is a factor of the
determinant of a Macaulay style matrix and gave an efficient
algorithm to compute the sparse resultant based on this matrix
representation \cite{emiris1}.
%
%
A determinant representation for the sparse resultant was given by
D'Andrea \cite{dandrea1}. Recently,  a rigorous definition for the
differential resultant of $n+1$ generic differential polynomials in
$n$ variables was presented \cite{gao} and also the theory of sparse
differential resultants for Laurent differentially essential systems
was developed \cite{li1,li}. It is meaningful to generalize the
theory of sparse resultant to difference polynomial systems.

In this paper,  the concept of sparse difference resultant for a
Laurent transformally essential system consisting of $n+1$ Laurent
difference polynomials in $n$ difference variables is introduced and
its basic properties are proved.
A criterion is given to check whether a Laurent difference system is
essential in terms of their supports, which is conceptually and
computationally simpler than the naive approach based on the
characteristic set method.
The concept of transformally homogeneous is introduced and it is
proved that the sparse difference resultant is transformally
homogeneous.
It is shown that the vanishing of the sparse difference resultant
gives a necessary condition for the corresponding difference
polynomial system to have nonzero solutions, which is also
sufficient in certain sense.
It is also shown that the sparse difference resultant is equal to
the algebraic sparse resultant of a generic sparse polynomial
system, and hence has a determinant representation.

We also give order and degree bounds for the sparse difference
resultant. It is shown that the order and effective order of the
sparse difference resultant can be bounded by the Jacobi number of
the corresponding difference polynomial system and the degree can be
bounded by a Bezout type bound.
Based on these bounds, an algorithm is given to compute the sparse
difference resultant. The complexity of the algorithm in the worst
case is single exponential of the form
$O(m^{O(nlJ^2)}(nJ)^{O(lJ)})$, where $n,m,J,$ and $l$ are  the
number of variables, the degree, the Jacobi number, and the size of
the Laurent transformally essential system, respectively.

For the difference resultant, which is non-sparse, more and better
properties are proved including its precise order and degree, a
determinant representation, and a Poisson-type product formula.
%

Although most properties for sparse difference resultants and
difference resultants are similar to their differential counterparts
given in \cite{li1,li,gao}, some of them are quite different in
terms of descriptions and proofs due to the distinct nature of the
differential and difference operators.
Firstly, the definition for difference resultant is more subtle than
the differential case as illustrated by Problem
\ref{problem-generator} in this paper.
Secondly, the criterion for Laurent transformally essential systems
given in Section 3.3 is quite different and much simpler than its
differential counterpart given in \cite{li}. Also, determinant
representations for the sparse difference resultant and the
difference resultant are given in Section \ref{sec-sares} and
Section \ref{differenceresultant}, but such a representation is
still not known for differential resultants
\cite{dres-matrix,sonia-arxiv,sonia-raf}.
Finally,
%
%
there does not exist a definition for homogeneous difference
polynomials, and the definition we give in this paper is different
from its differential counterpart \cite{li-pdcf}.
%

The  rest of the paper is organized as follows.
In Section \ref{sec-preliminaries}, we prove some preliminary
results. In Section \ref{sec-sres}, we first introduce the concepts
of Laurent difference polynomials and Laurent transformally
essential systems, and then define the sparse difference resultant
for Laurent transformally essential systems. Basic properties of
sparse difference resultant are proved in Section
\ref{sec-property}.
In Section \ref{sec-sares}, the sparse difference resultant is shown
to be the algebraic sparse resultant for certain generic polynomial
system.
In  Section \ref{sec-algorithm}, we present an algorithm to compute
the sparse difference resultant.
In Section~\ref{differenceresultant}, we introduce the notion of
difference resultant and prove its basic properties.
In Section \ref{sec-conc}, we conclude the paper by proposing
several problems for future research.
An extended abstract of this paper appeared in the proceedings of
ISSAC2013 \cite{li-2013}. Section \ref{sec-eord} and Section
\ref{sec-sares} are newly added.

\section{Preliminaries}\label{sec-preliminaries}
In this section, some basic notations and preliminary results in
 difference algebra will be given. For more details about difference algebra,
 please refer to \cite{cohn,Hrushovski1,levin,wibmer}.

\subsection{Difference polynomial ring}

An ordinary difference field $\F$ is a field with a third unitary
operation $\sigma$ satisfying that for any $a,b\in\F$,
$\sigma(a+b)=\sigma(a)+\sigma(b)$, $\sigma(ab)=\sigma(a)\sigma(b)$,
and $\sigma(a)=0$ if and only if $a=0$. We call $\sigma$ the {\em
transforming operator} of $\F$. If $a\in\F$, $\sigma(a)$ is called
the transform of $a$ and is denoted by $a^{(1)}$. And for
$n\in\mathbb{Z}^{+}$, $\sigma^n(a)=\sigma^{n-1}(\sigma(a))$ is
called the $n$-th transform of $a$ and denoted by $a^{(n)}$, with
the usual assumption $a^{(0)}=a$. By $a^{[n]}$ we mean the set
$\{a,a^{(1)},\ldots,a^{(n)}\}$.
If $\sigma^{-1}(a)$ is defined for each $a\in\F$, we say that $\F$
is inversive. All difference fields in this paper are assumed to be inversive. 
A typical example of difference field is $\Q(x)$ with
$\sigma(f(x))=f(x+1)$.

Let $S$ be a subset of a  difference field $\mathcal{G}$ which
contains $\mathcal {F}$.   We will   denote respectively by
$\mathcal {F}[S]$, $\mathcal {F}(S)$, $\mathcal {F}\{S\}$, and
$\mathcal {F}\langle S\rangle$    the smallest subring, the smallest
subfield, the smallest difference subring, and the smallest
difference subfield of $\mathcal{G}$ containing $\mathcal {F}$ and
$S$.  If we denote $\Theta(S)=\{\sigma^ka|k\geq0,a\in S\}$,  then we
have $\mathcal {F}\{S\}=\mathcal    {F}[\Theta(S)]$ and $\mathcal
{F}\langle
S\rangle=\mathcal    {F}(\Theta(S))$. 

A subset $\mathcal {S}$ of a  difference extension field $\mathcal
{G}$ of $\mathcal {F}$ is said to be {\em transformally dependent}
over $\mathcal {F}$ if the set $\{\sigma^{k}a\big|a\in\mathcal
{S},k\geq0\}$ is algebraically dependent over $\mathcal {F}$, and is
said to be {\em transformally independent} over $\mathcal {F}$, or
to be a family of {\em difference indeterminates} over $\mathcal
{F}$ in the contrary case.
In the case $\mathcal{S}$ consists of one element $\alpha$, we say
that $\alpha$ is {\em transformally algebraic }or {\em transformally
transcendental} over $\mathcal {F}$, respectively. The maximal
subset $\Omega$ of $\mathcal {G}$ which are transformally
independent over $\mathcal {F}$ is said to be a transformal
transcendence basis of $\mathcal {G}$ over $\mathcal {F}$. We use
$\dtrdeg \,\mathcal {G}/\mathcal {F}$  to denote the {\em difference
transcendence degree} of $\mathcal {G}$ over $\mathcal {F}$, which
is the cardinal number of $\Omega$. Considering $\mathcal {F}$ and
$\mathcal {G}$ as ordinary algebraic fields, we denote the algebraic
transcendence degree of $\mathcal {G}$ over $\mathcal {F}$ by
$\trdeg\,\mathcal {G}/\mathcal {F}$.

Now suppose $\Y=\{y_{1}, y_{2}, \ldots, y_{n}\}$ is a set of
difference indeterminates over $\ff$.   The elements of $\mathcal
{F}\{\Y\}=\mathcal {F}[y_j^{(k)}:j=1,\ldots,n;k\in \mathbb{N}_0]$
are called {\em difference polynomials} over $\ff$ in $\Y$, and
$\mathcal {F}\{\Y\}$ itself is called the {\em difference polynomial
ring } over $\ff$ in $\Y$. A {\em difference ideal} $\mathcal {I}$
in $\mathcal {F}\{\Y\}$ is an ordinary algebraic ideal which is
closed under transforming, i.e. $\sigma(\mathcal {I})\subset\mathcal
{I}$. If $\mathcal{I}$ also has the property that
$a^{(1)}\in\mathcal{I}$ implies that $a\in\mathcal{I}$, it is called
a {\em reflexive difference ideal}.
A prime  difference ideal is a difference ideal which is prime as an
ordinary algebraic polynomial ideal. For convenience, a prime
difference ideal is assumed not to be the unit ideal in this paper.
If $S$ is a finite set of difference polynomials, we use $(S)$ and
$[S]$ to denote the algebraic ideal and the difference ideal in
$\ff\{\Y\}$ generated by $S$.

An $n$-tuple over $\ff$ is an $n$-tuple of the form
$\textbf{a}=(a_1,\ldots,a_n)$ where the $a_i$ are selected from a
difference overfield of $\ff$. For a difference polynomial
$f\in\ff\{y_1,\ldots,y_n\}$, $\textbf{a}$ is called a difference
zero of $f$ if when substituting $y_i^{(j)}$ by $a_i^{(j)}$ in $f$,
the result is $0$.
 An $n$-tuple $\eta$ is called a {\em generic zero} of a difference ideal $\CI\subset\ff\{\Y\}$ if for any polynomial $P\in\ff\{\Y\}$ we
have $P(\eta)=0 \Leftrightarrow P\in\CI$. It is well known that
\begin{lem}\cite[p.77]{cohn}\label{lm-gp}
A difference ideal possesses a generic zero if and only if it is a
reflexive prime difference ideal other than the unit ideal.
\end{lem}

Let $\CI$ be a reflexive prime difference ideal and $\eta$ a generic
zero of $\CI.$ The {\em dimension} of $\CI$ is defined to be
$\dtrdeg\ff\langle\eta\rangle/\ff$.

Given two $n$-tuples $\textbf{a}=(a_1,\ldots,a_n)$ and
$\bar{\textbf{a}}=(\bar{a}_1,\ldots,\bar{a}_n)$ over $\ff$.
$\bar{\textbf{a}}$ is called a {\em specialization} of $\textbf{a}$
over $\ff$, or $\textbf{a}$ specializes to $\bar{\textbf{a}}$, if
for any difference polynomial $P\in\ff\{\Y\}$, $P(\textbf{a})=0$
implies that $P(\bar{\textbf{a}})=0$. The following property about
difference specialization will be needed in this paper.
\begin{lem}\label{lm-special}
Let  $P_{i}(\U, \Y)\in \mathcal {F}\langle \Y\rangle\{\U\}$ $(i=1,
\ldots, m)$ where $\U=(u_{1},\ldots,u_{r})$ and $\Y=(y_{1}, \ldots,
y_{n})$ are sets of difference indeterminates. If $P_{i}(\U, \Y)$
$(i=1, \ldots, m)$ are transformally dependent over $\mathcal
{F}\langle \U \rangle$, then for any difference specialization $\U$
to $\overline{\U}$ which are elements in $\mathcal {F}$,
$P_{i}(\overline{\U},\Y) \, (i=1, \ldots,  m)$ are transformally
dependent over $\mathcal {F}$.
\end{lem}
\proof It suffices to show the case $r=1$. Denote $u=u_1$. 
Since  $P_{i}(u, \Y)$ $(i=1, \ldots, m)$ are transformally dependent
over $\mathcal {F}\langle u \rangle$, there exist natural numbers
$s$ and $l$ such that $\P_i^{(k)}(u,\Y)\,(k\leq s)$ are
algebraically dependent over $\ff(u^{(k)}|k\leq s+l)$. When $u$
specializes to $\bar{u}\in\ff$,  $u^{(k)}\,(k\geq 0)$ are
correspondingly algebraically specialized to $\bar{u}^{(k)}\in\ff$.
By \cite[p.161]{wu}, $\P_i^{(k)}(\bar{u},\Y)\,(k\leq s)$ are
algebraically dependent over $\ff$. Thus, $P_{i}(\bar{u},\Y) \,
(i=1, \ldots,  m)$ are transformally dependent over $\mathcal {F}$.
\qed

\subsection{Characteristic set for a  difference polynomial system}
In this section, we prove several preliminary results about the
characteristic set for a difference polynomial system. For details
on difference characteristic set method, please refer to
\cite{gao-dcs}.

Let $f$ be a difference polynomial in $\ff\{\Y\}$.  The order of $f$
w.r.t. $y_i$ is defined to be the greatest number $k$ such that
$y_{i}^{(k)}$ appears effectively in $f$, denoted by
$\ord(f,y_{i})$. And if $y_{i}$ does not appear in $f$, then we set
$\ord(f,y_{i})=-\infty$. The {\em order} of $f$ is defined to be
$\max_{i}\,\ord(f,y_{i})$, that is,
$\ord(f)=\max_{i}\,\ord(f,y_{i})$.

     A {\em ranking} $\mathscr{R}$ is a total order over $\Theta (\Y)=\{\sigma^ky_i|1\leq i\leq n,k\geq0\}$, which satisfies the following properties:

     1) $\sigma(\theta) >\theta $ for all derivatives $\theta \in\Theta (\Y)$.

     2) $\theta_{1} >\theta_{2} $ $\Longrightarrow$ $\sigma(\theta_{1} ) >\sigma(\theta_{2} )$
for $\theta_{1}, \theta_{2}\in \Theta (\Y)$.


%
%
%

    Let $f$ be a difference polynomial in
    $\mathcal {F}\{\Y\}$ and $\mathscr{R}$  a ranking
    endowed on it.  The greatest $y_j^{(k)}$ w.r.t.  $\mathscr{R}$ which  appears effectively in $f$ is called the {\em leader} of $p$,
    denoted by  $\lead(f)$ and correspondingly $y_j$ is called the {\em leading variable }of $f$, denoted by $\lv(f)=y_j$.
%
   The leading coefficient of $f$  as a univariate polynomial in $\lead(f)$ is called the {\em initial} of $f$ and is denoted by $\init_{f}$.

    Let $p$ and $q$ be two difference polynomials in $\ff\{\Y\}$. $q$ is said to be of higher rank than $p$ if

    1) $\lead(q)>\lead(p)$, or

    2) $\lead(q)=\lead(p)=y_j^{(k)}$ and $\deg(q,y_j^{(k)})>\deg(p, y_j^{(k)})$.

Suppose $\lead(p)=y_j^{(k)}$. Then  $q$ is said to be
  {\em reduced} w.r.t. $p$ if  $\deg(q,y_j^{(k+l)})<\deg(p,y_j^{(k)})$ for all $l\in\mathbb{N}_0$.

  A finite chain of nonzero difference polynomials $\mathcal {A}=A_1,\ldots,A_m$  is said to be an
    {\em ascending chain} if

    1) $m=1$ and $A_1\neq0$ or

    2) $m>1$, $A_j>A_i$ and $A_j$ is reduced
    w.r.t. $A_i$ for $1\leq i<j\leq m$.

Let $\mathcal {A}=A_{1},A_{2},\ldots,A_{t}$ be an ascending chain
with  $\init_{i}$ as the initial of $A_{i}$, and $f$ any  difference
polynomial.
   Then there exists an
   algorithm, which reduces
   $f$ w.r.t. $\mathcal {A}$ to a  polynomial $r$ that is
   reduced w.r.t. $\mathcal {A}$, satisfying the relation
   $$\prod_{i=1}^t\prod_{k=0}^{d_{i}}(\sigma^k\init_{i}
   )^{e_{ik}} \cdot f \equiv
   r, \mod\, [\mathcal {A}],$$
   where the $e_{ik}$ are nonnegative integers.
  The difference polynomial $r$ is called the {\em difference remainder} of $f$ w.r.t. $\A$~\cite{gao-dcs}.

Let $\mathcal {A}$ be an ascending chain. Denote
$\mathbb{I}_{\mathcal {A}}$ to be  the minimal multiplicative set
containing the initials of elements of $\mathcal{A}$ and their
transforms.
    The {\em saturation ideal} of $\A$ is defined to be
    $$\sat(\A)=[\mathcal
   {A}]:\mathbb{I}_{\mathcal
{A}} = \{p: \exists h\in \mathbb{I}_{\mathcal {A}}, \,{\text s. t.
}\, hp\in[A]\}.$$ And the {\em algebraic saturation ideal} of $\A$
is $\asat(\A)=(\A):\init_{\A}$, where $\init_{\A}$ is the minimal
multiplicative set containing the initials of elements of
$\mathcal{A}$.

 An ascending chain $\mathcal {C}$ contained in a difference polynomial set
 $\mathcal {S}$ is said to be a {\em characteristic set} of $\mathcal {S}$,
 if  $\mathcal {S}$ does not contain any nonzero element reduced w.r.t.
$\mathcal {C}$. A characteristic set $\mathcal{C}$ of a difference
ideal $\mathcal {J}$ reduces all elements of $\mathcal {J}$ to zero.

Let $\mathcal{A}$ be a characteristic set of a reflexive prime
difference ideal $\mathcal{I}$. We rewrite $\mathcal{A}$ in the
following form
\[\mathcal{A}=\left\{\begin{array}{l}A_{11},\ldots,A_{1k_1}\\ \cdots \\ A_{p1},\ldots,A_{pk_p} \end{array}
\right.\] where $\lv(A_{ij})=y_{c_i}$ for $j=1,\ldots,k_i$ and
$\ord(A_{ij},y_{c_i})<\ord(A_{il},y_{c_i})$ for $j<l$. In terms of
the characteristic set of the above form, $p$ is equal to the {\em
codimension} of  $\I$, that is $n-\dim(\I)$. Unlike the differential
case, here even though $\I$ is of codimension one, there may be more
than one difference polynomials in a characteristic set of $\I$ as
shown by the following example.
\begin{exmp}\label{ex-a}
Let $A_{11} = (y_1^{(1)})^2 +y_1^2 +1$, $A_{12} = y_1^{(2)} - y_1$.
Then $\I=[A_{11},\A_{12}]$ is a reflexive prime difference ideal
whose characteristic set is $\mathcal{A} = A_{11}, A_{12}$ and $\I =
\sat(\mathcal{A})$ \cite {gao-dcs}. Note that $[A_{11}]$ is not a
prime difference ideal, because $\sigma(A_{11}) - A_{11} =
(y_1^{(2)} -y_1)(y_1^{(2)} +y_1)\in [A_{11}]$ and both $y_1^{(2)}
-y_1$ and $y_1^{(2)} +y_1$ are not in $[A_{11}]$.
\end{exmp}

Now we proceed to show that  a  property of uniqueness still exists
in characteristic sets of a reflexive prime difference ideal in some
sense. Firstly, we need several algebraic results.

Let $\mathcal{B}=B_1,\ldots,B_m$ be an algebraic triangular set in
$\ff[x_1,\ldots,x_n]$ with $\lv(B_i)=y_i$ and
$U=\{x_1,\ldots,x_n\}\big\backslash \{y_1,\ldots,y_m\}$. We assume
$U < y_1 < y_2 < \ldots < y_m$. A polynomial $f$ is said to be
invertible w.r.t. $\mathcal{B}$ if $(f,B_1,\ldots,B_s)\cap K[U]\neq
\{0\}$ where $\lv(f)=\lv(B_s)$. We call $\mathcal{B}$ a {\em regular
chain} if for each $i>1$, the initial of $B_i$ is invertible w.r.t.
$B_1,\ldots,B_{i-1}$. For a regular chain $\mathcal{B}$, we say that
$f$ is invertible w.r.t. $\asat(\mathcal{B})$ if
$(f,\asat(\mathcal{B}))\cap\ff[U]\neq\{0\}$.
The next two lemmas use the notations introduced in this paragraph.

\begin{lem}\label{lm-asatinvertible}
Let  $\mathcal{B}$ be a regular chain in $\ff[x_1,\ldots,x_n]$. If
$\sqrt{\asat(\mathcal{B})}=\bigcap_{i=1}^{m}\mathcal {P}_i$ is an
irredundant prime decomposition of $\sqrt{\asat(\mathcal{B})}$, then
a polynomial $f$ is invertible w.r.t. $\asat(\mathcal{B})$ if and
only if $f\notin\mathcal{P}_i$ for all $i=1,\ldots,m$.
\end{lem}
\proof  Since $\sqrt{\asat(\mathcal{B})}=\bigcap_{i=1}^{m}\mathcal
{P}_i$ is an irredundant prime decomposition of
$\sqrt{\asat(\mathcal{B})}$, $U$ is a parametric set of
$\mathcal{P}_i$ for each $i$ by paper \cite{gao-chou}. And for prime
ideals $\mathcal{P}_i$, $f\notin\mathcal{P}_i$ if and only if
$(f,\mathcal{P}_i)\cap\ff[U]\neq\{0\}$. If $f$ is invertible w.r.t.
$\asat(\mathcal{B})$,
$\{0\}\neq(f,\asat(\mathcal{B}))\cap\ff[U]\subset(f,\mathcal{P}_i)\cap\ff[U]$.
Thus, $f\notin\mathcal{P}_i$ for each $i.$ For the other side,
suppose $f\notin\mathcal{P}_i$ for all $i$, then there exist nonzero
polynomials $h_i(U)$ such that $h_i(U)\in(f,\mathcal{P}_i)$. Thus,
there exists $t\in\mathbb{N}$ such that
$(\prod_{i=1}^mh_i(U))^t\in(f,\asat(\mathcal{B}))$. So $f$ is
invertible w.r.t. $\asat(\mathcal{B})$.\qed

\begin{lem}\cite{bouziane}\label{lm-regular-unmixed}
Let  $\mathcal{B}$ be a regular chain in $\ff[U,Y]$. Let $f$ be a
polynomial in $\ff[U,Y]$ and $L$ in $\ff[U]\backslash\{0\}$ such
that $Lf\in(\mathcal{B})$. Then $f\in\asat(\mathcal{B})$.
\end{lem}

\begin{lem}\label{lm-sigmainvertible}
Let $A$  be an irreducible difference polynomial in $\ff\{\Y\}$ with
$\deg(A,y_{i_0})>0$ for some $i_0$. If $f$ is invertible w.r.t.
$A^{[k]}=A,A^{(1)},\ldots,A^{(k)}$ under some ranking $\mathscr{R}$,
then $\sigma(f)$ is invertible w.r.t.
$A^{[k+1]}=A,\ldots,A^{(k+1)}$. In particular, $A^{[k]}$ is a
regular chain for any $k\geq0.$
\end{lem}
\proof Since as a difference ascending chain, $A$ is coherent and
proper irreducible, by Theorem 4.1 in paper \cite{gao-dcs}, $A$ is
difference  regular. As a consequence, $A^{[k]}$ is regular  for any
$k\geq0.$\qed
%
%
%
%
%
%

The following fact is needed to define sparse difference resultant.
\begin{lem} \label{le-char-codim1}
Let $\I$ be a reflexive prime difference ideal of codimension one in
$\ff\{\Y\}$.  The first element in any characteristic set of
$\mathcal{I}$ w.r.t. any ranking, when taken irreducible, is unique
up to a factor in $\ff$.
\end{lem}
\proof Let $\mathcal{A}=A_1,\ldots,A_m$ be a characteristic set of
$\I$ w.r.t. some ranking $\mathscr{R}$ with $A_1$ irreducible.
Suppose $\lv(\mathcal{A})=y_1$. Given another  characteristic set
$\mathcal{B}=B_1,\ldots,B_l$ of $\mathcal{I}$ w.r.t. some other
ranking $\mathscr{R}'$ ($B_1$ is irreducible), we need to show that
there exists $c\in\ff$ such that $B_1=c\cdot A_1$. It suffices to
consider the case $\lv(\mathcal{B})\neq y_1$. Suppose
$\lv(B_1)=y_2.$ Clearly, $y_2$ appears effectively in $A_1$ for
$\mathcal{B}$ reduces $A_1$ to $0$. And since $\I$ is reflexive,
there exists some $i_0$ such that $\deg(A_1,y_{i_0})>0.$

Suppose $\ord(A_1,y_2)=o_{2}$. Take another ranking under which
$y_2^{(o_{2})}$ is the leader of $A_1$ and we use $\widetilde{A}_1$
to distinguish it from the $A_1$ under $\mathscr{R}$. By
Lemma~\ref{lm-sigmainvertible}, for each $k$, $A_1^{[k]}$ and
$\widetilde{A}_1^{[k]}$ are regular chains.

Now we claim that $\asat(A_1^{[k]})=\asat(\widetilde{A}_1^{[k]})$
for any $k$. On the one hand, for any polynomial
$f\in\asat(A_1^{[k]})$, we have
$(\prod_{i=0}^{k}\sigma^i(\init_{A_1}))^{a}f\in(A_1^{[k]})$. Since
$\init_{A_1}$ is invertible w.r.t. $\widetilde{A}_1$, by
Lemma~\ref{lm-sigmainvertible}, 
$(\prod_{i=0}^{k}\sigma^i(\init_{A_1}))^{a}$ is invertible w.r.t.
$\widetilde{A}_1^{[k]}$. Denote the parameters of
$\widetilde{A}_1^{[k]}$ by $\widetilde{U}$. So there exists a
nonzero polynomial $h(\widetilde{U})$ such that
$h(\widetilde{U})\in((\prod_{i=0}^{k}\sigma^i(\init_{A_1}))^{a},\widetilde{A}_1^{[k]})$.
Thus, $h(\widetilde{U})f\in(\widetilde{A}_1^{[k]})$. Since
$\widetilde{A}_1^{[k]}$ is a regular chain, by
Lemma~\ref{lm-regular-unmixed}, $f\in \asat(\widetilde{A}_1^{[k]})$.
So $\asat(A_1^{[k]})\subseteq\asat(\widetilde{A}_1^{[k]})$.
Similarly, we can show that
$\asat(\widetilde{A}_1^{[k]})\subseteq\asat(A_1^{[k]})$. Thus,
$\asat(A_1^{[k]})=\asat(\widetilde{A}_1^{[k]})$.

Suppose $\ord(B_1,y_2)=o_{2}'$. Clearly, $o_{2}\geq o_{2}'$.
 We now proceed to show that it is impossible for $o_{2}> o_{2}'$.
Suppose the contrary, i.e. $o_{2}> o_{2}'$. Then $B_1$ is invertible
w.r.t. $\asat(\widetilde{A}_1^{[k]})$. Suppose
$\sqrt{\asat(\widetilde{A}_1^{[k]})}=\bigcap_{i=1}^t \mathcal{P}_i$
is an irredundant prime decomposition. By
Lemma~\ref{lm-asatinvertible}, $B_1\notin \mathcal{P}_i$ for each
$i$. Since $\asat(A_1^{[k]})=\asat(\widetilde{A}_1^{[k]})$, using
Lemma~\ref{lm-asatinvertible} again, $B_1$ is invertible w.r.t.
$\asat(A_1^{[k]})$. Thus, there exists a nonzero difference
polynomial $H$ with $\ord(H,y_1)<\ord(A_1,y_1)$ such that
$H\in(B_1,\asat(A_1^{[k]}))\subset\I$, which is a contradiction.
Thus, $o_2=o_2'$. Since $\mathcal{B}$ reduces $A_1$ to zero and
$A_1$ is irreducible, there exists $c\in\ff$ such that $B_1=c\cdot
A_1$. \qed

\section{Sparse difference resultant} \label{sec-sres}
In this section, the concepts of Laurent difference polynomials and
Laurent transformally essential systems are first introduced, and
then the sparse difference resultant for Laurent  transformally
essential systems is defined. A criterion for a Laurent polynomial
system to be Laurent  transformally essential in terms of the
support of the given system is also given.

\subsection{Laurent difference polynomial}\label{subsec-laurent}
Let $\F$ be an ordinary difference field with a transforming
operator $\sigma$ and $\F\{\Y\}$ the ring of difference polynomials
in the difference indeterminates $\Y=\{y_1,\ldots,y_n\}$. Before
defining sparse  difference  resultant, we first introduce the
concept of Laurent difference polynomials.
\begin{defn}
A Laurent difference monomial of order $s$ is a Laurent monomial in
variables $\Y^{[s]}=(y_i^{(k)})_{1\leq i\leq n;0\leq k\leq s}$. More
precisely, it has the form
$\prod_{i=1}^n\prod_{k=0}^s(y_i^{(k)})^{d_{ik}}$ where $d_{ik}$ are
integers which can be negative. A {\em Laurent difference
polynomial} over $\ff$ is a finite linear combination of Laurent
difference monomials with coefficients in $\ff$.
\end{defn}

Clearly, the collections of all Laurent difference polynomials form
a commutative difference ring under the obvious sum, product, and
the usual transforming operator $\sigma$, where all Laurent
difference monomials are invertible. 
We denote the difference ring of Laurent difference polynomials with
coefficients in $\mathcal {F}$ by $\mathcal
{F}\{y_1,y_1^{-1},\ldots,y_n,y_n^{-1}\}$, or simply by
$\ff\{\Y,\Y^{-1}\}$.

\begin{defn}
For every  Laurent difference polynomial $F\in\ff\{\Y,\Y^{-1}\}$,
there exists a unique Laurent difference monomial $M$ such that 1)
$M\cdot F\in\ff\{\Y\}$ and 2) for any Laurent difference monomial
$T$ with $T\cdot F\in\ff\{\Y\}$, $T\cdot F$ is divisible by $M\cdot
F$ as  polynomials. This $M\cdot F$ is defined to be the {\em norm
form} of $F$, denoted by $\norm(F)$. The order and degree of
$\norm(F)$ is defined to be the {\em  order} and {\em degree} of
$F$, denoted by $\ord(F)$ and $\deg(F)$.
\end{defn}

In the following, we consider zeros for Laurent difference
polynomials.
\begin{defn}
 Let $F$ be a Laurent difference polynomial in $\ff\{\Y,\Y^{-1}\}$. An $n$-tuple
$(a_1,$ $\ldots,a_n)$ over $\ff$ with each $a_i\neq0$  is called a
{\em nonzero difference solution} of $F$  if  $F(a_1,\ldots,a_n)=0.$
\end{defn}

For an ideal  $\I\in\ff\{\Y,\Y^{-1}\}$, the difference zero set of
$\I$ is the set of common nonzero difference zeros of all Laurent
difference polynomials in $\I$. We will see later in Example
\ref{ex-sp12}, how nonzero difference solutions are naturally
related with the sparse difference resultant.

\subsection{Definition of sparse difference resultant}

In this section, the definition of the sparse difference resultant
will be given. Similar to the study of sparse resultants and sparse
differential resultants, we first define sparse difference
resultants for Laurent difference polynomials whose coefficients are
difference indeterminates. Then the sparse difference resultant for
a given Laurent difference polynomial system with concrete
coefficients is the value which the resultant in the generic case
assumes for the given case.

Suppose $\mathcal
{A}_i=\{M_{i0},M_{i1},\ldots,M_{il_i}\}\,(i=0,1,\ldots,n)$ are
finite sets of Laurent difference monomials
in $\Y.$ 
Consider $n+1$ {\em generic Laurent difference polynomials} defined
over $\mathcal {A}_0,\ldots,\mathcal{A}_n$:
\begin{equation} \label{eq-sparseLaurent}
\P_i=\sum\limits_{k=0}^{l_i}u_{ik} M_{ik}\,(i=0,\ldots,n),
\end{equation} where all the $u_{ik}$ are  transformally independent over
the rational number field $\Q$.
 Denote
\begin{equation} \label{eq-uu}
 \bu_i=(u_{i0},u_{i1},\ldots,u_{il_i})\,\,(i=0,\ldots,n) \hbox{ and }
   \bu=\bigcup_{i=0}^n\bu_i\backslash\{u_{i0}\}.
\end{equation}
The number $ l_i +1$ is called the {\em size} of $\P_i$ and
$\mathcal{A}_i$ is called the {\em support} of $\P_i$. To avoid the
triviality, $l_i\geq1\,(i=0,\ldots,n)$ are always assumed in this
paper.

\begin{defn}\label{def-tdes} A set of Laurent difference
polynomials of the form \bref{eq-sparseLaurent} is called  {\em
Laurent transformally essential} if there exist
$k_i\,(i=0,\ldots,n)$ with $1\leq k_i\leq l_i$ such that
$\dtrdeg\,\Q\langle\frac{M_{0k_0}}{M_{00}},$
$\frac{M_{1k_1}}{M_{10}},\ldots,\frac{M_{nk_n}}{M_{n0}}\rangle/\Q=n.$
In this case, we also say that $\mathcal{A}_0,\ldots,\mathcal{A}_n$
form a Laurent transformally essential system.
\end{defn}

Although  $M_{i0}$ are used as denominators to define transformally
essential system, the following lemma shows that the definition does
not depend on the choices of $M_{i0}$.

\begin{lem} \label{le-defsparse}
The following two conditions are equivalent.
\begin{enumerate}
\item  There exist
$k_i\,(i=0,\ldots,n)$ with $1\leq k_i\leq l_i$ such that
$\dtrdeg\,\Q\langle\frac{M_{0k_0}}{M_{00}},$
$\ldots,\frac{M_{nk_n}}{M_{n0}}\rangle/\Q$ $=n.$
\item  There exist pairs
$(k_i,j_i)\,(i=0,\ldots,n)$ with $k_i\neq j_i\in\{0,\ldots,l_i\}$
such that\\ $\dtrdeg\,\Q\langle\frac{M_{0k_0}}{M_{0j_0}},$
$\ldots,\frac{M_{nk_n}}{M_{nj_n}}\rangle/\Q=n.$
\end{enumerate}

\end{lem}
\proof Similar to the proof of \cite[Lemma 3.7]{li}, it can be
easily shown.\qed

Let $\mathbbm{m}$ be the set of all difference monomials in $\Y$ and
$[\norm(\P_0),\ldots,\norm(\P_n)]$ the difference ideal generated by
$\norm(\P_i)$ in $\Q\{\Y,\bu_0,\ldots,\bu_n\}$.
Let
\begin{eqnarray}
\mathcal{I}_{\Y,\bu}&=&([\norm(\P_0),\ldots,\norm(\P_n)]:\mathbbm{m}),\label{eq-I}\\
\mathcal{I}_{\bu}&=& \mathcal{I}_{\Y,\bu}\cap\Q\{
\bu_0,\ldots,\bu_n\}.\label{eq-IU}
 \end{eqnarray}
The following result is a foundation for defining sparse difference
 resultants.
\begin{thm} \label{th-Mcodim1}
Let $\P_0,\ldots,\P_n$ be the Laurent difference polynomials defined
in (\ref{eq-sparseLaurent}). Then the following assertions hold.
\begin{enumerate}
\item $\mathcal{I}_{\Y,\bu}$ is a reflexive prime
difference ideal in $\Q\{\Y,\bu_0,$ $\ldots,\bu_n\}$.
\item
$\mathcal{I}_{\bu}$ is of codimension one if and only if
$\P_0,\ldots,\P_n$ form a Laurent transformally essential system.
\end{enumerate}
\end{thm}
\proof Let $\eta=(\eta_1,\ldots,\eta_n)$ be a sequence of
transformally independent elements over  $\Q\langle\bu\rangle$,
where $\bu$ is defined in \bref{eq-uu}. Let
\begin{equation}\label{eq-zeta}
\zeta_i=-\sum_{k=1}^{l_i}u_{ik}\frac{M_{ik}(\eta)}{M_{i0}(\eta)}\,\,(i=0,1,\ldots,n).
\end{equation}
We claim that
$\theta=(\eta;\zeta_0,u_{01},\ldots,u_{0l_0};\ldots;\zeta_n,u_{n1},\ldots,u_{nl_n})$
is a generic zero of $\mathcal{I}_{\Y,\bu}$, which follows that
$\mathcal{I}_{\Y,\bu}$ is a reflexive prime difference ideal.

Denote $\norm(\P_i)=M_i\P_i\,(i=0,\ldots,n)$ where $M_i$ are Laurent
difference monomials. Clearly, $\norm(\P_i)=M_i\P_i$ vanishes at
$\theta$\,$(i=0,\ldots,n)$. For any
 $f\in\mathcal{I}_{\Y,\bu}$,
there exists an $M\in\mathbbm{m}$ such that
$Mf\in[\norm(\P_0),\ldots,\norm(P_n)]$. It follows that
$f(\theta)=0$. Conversely, let $f$ be any difference polynomial in
$\Q\{\Y,\bu_0,\ldots,\bu_n\}$ satisfying $f(\theta)=0$. Clearly,
$\norm(\P_0),\norm(\P_1),\ldots,\norm(\P_n)$ constitute  an
ascending chain with $u_{i0}$ as leaders. Let $f_1$ be the
difference remainder of $f$ w.r.t. this ascending chain. Then $f_1$
is free from $u_{i0}\,(i=0,\ldots,n)$ and there exist
$a,s\in\mathbb{N}$ such that
$(\prod_{i=0}^n\prod_{l=0}^s(\sigma^l(M_iM_{i0})))^{a}\cdot f\equiv
f_1,\mod\,[\norm(\P_0),\ldots,$ $\norm(P_n)]$. Clearly,
$f_1(\theta)=0$. Since $f_1\in\Q\{\bu,\Y\}$, $f_1=0$. Thus,
$f\in\mathcal{I}_{\Y,\bu}$. So $\mathcal{I}_{\Y,\bu}$ is a reflexive
prime  difference ideal with a generic zero $\theta$.

Consequently,
$\mathcal{I}_{\bu}=\mathcal{I}_{\Y,\bu}\cap\Q\{\bu_0,\ldots,\bu_n\}$
is a reflexive prime difference ideal with a generic zero
$\zeta=(\zeta_0,u_{01},\ldots,$ $u_{0l_0};\ldots;$
$\zeta_n,u_{n1},\ldots,u_{nl_n})$. From \bref{eq-zeta}, it is clear
that $\dtrdeg\,\Q\langle \zeta\rangle/\Q\leq \sum_{i=0}^nl_i+n$. If
there exist pairs $(i_k,j_k)$\, $(k=1,\ldots,n)$ with $1\leq j_k\leq
l_{i_k}$ and $i_{k_1}\neq i_{k_2}$ $(k_1\neq k_2)$ such that
$\frac{M_{i_1j_1}}{M_{i_10}},\ldots,\frac{M_{i_nj_n}}{M_{i_n0}}$ are
transformally independent over $\Q$, then by Lemma~\ref{lm-special},
$\zeta_{i_1},\ldots,\zeta_{i_n}$ are transformally independent over
$\Q\langle\bu\rangle$. It follows that $\dtrdeg\,\Q\langle
\zeta\rangle/\Q= \sum_{i=0}^nl_i+n$. Thus, $\mathcal{I}_{\bu}$ is of
codimension 1.

Conversely, let us assume that $\mathcal{I}_{\bu}$ is of codimension
1. That is,  $\dtrdeg\,\Q\langle \zeta\rangle/\Q=
\sum_{i=0}^nl_i+n$. We want to show that there exist pairs
$(i_k,j_k)$\, $(k=1,\ldots,n)$ with $1\leq j_k\leq l_{i_k}$ and
$i_{k_1}\neq i_{k_2}$ $(k_1\neq k_2)$
 such that
$\frac{M_{i_1j_1}}{M_{i_10}},\ldots,\frac{M_{i_nj_n}}{M_{i_n0}}$ are
transformally independent over $\Q$. Suppose the contrary,
 i.e.,
$\frac{M_{i_1j_1}(\eta)}{M_{i_10}(\eta)},\ldots,\frac{M_{i_nj_n}(\eta)}{M_{i_n0}(\eta)}$
are transformally dependent for any $n$ different $i_k$ and
$j_k\in\{1,\ldots,l_{i_k}\}$. Since each $\zeta_{i_k}$ is a linear
combination of $\frac{M_{i_kj_k}(\eta)}{M_{i_k0}(\eta)}$
$(j_k=1,\ldots,l_{i_k})$, it follows that
$\zeta_{i_1},\ldots,\zeta_{i_n}$ are transformally dependent over
$\Q\langle\bu\rangle$. Thus, we have $\dtrdeg\,\Q\langle
\zeta\rangle/\Q< \sum_{i=0}^nl_i+n$, a contradiction to the
hypothesis.\qed

Let $[\P_0,\ldots,\P_n]$ be the difference ideal in
$\Q\{\Y,\Y^{-1};\bu_0,\ldots,\bu_n\}$ generated by $\P_i$.  Then we
have
\begin{cor}
$\I_\bu=[\P_0,\P_1,\ldots,\P_n]\cap\Q\{\bu_0,\ldots,\bu_n\}$ is a
reflexive prime difference ideal of codimension one if and only if
$\{\P_i: i=0,\ldots,n\}$ is a Laurent transformally essential
system.
\end{cor}
\proof It is easy to show that
$[\P_0,\P_1,\ldots,\P_n]\cap\Q\{\bu_0,\ldots,\bu_n\}=\mathcal{I}_{\Y,\bu}\cap\Q\{\bu_0,\ldots,\bu_n\}
=\I_\bu.$ And the result is a direct consequence of
Theorem~\ref{th-Mcodim1}. \qed

Now suppose $\{\P_0,\ldots,\P_n\}$ is a Laurent transformally
essential system.
%
Since $\I_{\bu}$ defined in \bref{eq-IU} is a reflexive prime
difference ideal of codimension one, by Lemma~\ref{le-char-codim1},
there exists a unique irreducible difference polynomial $\SR(\bu;
u_{00},\ldots,u_{n0})=\SR(\bu_{0},\ldots,\bu_{n})$
$\in\Q\{\bu_{0},\ldots,\bu_{n}\}$ such that $\SR$ can serve as the
first polynomial in each characteristic set of $\I_{\bu}$ w.r.t. any
ranking endowed on $\bu_{0},\ldots,\bu_{n}$. That is, if $u_{i0}$
appears in $\SR$, then among all the difference polynomials in
$\I_{\bu}$, $\SR$ is of minimal order in $u_{i0}$ and of minimal
degree with the same order.

Now  the definition of sparse difference resultant is given as
follows:
\begin{defn}\label{def-sparse}
The above
$\SR(\bu_{0},\ldots,\bu_{n})\in\Q\{\bu_{0},\ldots,\bu_{n}\}$ is
defined to be the {\em sparse difference resultant} of the Laurent
transformally essential system $\P_0,\ldots,\P_n$, denoted by
$\Res_{\mathcal{A}_0,\ldots,\mathcal{A}_n}$ or
$\Res_{\P_0,\ldots,\P_n}$. When all the $\mathcal{A}_i$ are equal to
the same $\mathcal{A}$, we simply denote it by $\Res_\mathcal{A}$.
\end{defn}

The following lemma gives another description of sparse difference
resultant from the perspective of generic zeros.
\begin{lem}\label{lm-genericresultant}
 Let
$\zeta_i=-\sum_{k=1}^{l_i}u_{ik}\frac{M_{ik}(\eta)}{M_{i0}(\eta)}\,\,(i=0,1,\ldots,n)$
be defined as in equation~(\ref{eq-zeta}), where
 $\eta=(\eta_1,\ldots,\eta_n)$ is a generic zero of
$[0]$ over $\Q\langle\bu\rangle$. Then among all the polynomials in
$\Q\{\bu_{0},\ldots,\bu_{n}\}$ vanishing at $(\bu;\zeta_{0
},\ldots,\zeta_{n })$,
$\SR(\bu_{0},\ldots,\bu_{n})=\SR(\bu;u_{00},\ldots,u_{n0})$ is of
minimal order and degree in each $u_{i0}\,(i=0,\ldots,n)$.
%
%
Equivalently, among all the polynomials in $\mathcal{I}_{\bu}$,
$\SR$ is of minimal order and degree in each
$u_{i0}\,(i=0,\ldots,n)$.
\end{lem}
\proof It is a direct consequence of Theorem~\ref{th-Mcodim1} and
Definition~\ref{def-sparse}.\qed

\begin{rem}
From its definition, the sparse difference resultant can be computed
as follows. With the characteristic set method given in
\cite{gao-dcs}, we can compute a proper irreducible ascending chain
$\mathcal{A}$ which is a characteristic set for the difference
polynomial system
 $\{\P_0,\P_1,\ldots,\P_n\}$ under a ranking such that $u_{ij} < y_k$.
Then the first difference polynomial in $\mathcal{A}$ is the sparse
difference resultant.
This algorithm does not have a complexity analysis. In Section 5, we
will give a single exponential algorithm to compute the sparse
difference resultant.
\end{rem}

We give several examples which will be used throughout the paper.

\begin{exmp}\label{ex-0}
Let $n=1$ and $\P_0=u_{00}+u_{01}y_1^2,\,\,
\P_1=u_{10}y_1^{(1)}+u_{11}y_1.$ Clearly, $\P_0,\P_1$ are Laurent
transformally essential. The sparse difference resultant of
$\P_0,\P_1$ is
$$\SR=u_{10}^2u_{01}u_{00}^{(1)}-u_{11}^2u_{00}u_{01}^{(1)}.$$
\end{exmp}

\begin{exmp} \label{ex-1}
Let $n=2$ and the $\P_i$ have the form
$$\P_i=u_{i0}y_1^{(2)}+u_{i1}y_1^{(3)}+u_{i2}y_2^{(3)}\,(i=0,1,2).$$
It is easy to show that $y_1^{(3)}/y_1^{(2)}$ and
$y^{(3)}_2/y^{(2)}_1$ are transformally independent over $\Q$. Thus,
$\P_0,\P_1,\P_2$ form a Laurent transformally essential system. The
sparse difference resultant is \[\SR=
\Res_{\P_0,\P_1,\P_2}=\left|\begin{array}{lll}
u_{00}&u_{01}&u_{02}\\
u_{10}&u_{11}&u_{12}\\
u_{20}&u_{21}&u_{22}
\end{array}  \right| .\]
 \end{exmp}

The following example shows that for a Laurent transformally
essential system, its sparse difference resultant may not involve
the coefficients of some $\P_i$.
\begin{exmp}\label{ex-2}
Let $n=2$ and the $\P_i$ have the form
$$\P_0=u_{00}+u_{01}y_1y_2,\,
  \P_1=u_{10}+u_{11}y_1^{(1)}y_2^{(1)},\,
\P_2=u_{20}+u_{2 1}y_2.$$
Clearly, $\P_0,\P_1,\P_2$ form a Laurent transformally essential
system. The sparse difference resultant of $\P_0,\P_1,\P_2$ is
$$\SR=u_{00}^{(1)}u_{11}-u_{01}^{(1)}u_{10},$$
which is free from the coefficients of $\P_2.$
\end{exmp}

Example \ref{ex-2} can be used to illustrate the difference between
the differential and difference cases. If $\P_0,\P_1,\P_2$ in
Example \ref{ex-2} are differential polynomials, then the sparse
differential resultant is
$u_{01}^2u_{10}u_{20}^2u_{21}^2-u_{01}u'_{00}u_{11}u_{20}u_{21}^2u'_{20}+u_{00}u'_{01}u_{11}u_{20}u_{21}^2u'_{20}
+u_{01}u_{00}u_{11}u_{20}^2$ $(u'_{21})^2
+u_{00}u_{01}u_{11}u_{21}^2(u'_{20})^2-2u_{01}u_{00}u_{11}u_{20}u_{21}u'_{20}u'_{21}
+u_{01}u'_{00}u_{11}u_{20}^2u'_{21}u_{21}-u_{00}u'_{01}u_{11}$
$u_{21}u'_{21}u_{20}^2$ which contains all the coefficients of
$\P_0,\P_1,\P_2$.

\begin{rem}
When all the $\mathcal{A}_i\,(i=0,\ldots,n)$ are sets of difference
monomials,  unless explicitly mentioned, we always consider $\P_i$
as Laurent difference polynomials. But when we regard $\P_i$ as
difference polynomials, $\Res_{\mathcal{A}_0,\ldots,\mathcal{A}_n}$
is also called the sparse difference resultant of the difference
polynomials $\P_i$ and we call $\P_i$ a {\em transformally essential
system}. In this paper, sometimes we regard $\P_i$ as difference
polynomials where we will highlight it.
\end{rem}

We now define the sparse difference resultant for any set of
specific Laurent difference polynomials over a Laurent transformally
essential system.
For any finite set $\mathcal{A}=\{M_0,M_1,\ldots,M_l\}$ of Laurent
difference monomials in $\Y$, we use
 \begin{equation}\label{eq-L}
  \mathcal {L}(\mathcal {A}) = \big\{ \sum_{i=0}^l a_{i} M_i \big\}
 \end{equation}
to denote the set of all Laurent difference polynomials with
{support} $\mathcal{A}$, where the $a_{i}$ are in some difference
extension field of $\Q$.
%

\begin{defn}\label{def-sparse1}
Let $\mathcal
{A}_i=\{M_{i0},M_{i1},\ldots,M_{il_i}\}\,(i=0,1,\ldots,n)$ be  a
Laurent transformally essential system. Consider $n+1$ Laurent
difference polynomials
$(F_0,F_1,\ldots,F_n)\in\prod_{i=0}^n\mathcal{L}(\mathcal{A}_i)$.
The sparse difference resultant of $F_0,F_1,\ldots,F_n$, denoted as
$\Res_{F_0,\ldots,F_n}$, is obtained by replacing $\bu_i$ with the
corresponding coefficient vector of $F_i$ in
$\Res_{\mathcal{A}_0,\ldots,\mathcal{A}_n}(\bu_{0},$
$\ldots,\bu_{n})$.
\end{defn}

A major unsolved problem about difference resultant is whether $\SR$
defined above contains all the information about the elimination
ideal $\mathcal{I}_{\bu}$ defined in \bref{eq-IU}. More precisely,
we propose the following problem.
\begin{prob}\label{problem-generator}
As shown by Example \ref{ex-a}, the characteristic set for a
reflexive prime difference ideal of codimension one could contain
more than one elements.
Let $\mathcal{I}_{\bu}$ be the ideal defined in \bref{eq-IU}. Then
$\mathcal{I}_{\bu}$ is a reflexive prime difference ideal of
codimension one and
\begin{eqnarray}\label{ex-I}
\mathcal{I}_{\bu}=
\mathcal{I}_{\Y,\bu}\cap\Q\{\bu_0,\ldots,\bu_n\}=\sat(\SR,R_1,\ldots,R_m),
\end{eqnarray}
where $\SR$ is the sparse difference resultant of
$\{\P_0,\ldots,\P_n\}$ and $\SR,R_1,\ldots,R_m$ is a characteristic
set of $\mathcal{I}_{\bu}$. We conjecture that $m=0$, or
equivalently $\mathcal{I}_{\bu}=\sat(\SR)$. If this is valid, then
better properties can be shown for sparse difference resultant as we
will explain later.
It is easy to check that for Examples \ref{ex-0}, \ref{ex-1}, and
\ref{ex-2}, $\mathcal{I}_{\bu}=\sat(\SR)$.
\end{prob}

\subsection{A criterion for Laurent transformally essential system in terms of supports}

Let $\A_i\,(i=0,\ldots,n)$ be finite sets of  Laurent difference
monomials. According to Definition~\ref{def-tdes}, in order to check
whether they form a Laurent transformally essential system, we need
to check whether there exist  $M_{ik_i}, M_{ij_i}\in\A_i
(i=0,\ldots,n)$ such that $\dtrdeg\, \Q\langle M_{0k_0}/M_{0j_0},$
$\ldots,M_{nk_n}/M_{nj_n}\rangle/\Q=n$. This can be done with the
difference characteristic set method given in paper \cite{gao-dcs}.
In this section, a criterion will be given to check whether a
Laurent difference system is essential in terms of their supports,
which is conceptually and computationally simpler than the naive
approach based on the characteristic set method.

Let $B_i = \prod_{j=1}^n\prod_{k = 0}^s
(y_{j}^{(k)})^{d_{ijk}}\,(i=1,\ldots,m)$ be $m$ Laurent difference
monomials. We now introduce a new algebraic indeterminate $x$ and
let
$$d_{ij} = \sum_{k=0}^{s} d_{ijk}x^{k}\, (i=1,\ldots,m ,j=1,\ldots,n)$$
be univariate polynomials in $\mathbb{Z}[x]$. If $\ord(B_i,y_j) =
-\infty$, then set $d_{ij}=0$.
The vector  $(d_{i1},d_{i2},$ $\ldots,d_{in})$ is called the {\em
symbolic support vector} of $B_i$.
The matrix $M=(d_{ij})_{m\times n}$
is called the {\em symbolic support matrix} of $B_1,\ldots, B_m$.

Note that there is a one-to-one correspondence between Laurent
difference monomials and their symbolic support vectors, so we will
not distinguish these two concepts in case there is no confusion.
The same is true for a set of Laurent difference monomials and its
symbolic support matrix.

\begin{defn}
A  matrix $M=(d_{ij})_{m\times n}$ over $\Q[x]$ is called   {\em
normal upper-triangular of rank $r$} if for each $i\leq r$,
$d_{ii}\neq0$ and $d_{i,i-k}=0\,(1\leq k\leq i-1)$, and the last
$m-r$ rows are zero vectors.
\end{defn}
A normal upper-triangular matrix is of the following form:
\[
\left(
\begin{array}{llllll}
a_{11}&*&\cdots&*&\cdots&*\\
0&a_{22}&\cdots&*&\cdots&*\\
\vdots&\vdots&\ddots&&&\vdots\\
0&0&\cdots&a_{rr}&\cdots&*\\
0&0&\cdots&0&\cdots& 0\\
\multicolumn{6}{c}{\dotfill}\\
0&0&\cdots&0&\cdots& 0\\
\end{array}
\right)
\]
\begin{defn}
A set of Laurent difference monomials $B_1, B_2, \ldots, B_m$ is
said to be in {\em $r$-upper-triangular} form if its symbolic
support matrix $M$ is a normal upper triangular matrix of rank $r$.
\end{defn}

The following lemma shows that it is easy to compute the difference
transcendence degree of a set of Laurent difference monomials in
upper-triangular form.

\begin{lem} \label{lm-triangular}
Let $B_1, \ldots, B_m$ be a set of Laurent difference monomials in
$r$-upper-triangular form. Then $\dtrdeg\,\Q\langle B_1, \ldots,
B_m\rangle/\Q=r.$
\end{lem}
\proof From the structure of the symbolic support matrix,
 $B_i=\prod _{j=i}^n\prod _{k\geq0}(y_j^{(k)})^{d_{ijk}}\,(i=1,\ldots,r)$ with $\ord(B_i,y_i)\geq0$ and $B_{r+1}=\cdots=B_m=1$.
 Let  $B'_i=\prod _{j=i}^r\prod _{k\geq0}(y_j^{(k)})^{d_{ijk}}$. Then
 \begin{eqnarray}
&&\dtrdeg\,\Q\langle B_1, \ldots, B_m\rangle/\Q \nonumber\\&=&\dtrdeg\,\Q\langle B_1, \ldots, B_r\rangle/\Q \nonumber\\
&\geq& \dtrdeg\,\Q\langle y_{r+1},\ldots,y_n\rangle\langle B_1, \ldots, B_r\rangle/\Q\langle y_{r+1},\ldots,y_n\rangle \nonumber\\
&=&\dtrdeg\,\Q\langle B'_1, \ldots, B'_r\rangle/\Q.   \nonumber
\end{eqnarray}
So it suffices to prove $\dtrdeg\,\Q\langle B'_1, \ldots,
B'_r\rangle/\Q=r$.

 If $r=1$, $B'_1$ is a nonconstant Laurent difference monomial in $y_1$, so $\dtrdeg\,\Q\langle B'_1 \rangle/\Q=1.$
 Suppose we have proved for the case $r-1$. Let  $B''_i=\prod _{j=i}^{r-1}\prod _{k\geq0}(y_j^{(k)})^{d_{ijk}}$, then
 by the hypothesis, $\dtrdeg\,\Q\langle B''_1, \ldots, B''_{r-1}\rangle/\Q=r-1$.  Since $B'_r \in \Q\{y_r\}$, we have
  \begin{eqnarray}
r&\geq&\dtrdeg\,\Q\langle B'_1, \ldots, B'_r\rangle/\Q\nonumber\\
&=&\dtrdeg\,\Q\langle B'_r \rangle/\Q+\dtrdeg\,\Q\langle B'_1, \ldots, B'_r\rangle/\Q\langle B'_r \rangle  \nonumber\\
&\geq& 1+ \dtrdeg\,\Q\langle y_{r} \rangle\langle B'_1, \ldots, B'_{r-1}\rangle/\Q\langle y_{r} \rangle \nonumber\\
&=&1+\dtrdeg\,\Q\langle B''_1, \ldots, B''_{r-1}\rangle/\Q=r.
\nonumber
\end{eqnarray}
 So $\dtrdeg\,\Q\langle B_1, \ldots, B_m\rangle/\Q=r.$
\qed

In the following, we will show that each set of Laurent difference
monomials can be transformed to an upper-triangular set with the
same difference transcendence degree. Here we use three types of
elementary matrix transformations. For a matrix $M$ over $\Q[x]$,
\begin{itemize}
\item
Type 1 operations consist of interchanging two rows of $M$, say the
$i$-th and $j$-th rows, denoted by $r[i,j]$;

\item Type 2 operations consist of adding an $f(x)$-multiple of the $j$-th
row to the $i$-th row, where $f(x)\in\Q[x]$, denoted by
$[i+j(f(x))]$;

\item
Type 3 operations consist of interchanging  two columns, say the
$i$-th and $j$-th columns, denoted by $c[i,j]$.
\end{itemize}
In this section, by {\em elementary transformations}, we mean the
above three types of transformations.

Let $B_1,\ldots,B_m$ be Laurent difference monomials and $M$ their
symbolic support matrix. Then the above three types of elementary
transformations of $M$ correspond to certain transformations of the
difference monomials. Indeed, interchanging the $i$-th and the
$j$-th rows  of $M$ means interchanging $B_i$ and $B_j$, and
interchanging the $i$-th and the $j$-th columns of $M$ means
interchanging $y_i$ and $y_j$ in $B_1,\ldots,B_m$(or in the variable
order). Multiplying the $i$-th row of $M$ by a polynomial
$f(x)=a_dx^d+a_{d-1}x^{d-1}+\cdots+a_0\in\Q[x]$ and adding the
result to the $j$-th row means changing $B_j$ to
$\prod_{k=0}^d(\sigma^kB_i)^{a_k}B_j$.

\begin{lem} \label{lm-keeptransdegree}
Let $B_1,\ldots,B_m$ be Laurent difference monomials and
$C_1,\ldots,C_m$ obtained by successive elementary transformations
defined above. Then $\dtrdeg\,\Q\langle B_1, \ldots, B_m\rangle/\Q$
$=\dtrdeg\, \Q\langle C_1,$ $\ldots, C_m\rangle/\Q$.
\end{lem}
\proof It suffices to show that Type 2 operations do not change  the
difference transcendence degree. That is, for $\sum_{i=0}^da_{
i}x^i\in\Q[x],$ $\dtrdeg\,\Q\langle B_1, B_2 \rangle/\Q=\dtrdeg\,
\Q\langle B_1,$ $\prod_{k=0}^d(\sigma^kB_1)^{a_k}B_2 \rangle/\Q$.

Suppose $a_i=p_i/q$ where $p_i,q\in\mathbb{Z}^{\ast }$. Then,
clearly, $\dtrdeg\,\Q\langle B_1\rangle/\Q=\dtrdeg\, \Q\langle
\prod_{k=0}^d$ $(\sigma^kB_1)^{p_k}\rangle/\Q$.
Thus,
 $\dtrdeg\, \Q\langle B_1, $ $\prod_{k=0}^d(\sigma^kB_1)^{a_k}B_2 \rangle/\Q
 =\dtrdeg\, \Q\langle $ $\prod_{k=0}^d(\sigma^kB_1)^{p_k}, $ $\prod_{k=0}^d$ $(\sigma^kB_1)^{p_k}B_2^q \rangle/\Q
 =\dtrdeg\, \Q\langle \prod_{k=0}^d(\sigma^kB_1)^{p_k} ,$ $ B_2^q \rangle/\Q
 =\dtrdeg\, \Q\langle B_1 , B_2 \rangle/\Q
 $.
 \qed

 \begin{thm} \label{th-symbolicmonomial}
 Let $B_1,\ldots,B_m$ be a set of Laurent difference monomials with  symbolic support matrix  $M$.
 Then $\dtrdeg\,\Q\langle B_1,
\ldots, B_m\rangle/\Q=\rk(M)$.
 \end{thm}
 \proof By Lemma~\ref{lm-triangular} and Lemma~\ref{lm-keeptransdegree}, it suffices to show that $M$ can be
 reduced to a normal upper-triangular matrix by performing a series of elementary transformations.
 This can be done since $\Q[x]$ is an Euclidean domain.

Suppose $M=(d_{ij})\neq\textbf{0}_{m\times n}$ and we denote the new
matrix obtained after performing elementary transformations also by
$M$. Firstly, perform Type 1 and Type 3 operations when necessary to
make $d_{11}\neq0$ have the minimum degree among all $d_{ij}$.
Secondly, try to use $d_{11}(x)$ to reduce other elements in the
first column to $0$ by performing Type 2 operations.
Let $d_{k1}\ne0$ and suppose $d_{k1}(x)=d_{11}(x)q(x)+r(x)$ where
$\deg(r(x))<\deg(d_{11}(x))$. Performing the transformation
$[k+1(-q(x))]$ and then the transformation  $r[1,k]$ if $r(x)\ne0$,
we obtain a new matrix in which the degree of $d_{11}$ strictly
decreases. Repeat this process when necessary, then after a finite
number of steps, we obtain a new matrix $M$ such that $d_{k1}(x)=0$
for $k>1$. That is,
 \[ M=
\left(
\begin{array}{ll}
d_{11}&*\\
\textbf{0}&M_1
\end{array}
\right). \] Now we repeat the above process for $M_1$ and whenever
Type 3 operations are performed for $M_1$, we assume the same
transformations are performed for the whole matrix $M$. In this way,
after a finite number of steps, we obtain a normal upper-triangular
matrix $M$. \qed

\begin{rem}
In the proof of Theorem~\ref{th-symbolicmonomial},  the Euclidean
algorithm plays a crucial role. That is why we work with $\Q[x]$,
even if the symbolic support matrix of $B_1,\ldots,B_m$ is a matrix
over $\mathbb{Z}[x]$.
\end{rem}

\begin{exmp}
Let $B_1=y_1y_2$ and $B_2=y_1^{(a)}y_2^{(b)}$. Then the symbolic
support matrix of $B_1$ and $B_2$ is $ M= \left(
\begin{array}{cc}
1&1\\
x^a&x^b
\end{array}
\right).$ Then $\rank(M)=\left\{
\begin{array}{ll}
1& \text{if}\,a=b\\
2& \text{if}\,a\neq b.
\end{array}
\right.$ Thus, by Theorem~\ref{th-symbolicmonomial}, if $a\neq b$,
$B_1$ and $B_2$ are transformally independent over $\Q$. Otherwise,
they are transformally dependent over $\Q$.
\end{exmp}
%
%

We now extend Theorem \ref{th-symbolicmonomial} to generic
difference polynomials in \bref{eq-sparseLaurent}. Let $I \subseteq
\{0,\ldots,n\}$  and for any $i\in I$, let $\beta_{ik}$ be the
symbolic support vector of $M_{ik}/M_{i0}$. Then the vector
 $$w_i = \sum_{k=0}^{l_i} u_{ik}\beta_{ik}$$
 is called the {\em symbolic support vector} of $\P_i$ and the matrix $M_I$ whose rows are
$w_i$ for $i\in I$ is called the {\em symbolic support matrix} of
$\P_i$ for $i\in I$.  Similar to Theorem 4.17 in \cite{li}, we have
\begin{lem}\label{lm-rank111}
Use the notations introduced above. We have
$\dtrdeg\,\Q\langle\cup_{i\in I} \bu_i\rangle\langle
\P_i/M_{i0}\,:\, i\in I\rangle/\Q\langle\cup_{i\in I}\bu_i\rangle $
$= \rank(M_I)$, where $\bu_i$ $=(u_{i0},\ldots,u_{il_i})$.
\end{lem}
%

Now, we have the following criterion for Laurent transformally
essential system.
\begin{thm}\label{th-cri}
Consider the set of generic Laurent difference polynomials defined
in \bref{eq-sparseLaurent}. The following three conditions are
equivalent.
\begin{enumerate}
\item $\P_0,\ldots,\P_n$ form a Laurent transformally essential system.
\item There exist $M_{ik_i}\,(i=0,\ldots,n)$ with $1\leq k_i\leq l_{i}$ such that the symbolic support matrix of
 $M_{0k_0}/M_{00},\ldots,M_{nk_n}/M_{n0}$
 is of rank $n$.
\item The rank of $M_I$ is equal to $n$, where $I=
=\{0,1,\ldots,n\}$.
\end{enumerate}
\end{thm}
\proof The equivalence of 1) and 2) is a direct consequence of
Theorem~\ref{th-symbolicmonomial} and Definition \ref{def-tdes}. The
equivalence of 1) and 3) follows from Lemma \ref{lm-rank111}.\qed
%

Both Theorem \ref{th-symbolicmonomial} and Theorem \ref{th-cri} can
be used to check whether a system is transformally essential.
\begin{exmp}\label{ex-22} Continue from Example \ref{ex-2}.
Let $B_0 = M_{01}/M_{00}=y_1y_2,$ $
B_1=M_{11}/M_{10}=y_1^{(1)}y_2^{(1)},$ and $B_2=M_{21}/M_{20}=y_2$.
Then the symbolic support matrix for $\{B_0,B_2\}$ is
$ M= \left(
\begin{array}{cc}
1&1\\
0&1
\end{array}
\right)$. We have $\rank(M)=2$ and by
Theorem~\ref{th-symbolicmonomial}, the system
$\P=\{\P_0,\P_1,\P_2\}$ is transformally essential.
Also, the symbolic support matrix for $\P$ is
$M_\P = \left(
\begin{array}{cc}
 u_{01}&u_{01}\\
 u_{11}x&u_{11}x\\
 0&u_{21}
\end{array}
\right)$. We have $\rank(M_\P)=2$ and by Theorem \ref{th-cri}, $\P$
is transformally essential.
\end{exmp}

\vskip10pt We will end this section by introducing a new concept,
namely super-essential systems, through which one can identify
certain $\P_i$ such that their coefficients will not occur in the
sparse difference resultant. This will lead to the simplification in
the computation of the resultant.
Let $\TT\subset\{0,1,\ldots,n\}$. We denote by $\P_\TT$ the Laurent
difference polynomial set consisting of $\P_i\,( i\in\TT)$, and
$M_{\P_\TT}$ its symbolic support matrix. For a subset
$\TT\subset\{0,1,\ldots,n\}$, if $\card(\TT) = \rank(M_{\P_\TT})$,
then $\P_\TT$, or $\{\mathcal{A}_i:\,i\in\TT\}$, is called a {\em
transformally independent set}.
\begin{defn}
Let $\TT\subset\{0,1,\ldots,n\}$. Then we call $\TT$ or $\P_\TT$
{\em super-essential} if the following conditions hold: (1)
$\card(\TT) - \rank(M_{\P_\TT}) = 1$ and (2) $\card(\JJ) =
\rank(M_{\P_\JJ})$ for each proper subset $\JJ$ of $\TT$.
\end{defn}

Note that super-essential systems are the difference analogue of
essential systems introduced in paper \cite{sturmfels2} and also
that of rank essential systems introduced in \cite{li} . Using this
definition, we have the following property, which is similar to
Corollary 1.1 in \cite{sturmfels2}.
\begin{thm}\label{th-rankessential}
If $\{\P_0,\ldots,\P_n\}$ is a Laurent transformally essential
system, then for any $\TT\subset\{0, 1,\ldots, n \}$, $\card(\TT) -
\rank(M_{\P_\TT}) \leq 1$ and there exists a unique $\TT$ which is
super-essential. In this case, the sparse difference resultant of
$\P_0,\ldots,\P_n$ involves only the coefficients of
$\P_i\,(i\in\TT)$.
\end{thm}
\proof Since $n = \rank(M_{\P}) \leq \rank(M_{\P_\TT}) + \card(\P) -
\card(\P_\TT) = n+1 + \rank(M_{\P_\TT}) - \card(\TT)$, we have
$\card(\TT) - \rank(M_{\P_\TT}) \leq 1$.
Since $\card(\TT) - \rank(M_{\P_\TT})\ge 0$, for any $\TT$, either
$\card(\TT) - \rank(M_{\P_\TT})=0$ or $\card(\TT) -
\rank(M_{\P_\TT})=1$. From this fact, it is easy to show the
existence of a super-essential set $\TT$.
For the uniqueness, we assume that there exist two subsets
$\TT_1,\TT_2\subset\{1,\ldots,m\}$ which are super-essential. Then,
we have
\begin{equation*}\begin{array}{lll}
\rank(M_{\P_{\TT_1\cup\TT_2}}) &\leq& \rank(M_{\P_{\TT_1}}) + \rank(M_{\P_{\TT_2}}) - \rank(M_{\P_{\TT_1\cap\TT_2}})\\
&= & \card(\TT_1) - 1 + \card(\TT_2) -1 - \card(\TT_1\cap\TT_2)
\\&=& \card(\TT_1\cup\TT_2) - 2,
\end{array}
\end{equation*}
which is a contradiction.

Let $\TT$ be a super-essential set. Similar to the proof of
Theorem~\ref{th-Mcodim1}, it is easy to show that
$[\P_i]_{i\in\TT}\cap\Q\{\bu_i\}_{i\in\TT}$ is of codimension one,
which means that the sparse difference resultant of
$\P_0,\ldots,\P_n$ only involves the coefficients of
$\P_i\,({i\in\TT})$. \qed

\begin{rem}
If $\P_\TT$ is the super-essential subsystem of a Laurent
transformally essential system $\P=\{\P_0,\ldots,\P_n\}$, then
clearly
$[\P_\TT]\cap\Q\{\bu_i:i\in\TT\}=[\P]\cap\Q\{\bu_i:i\in\TT\}=\sat(\SR,\ldots)$.
For convenience, sometimes we will not distinguish $\P$ and $\P_\TT$
and also call $\SR$ the sparse difference resultant of $\P_\TT.$
\end{rem}

Using this property, one can determine which polynomial is needed
for computing the sparse difference resultant, which  will
eventually reduce the computation complexity.
\begin{exmp} Continue from Example~\ref{ex-2}.
It is easy to show that $\P=\{\P_0,\P_1,\P_2\}$ is a Laurent
transformally essential system and $\P_0,\P_1$ constitute a
super-essential system. Recall that the sparse difference resultant
of $\P $ is
  free from the coefficients of $\P_2.$
\end{exmp}

\section{Basic properties of sparse difference resultant}\label{sec-property}
In this section, we will prove some basic properties for the sparse
difference resultant.

\subsection{Sparse difference resultant is transformally homogeneous }
We first introduce the concept of transformally homogeneous
polynomials.
\begin{defn} \label{d-homogeneous}
A difference polynomial $f \in \mathcal {F}\{y_{0},\ldots,y_{n}\}$
is called {\em transformally homogeneous} if for a new difference
indeterminate $\lambda$, there exists a difference monomial $
M(\lambda)$ in $\lambda$ such that $f(\lambda y_{0}, \ldots,\lambda
y_{n})=M(\lambda)p(y_{0}, \ldots,y_{n})$. If $\deg(M(\lambda))=m$,
$f$ is called transformally homogeneous of degree $m$.
\end{defn}

The difference analogue of Euler's theorem related to homogeneous
polynomials is valid.
\begin{lem}\label{lm-dhomo}\,
 $f \in \mathcal{F}\{y_{0},y_{1},\ldots,y_{n}\}$ is transformally
homogeneous if and only if for each $r\in\mathbb{N}_0$, there exists
$m_r\in\mathbb{N}_0$ such that \[\sum_{i=0}^{n} y_{i}^{(r)}
\frac{\partial f(y_{0},\ldots,y_{n})}{\partial y_{i}^{(r)} } =
m_rf.\]
That is, $f$ is transformally homogeneous if and only if $f$ is
homogeneous in  $\{y_1^{(r)},\ldots,y_n^{(r)}\}$ for any
$r\in\mathbb{N}_0$.
\end{lem}
\proof ``$\Longrightarrow$" Denote $\Y=(y_0,\ldots,y_n)$
temporarily. Suppose $f$ is transformally homogeneous. That is,
there exists a difference monomial
$M(\lambda)=\prod_{r=0}^{r_0}(\lambda^{(r)})^{m_r}$ such that
$f(\lambda\Y)=M(\lambda)f(\Y)$. Then
$\sum_{i=0}^ny_i^{(r)}\frac{\partial f}{\partial y_{i}^{(r)}
}(\lambda\Y)=\sum_{i=0}^n\frac{\partial f}{\partial y_{i}^{(r)}
}(\lambda\Y)\frac{\partial (\lambda y_i)^{(r)}}{\partial
\lambda^{(r)} }=\frac{\partial f(\lambda\Y)}{\partial\lambda^{(r)}
}=\frac{\partial M(\lambda)f(\Y)}{\partial\lambda^{(r)}
}=\frac{m_rM(\lambda)}{\lambda^{(r)}}f(\Y)$. Substitute $\lambda=1$
into the above equality, we have
$\sum_{i=0}^ny_i^{(r)}\frac{\partial f}{\partial y_{j}^{(r)}}$
$=m_rf$.

``$\Longleftarrow$" Suppose $\ord(f,\Y)=r_0.$ Then for each $r\leq
r_0,$ $\lambda^{(r)}\frac{\partial
f(\lambda\Y)}{\partial\lambda^{(r)}}$
$=\lambda^{(r)}\sum\limits_{i=0}^ny_i^{(r)}\frac{\partial
f}{\partial y_{i}^{(r)} }(\lambda\Y)$ $=\sum_{i=0}^n(\lambda
y_i)^{(r)}\frac{\partial f}{\partial y_{i}^{(r)} }(\lambda\Y)$
 $=m_rf(\lambda\Y)$.
So $f(\lambda\Y)$ is homogeneous of degree $m_r$ in $\lambda^{(r)}$.
 Thus, $f(\lambda\Y)=f(\lambda y_0,\ldots,\lambda y_n;\lambda^{(1)}y_0^{(1)},\ldots,$
  $\lambda^{(1)}y_n^{(1)};\ldots;\lambda^{(r_0)}y_0^{(r_0)},\ldots,\lambda^{(r_0)}y_n^{(r_0)})
  =$\\ $\prod\limits_{r=0}^{r_0}(\lambda^{(r)})^{m_r}f(\Y)$. Thus,
  $f$ is transformally homogeneous. \qed

Sparse difference resultants have the following property.
\begin{thm}\label{th-homo}
The sparse difference resultant is transformally homogeneous in each
$\bu_i$ which is the coefficient set of  $\P_i$.
\end{thm}
\proof Suppose $\ord(\SR,\bu_i)=h_i\geq0$. Follow the notations used
in Theorem~\ref{th-Mcodim1}. By Lemma~\ref{lm-genericresultant},
$\SR(\bu;\zeta_0,\ldots,\zeta_n)=0$. Differentiating this identity
w.r.t. $u_{ij}^{(k)}\,(j=1,\ldots,l_i)$ respectively, due to
\bref{eq-zeta} we have
\begin{equation}\label{eq-partialdiff}\overline{\frac{\partial
\SR}{\partial u_{i j}^{(k)}}}+\overline{\frac{\partial \SR}{\partial
u_{i 0}^{(k)}}}\big(-\frac{M_{ij}(\eta)}{M_{i0}(\eta)}\big)^{(k)}=0.
\end{equation}
In the above equations,    $\overline{\frac{\partial \SR}{\partial
u_{i j}^{(k)}}}$ $(k=0,\ldots,h_i; j=0,\ldots,l_i)$ are  obtained by
replacing $u_{i0}$ by $\zeta_{i}\,(i=0, 1,    \ldots,    n)$ in each
$\frac{\partial \SR}{\partial u_{i j}^{(k)}}$ respectively.

Multiplying \bref{eq-partialdiff} by $u_{i j}^{(k)} $ and for $j$
from 1 to $l_i$, adding them together, we get
$\zeta_i^{(k)}\overline{\frac{\partial \SR}{\partial u_{i
0}^{(k)}}}+\sum_{j=1}^{l_i}u_{i j}^{(k)} \overline{\frac{\partial
\SR}{\partial u_{i j}^{(k)}}}=0.$ So the difference polynomial
$f_k=\sum_{j=0}^{l_i}u_{i j}^{(k)}  \frac{\partial \SR}{\partial
u_{i j}^{(k)}} $ vanishes at $(\zeta_0,\ldots,\zeta_n)$. Since
$\ord(f_k,u_{i0})\leq\ord(\SR,u_{i0})$ and $\deg(f_k)=\deg(\SR)$, by
Lemma~\ref{lm-genericresultant}, there exists an $m_k\in\mathbb{Z}$
such that $f_k=m_k\SR$. Thus, by  Lemma~\ref{lm-dhomo}, $\SR$ is
transformally homogeneous in $\bu_i.$ \qed

\subsection{Condition for existence of nonzero solutions}
In  this section, we will first give a condition for a system of
Laurent difference polynomials to have nonzero solutions in terms of
sparse difference resultant, and then study the structures of
nonzero solutions.

To be more precise, we first introduce some notations. Let
$\mathcal{A}=\{M_0,M_1,\ldots,M_l\}$ be a Laurent monomial set.
Then, there is a one to one correspondence between
$\mathcal{L}(\mathcal{A})$ defined in \bref{eq-L} and $\ee^{l+1}$
where $\ee$ is some difference extension field of $\Q$.
For $F=\sum_{i=0}^l c_i M_i \in\mathcal{L}(\mathcal{A})$ where
$c_i\in\ee$, denote the coefficient vector of $F$ by
$\mathbb{C}(F)=(c_0,\ldots,c_l)\in\ee^{l+1}$. Conversely, for any
$\textbf{c}=(c_0,\ldots,c_l)\in\ee^{l+1}$, denote the corresponding
Laurent difference polynomial by
$\mathbb{L}(\textbf{c})=\sum_{i=0}^l c_i M_i$.

Let $\mathcal{A}_0,\ldots,\mathcal{A}_n$ be a Laurent transformally
essential system of Laurent monomial sets. By
$(F_0,\ldots,F_n)\in\mathcal{L}(\mathcal{A}_0)\times\cdots\times\mathcal{L}(\mathcal{A}_n)$,
we always mean that there exists a common difference extension field
$\ee$ such that $\mathbb{C}(F_i)\in\ee^{l_i+1}\,(i=0,\ldots,n)$.
Clearly, each element
$(F_0,\ldots,F_n)\in\mathcal{L}(\mathcal{A}_0)\times\cdots\times\mathcal{L}(\mathcal{A}_n)$
can be represented by one and only one element
$(\mathbb{C}(F_0),\ldots,\mathbb{C}(F_n))\in
\hat{\ee}=\ee^{l_0+1}\times\cdots\times\ee^{l_n+1}$. Let $\mathcal
{Z}_0(\mathcal{A}_0,\ldots,\mathcal{A}_n)$ be the set consisting of
points $(\bv_0,\ldots,\bv_n)\in\hat{\ee}$ such that the
corresponding $\mathbb{L}(\bv_i)=0\,(i=0,\ldots,n)$ have nonzero
solutions. That is,
\begin{eqnarray}\label{eq-zA} \quad&\quad&\mathcal
{Z}_0(\mathcal{A}_0,\ldots,\mathcal{A}_n)=\bigcup_{\ee}\{(\bv_0,\ldots,\bv_n)\in\hat{\ee}:
\mathbb{L}(\bv_0)=\cdots=\mathbb{L}(\bv_n)=0 \,
\nonumber\\&\quad&\qquad\qquad \qquad\qquad\quad\text{ have a common
nonzero solution}\}.
\end{eqnarray}
Note that the sparse resultant $\SR(\bu_0,\ldots,\bu_n)$ has $L=
\sum_{i=0}^n (l_i+1)$ variables. In this section, each element
$\bv\in\ee^L$ is naturally treated as an element
$\bv=(\bv_0,\ldots,\bv_n)\in\ee^{l_0+1}\times\cdots\times\ee^{l_n+1}$
and $\SR(\bv)=\SR(\bv_0,\ldots,\bv_n)$. In this way,
${Z}_0(\mathcal{A}_0,\ldots,\mathcal{A}_n)$ and
$\V\big(\Res_{\mathcal{A}_0,\ldots,\mathcal{A}_n}\big)$ are in the
same affine space $\ee^L$ for any $\ee$.

The following result shows that the vanishing of sparse difference
resultant gives a necessary condition for the existence of nonzero
solutions.

\begin{lem} \label{le-necessarycondition}
${Z}_0(\mathcal{A}_0,\ldots,\mathcal{A}_n)\subseteq
\V\big(\Res_{\mathcal{A}_0,\ldots,\mathcal{A}_n}\big)$.
\end{lem}
\proof Let $\P_0,\ldots,\P_n$ be a generic Laurent transformally
essential system corresponding to
$\mathcal{A}_0,\ldots,\mathcal{A}_n$ with coefficient vectors
$\bu_0,\ldots,\bu_n$. By Definition~\ref{def-sparse},
$\Res_{\mathcal{A}_0,\ldots,\mathcal{A}_n}\in[\P_0,\ldots,\P_n]$
$\cap\,\Q\{\bu_0,\ldots,\bu_n\}$. For
  any point $(\bv_{0},\ldots,\bv_{n})\in
  {Z}_0(\mathcal{A}_0,\ldots,\mathcal{A}_n)$,
   let $(\overline{\P}_{0},\ldots,$ $\overline{\P}_{n})\in\mathcal{L}(\mathcal{A}_0)\times\cdots\times
\mathcal{L}(\mathcal{A}_n)$ be the difference polynomial system
represented by $(\bv_{0},\ldots,\bv_{n})$. Since
$\overline{\P}_{0},\ldots,\overline{\P}_{n}$ have a nonzero common
solution, $\Res_{\mathcal{A}_0,\ldots,\mathcal{A}_n}$ vanishes at
$(\bv_0,\ldots,\bv_n)$.
 \qed

\begin{exmp}\label{ex-sp12}
Continue from Example~\ref{ex-1}. Suppose $\F=\Q(x)$ and $\sigma
f(x)$ $= f(x+1)$. In this example, we have
$\Res_{\P_0,\P_1,\P_2}\ne0$. But $y_1=0, y_2=0$ constitute a zero
solution of $\P_0=\P_1=\P_2=0$.
This shows that Lemma \ref{le-necessarycondition} is not correct if
we do not consider nonzero solutions.
This example also shows why we need to consider nonzero difference
solutions, or equivalently why we consider Laurent difference
polynomials instead of the usual difference polynomials.
\end{exmp}

The following theorem shows that a particular principal component 
of the sparse difference resultant gives a sufficient and necessary
condition for a Laurent transformally essential system to have
nonzero solutions in certain sense.

\begin{thm}\label{th-nscond}
Let $ \mathcal{I}_{\bu}=
[\P_0,\ldots,\P_n]\cap\Q\{\bu_0,\ldots,\bu_n\}=\sat(\Res_{\mathcal{A}_0,\ldots,\mathcal{A}_n},
R_1, \ldots, R_m)$ as defined in \bref{ex-I}. Let
$\overline{\mathcal {Z}_0(\mathcal{A}_0,\ldots,\mathcal{A}_n)}$ be
the Cohn topological closure\footnote{For definition,
see\cite{wibmer}.}
 of $\mathcal {Z}_0(\mathcal{A}_0,\ldots,\mathcal{A}_n)$.
Then
$\overline{\mathcal{Z}_0(\mathcal{A}_0,\ldots,\mathcal{A}_n)}=\V\big(\sat(\Res_{\mathcal{A}_0,\ldots,\mathcal{A}_n},
R_1, \ldots, R_m)\big)$.
\end{thm}
\proof Similarly to the proof of Lemma~\ref{le-necessarycondition},
we can show that $\mathcal{I}_{\bu}$ vanishes at $\mathcal
{Z}_0(\mathcal{A}_0,\ldots,\mathcal{A}_n)$. So
$\overline{\mathcal{Z}_0(\mathcal{A}_0,\ldots,\mathcal{A}_n)}\subseteq\V\big(\sat(\Res_{\mathcal{A}_0,\ldots,\mathcal{A}_n},R_1,\ldots,R_m)\big)$.

For the other direction, follow the notations in the proof of
Theorem~\ref{th-Mcodim1}. By Theorem~\ref{th-Mcodim1},
$[\norm(\P_0),\ldots,\norm(\P_n)]:\mathbbm{m}$ is a reflexive prime
difference ideal with a generic point $(\eta,\zeta)$ where
$\eta=(\eta_1,\ldots,\eta_n)$ is a generic point of $[0]$ over
$\qq\langle(u_{ik})_{i=0,\ldots,n;k\neq0}\rangle$ and
$\zeta=(\zeta_0,u_{01},\ldots,$ $u_{0l_0};\ldots;$
$\zeta_n,u_{n1},\ldots,u_{nl_n})$.  Let $(F_0,\ldots,F_n)\in
\mathcal{L}(\mathcal{A}_0)\times\cdots\times
\mathcal{L}(\mathcal{A}_n)$ be a set of Laurent difference
polynomials represented by $\zeta$. Clearly, $\eta$ is a nonzero
solution of $F_i=0$. Thus, $\zeta\in\mathcal
{Z}_0(\mathcal{A}_0,\ldots,\mathcal{A}_n)\subset\overline{\mathcal
{Z}_0(\mathcal{A}_0,\ldots,\mathcal{A}_n)}$. Since
 $\zeta$ is a generic point of
$\sat(\Res_{\mathcal{A}_0,\ldots,\mathcal{A}_n},R_1,\ldots,R_m)$. It
follows that
$\V\big(\sat(\Res_{\mathcal{A}_0,\ldots,\mathcal{A}_n},R_1,\ldots,R_m)\big)\subseteq\overline{\mathcal
{Z}_0(\mathcal{A}_0,\ldots,\mathcal{A}_n)}$. As a consequence, the
theorem is proved. \qed

\begin{rem}
If Problem~\ref{problem-generator} can be solved positively, then
the vanishing of $\sat(\SR)$ also gives a sufficient condition for
$\P_0=\cdots=\P_n=0$ to have a nonzero solution in the sense of Cohn
topological closure. That is,
$\overline{\mathcal{Z}_0(\mathcal{A}_0,\ldots,\mathcal{A}_n)}=\V\big(\sat(\SR)\big)$.
\end{rem}

The following example shows that the vanishing of the sparse
difference resultant is not a sufficient condition for the given
system to have common nonzero solutions.
\begin{exmp}\label{ex-01}
Continue from Example~\ref{ex-0}. Suppose $\overline{\P}_0=y_1^2-4,
\overline{\P}_1=y_1^{(1)}+y_1$. Clearly,
$\Res(\overline{\P}_0,\overline{\P}_1)=0$ but
$\overline{\P}_0=\overline{\P}_1=0$ has no solution. Note that in
this example, Problem~\ref{problem-generator} has a positive answer,
that is,
 $\mathcal{I}_{\bu}=\sat(\SR)$.
Theorem \ref{th-nscond} shows that $\mathcal
{Z}_0(\mathcal{A}_0,\ldots,\mathcal{A}_n)$ is dense in
$\V(\sat(\SR))$. This example shows that for certain
$\mathcal{A}_i$, $\mathcal{Z}_0(\mathcal{A}_0,\ldots,\mathcal{A}_n)$
is a proper subset of $\V(\sat(\SR))$.
\end{exmp}
%

The following lemma reflects the structures of the nonzero
solutions.
\begin{lem} \label{lm-constructure}
Use the notations in \bref{eq-sparseLaurent}. Let
$\mathcal{A}_0,\ldots,\mathcal{A}_n$ be a Laurent transformally
essential system and
$\SR=\Res_{\mathcal{A}_0,\ldots,\mathcal{A}_n}$. Then there exists a
$\tau$ such that $\deg(\SR,u_{\tau0})>0.$ Suppose
$\overline{\P}_{i}=0$ is a system represented by $
(\bv_{0},\ldots,\bv_{n})\in
  {Z}_0(\mathcal{A}_0,\ldots,\mathcal{A}_n)$ and $\frac{\partial
\SR}{\partial u_{\tau0}}(\bv_0,\ldots,\bv_n)$ $\neq 0$. If $\xi$ is
a common nonzero difference solution of
$\overline{\P}_i=0(i=0,\ldots,n)$, then for each  $j$, we have
\begin{equation} \label{eq-root}
\frac{M_{\tau j}(\xi)}{M_{\tau 0}(\xi)}= \frac{\partial
\SR}{\partial u_{\tau j}}(\bv_0,\ldots,\bv_n)\Big/\frac{\partial
\SR}{\partial u_{\tau 0}}(\bv_0,\ldots,\bv_n).
\end{equation}
\end{lem}
\proof Since
$\mathcal{I}_{\Y,\bu}=[\norm(\P_0),\ldots,\norm(\P_n)]:\mathbbm{m}$
is a reflexive prime difference ideal and
$\SR\in\mathcal{I}_{\Y,\bu}$, there exists some $\tau$ and $j$ such
that $\deg(\SR,u_{\tau j})>0.$
 By equation~\bref{eq-partialdiff}, $\deg(\SR,u_{\tau0})>0$ and for each $j=1,\ldots,l_0$,
 the polynomial $\frac{\partial \SR}{\partial u_{\tau 0}}
M_\tau M_{\tau j}- \frac{\partial \SR}{\partial u_{\tau j }}M_\tau
M_{\tau 0}\in\mathcal{I}_{\Y,\bu}$, where
$\norm(\P_i)=M_i\P_i\,(i=0,\ldots,n)$.
 Thus, if $\xi$ is a common
nonzero difference solution of $\overline{\P}_i=0$, then
$\frac{\partial \SR}{\partial u_{\tau0}}(\bv_0,\ldots,\bv_n)\cdot
M_{\tau j}(\xi)-\frac{\partial \SR}{\partial u_{\tau j }
}(\bv_0,\ldots,\bv_n)M_{\tau0}(\xi)=0$. Since $\frac{\partial
\SR}{\partial u_{\tau0}}(\bv_0,\ldots,\bv_n)\neq 0$, \eqref{eq-root}
follows. \qed

The following result gives a condition for the system to have a
unique solution.
\begin{cor} \label{cor-sol1}
Assume that 1) 
for each $j=1,\ldots,n$, there exists $d_{jik}\in\mathbb{Z}$ such
that
$y_j=\prod_{i=0}^n\prod_{k=0}^{l_i}(\frac{M_{ik}}{M_{i0}})^{d_{jik}}$
and  2) for each $i$ and $k$, $\deg(\SR,u_{ik})>0$. Suppose
$\overline{\P}_{i}=0$ is a specialized  system represented by $
(\bv_{0},\ldots,\bv_{n})$ with $\SR(\bv_0,\ldots,\bv_n)=0$ and
$\frac{\partial \SR}{\partial u_{ik}}(\bv_0,\ldots,\bv_n)\neq
0\,(i=0,\ldots,n;k=0,\ldots,l_i)$. Then the system
$\overline{\P}_i=0\,(i=0,\ldots,n)$ could have at most one nonzero
solution.
Furthermore, if Problem~\ref{problem-generator} has a positive
answer, that is, $\mathcal{I}_{\bu}=\sat(\SR)$, then the system
$\overline{\P}_i=0\,(i=0,\ldots,n)$ has a unique nonzero solution.
\end{cor}
\proof Suppose $\xi$ is a nonzero  solution of $\overline{\P}_i=0$.
By \bref{eq-root},
for each $j$,
$\prod_{i=0}^n\prod_{k=0}^{l_i}(\frac{M_{ik}(\xi)}{M_{i0}(\xi)})^{d_{jik}}$
$= \prod_{i=0}^n\prod_{k=0}^{l_i}\big(\overline{\frac{\partial \SR
}{\partial u_{ik}}}\big/\overline{\frac{\partial \SR}{\partial u_{i
0}}}\big)^{d_{jik}}=\xi_j\ne0$, where $\overline{\frac{\partial \SR
}{\partial u_{ik}}}=\frac{\partial \SR }{\partial
u_{ik}}(\bv_0,\ldots,\bv_n)$. That is, $\xi$ is uniquely determined
by $\SR$ and $\bv_i$.
Suppose $\mathcal{I}_{\bu}=\sat(\SR)$. Let
$\mathcal{I}_{\Y,\bu}=[\norm(\P_0),\ldots,\norm(\P_n)]:\mathbbm{m}$.
Similar to the proof of Lemma \ref{lm-constructure}, $T_j =
\prod_{i=0}^n\prod_{k=0}^{l_i}M_{ik}^{d_{jik}} y_j -
\prod_{i=0}^n\prod_{k=0}^{l_i}M_{i0}^{d_{jik}} \in
\mathcal{I}_{\Y,\bu}$. Since $\mathcal{I}_{\bu}=\sat(\SR)$, $\A= \{
\SR, T_1,\ldots,T_n\}$ is a characteristic set for
$\mathcal{I}_{\bu}$ and $\mathcal{I}_{\bu}=\sat(\A)$, from which we
can deduce that $y_j =
\prod_{i=0}^n\prod_{k=0}^{l_i}\big(\overline{\frac{\partial \SR
}{\partial u_{ik}}}\big/\overline{\frac{\partial \SR}{\partial u_{i
0}}}\big)^{d_{jik}},j=1,\ldots,n$ constitute a nonzero solution of
$\overline{\P}_i=0$.\qed

\begin{exmp} Let $n=2$ and the $\P_i$ have the form
$$\P_0=u_{00}+u_{01}y_1y_2,\,
  \P_1=u_{10}+u_{11}y_1y_2^{(1)},\,
\P_2=u_{20}+u_{2 1}y_2.$$
Clearly, $\P_0,\P_1,\P_2$ form a super-essential system and the
sparse difference resultant of $\P_0,\P_1,\P_2$ is
$\SR=u_{21}u_{20}^{(1)}u_{11}u_{00}-u_{21}^{(1)}u_{20}u_{01}u_{10}$.
Moreover, $\P_0,\P_1,\P_2$ satisfy the conditions of Corollary
\ref{cor-sol1}, so given a specialized system  $\overline{\P}_i$
with   $\SR(\bv_0,\bv_1,\bv_2)=0$ and $\frac{\partial \SR}{\partial
u_{ik}}(\bv_0,\bv_1,\bv_2)\neq 0\,(i=0,1,2;k=0,1)$, the system
$\overline{\P}_i=0\,(i=0,\ldots,n)$ have a unique nonzero solution
$y_2 = -\frac{v_{20}}{v_{2 1}}$ and $y_1 =
-\frac{v_{00}}{v_{01}y_2}= \frac{v_{00}v_{2 1}}{v_{01}v_{20}}$.
\end{exmp}

\subsection{Order bound in terms of Jacobi number}
\label{sec-ord1} In this section, we will give an order bound for
the sparse difference resultant in terms of the Jacobi number of the
given system.

Consider a generic Laurent transformally essential system
$\{\P_0,\ldots,\P_n\}$ defined in \bref{eq-sparseLaurent} with
$\bu_i=(u_{i0},u_{i1},\ldots,u_{il_i})$ being the coefficient vector
of $\P_i\,(i=0,\ldots,n).$ Suppose $\SR$ is the sparse difference
resultant of $\P_0,\ldots,\P_n$. Denote $\ord(\SR,\bu_i)$ to be the
maximal order of $\SR$ in $u_{ik}\,(k=0,\ldots,l_i)$, that is,
$\ord(\SR,\bu_i)=\max_{k}\ord(\SR,u_{ik})$. If $\bu_i$ does not
occur in $\SR$,  then set $\ord(\SR,\bu_i)=-\infty$. Firstly, we
have the following result.
\begin{lem} \label{lm-order1}
For  fixed $i$ and $s$, if there exists  $k_0 $ such that
$\deg(\SR,u_{ik_0}^{(s)})>0$, then for all $k\in\{0,1,\ldots,l_i\}$,
$\deg(\SR,u_{ik}^{(s)})>0.$ In particular, if
$\ord(\SR,\bu_i)=h_i\geq0$, then
$\ord(\SR,u_{ik})=h_i\,(k=0,\ldots,l_i)$.
\end{lem}
\proof  Firstly, for each $k\in\{1,\ldots,l_i\}$, by differentiating
$\SR(\bu;\zeta_0,\ldots,\zeta_n)=0$ w.r.t. $u_{ik}^{(s)}$, we have
$\frac{\partial \SR}{\partial u_{i k}^{(s)}}
(\bu,\zeta_0,\ldots,\zeta_n)+ \frac{\partial \SR}{\partial u_{i
0}^{(s)}}(\bu,\zeta_0,\ldots,\zeta_n)\big(-\frac{M_{ik}(\eta)}{M_{i0}(\eta)}\big)^{(s)}=0.$
If $k_0=0$, then $\frac{\partial \SR}{\partial u_{i 0}^{(s)}}$ is a
nonzero difference polynomial not vanishing at
$(\bu,\zeta_0,\ldots,\zeta_n)$ by lemma~\ref{lm-genericresultant}.
So $\frac{\partial \SR}{\partial u_{ik}^{(s)}}\neq0$. Thus,
$\deg(\SR,u_{ik}^{(s)})>0$ for each $k$.
If $k_0\neq0$, then  $\frac{\partial \SR}{\partial u_{i
k_0}^{(s)}}(\bu,\zeta_0,\ldots,\zeta_n)\neq0$ and $\frac{\partial
\SR}{\partial u_{i0}^{(s)}}\neq0$ follows. So by the case $k_0=0$,
for all $k$, $\deg(\SR,u_{ik}^{(s)})>0$.

In particular, if $\ord(\SR,\bu_i)=h_i\geq0$, then there exists some
$k_0$ such that $\deg(\SR,u_{ik_0}^{(h_i)})>0$. Thus, for each
$k=0,\ldots,l_i$, $\deg(\SR,u_{ik}^{(h_i)})>0$ and
$\ord(\SR,u_{ik})=h_i$ follows. \qed

Let $A=(a_{ij})$ be an $n\times n$ matrix where $a_{ij}$ is an
integer or $-\infty$. A {\em diagonal sum} of $A$ is any sum
$a_{1\sigma(1)}+a_{2\sigma(2)}+\cdots+a_{n\sigma(n)}$ with $\sigma$
a permutation of $1,\ldots,n$. If $A$ is an $m\times n$ matrix with
$k=\min\{m,n\}$, then a diagonal sum of $A$ is a diagonal sum of any
$k\times k$ submatrix of $A$. The {\em Jacobi number}  of a matrix
$A$ is the maximal diagonal sum of $A$, denoted by $\Jac(A)$.
Refer to \cite{cohn2,Hrushovski1} for the concept of Jacobi number
and its relation with the order of a difference system.

Let $s_{ij}=\ord(\norm(\P_i),y_j)\,(i=0,\ldots,n;j=1,\ldots,n)$ and
$s_i=\ord(\norm(\P_i))$. We call the $(n+1)\times n$ matrix
$A=(s_{ij})$ the {\em order matrix} of $\P_0,\ldots,\P_n$. By
$A_{\hat{i}}$, we mean the submatrix of $A$ obtained by deleting the
$(i+1)$-th row from $A$. We use $\P$ to denote the set
$\{\norm(\P_0),\ldots,\norm(\P_n)\}$ and by $\P_{\hat{i}}$, we mean
the set $\P\backslash\{\norm(\P_i)\}$. We call
$J_i=\Jac(A_{\hat{i}})$ the {\em Jacobi number} of the system
$\P_{\hat{i}}$, also denoted by $\Jac(\P_{\hat{i}})$. Before giving
an order bound for sparse difference resultant in terms of the
Jacobi numbers, we first list several lemmas.

Given a vector
$\overrightarrow{K}=(k_0,k_1,\ldots,k_n)\in\mathbb{N}_0^{n+1}$, we
can obtain a prolongation of $\P$:
\begin{equation}\label{eq-pk1}
 \P^{[\overrightarrow{K}]} = \bigcup_{i=0}^n\norm(\P_i)^{[k_i]}.\end{equation}
Let $t_j=\max \{s_{0j}+k_0, s_{1j}+k_1, \ldots, s_{nj}+k_n\}$. Then
$\P^{[\overrightarrow{K}]}$ is contained in $\Q[\buk,\YK]$, where
$\buk=\cup_{i=0}^n \bu_i^{[k_i]}$ and $\YK = \cup_{j=1}^n
y_j^{[t_j]}$.

Denote $\nu(\PK)$ to be the number of $\Y$ and their transforms
appearing effectively in $ \PK$. In order to derive a difference
relation among $\bu_{i}\,(i=0,\ldots,n)$ from $\PK$, a sufficient
condition is
\begin{equation}\label{eq-psv1}|\PK|\geq \nu(\PK)+1.\end{equation}
Note that $ \nu(\PK)\leq|\YK|= \sum_{j=1}^{n} (t_j+1)$. Thus, if
$|\PK|\geq\YK+1$, or equivalently,
\begin{equation}\label{constraint} k_0 + k_1 + \cdots + k_n \geq
\sum\limits_{j=1}^{n} \max(s_{0j}+k_0, s_{1j}+k_1, \ldots,
s_{nj}+k_n)
\end{equation}
is satisfied,  then so is the inequality \bref{eq-psv1}.

\begin{lem}\label{le-order-cons}
Let $\P$ be a Laurent transformally  essential system and
$\overrightarrow{K}=(k_0,k_1,\ldots,k_n)\in\mathbb{N}_0^{n+1}$ a
vector satisfying \bref{constraint}. Then $\ord(\SR,\bu_i)\le k_i$
for each $i=0,\ldots,n$.
\end{lem}
\proof Denote $\mathbbm{m}^{[\overrightarrow{K}]}$ to be the set of
all monomials in variables $\YK$. Let
$\CI=(\P^{[\overrightarrow{K}]}):\mathbbm{m}^{[\overrightarrow{K}]}$
be an   ideal in the polynomial ring $\Q[\YK,\buk]$. Denote
$U=\buk\backslash\cup_{i=0}^nu_{i0}^{[k_i]}$.  Let
$\zeta_{il}=-(\sum_{k=1}^{l_i}u_{ik}M_{ik}/M_{i0})^{(l)}$ for
$i=0,1,\ldots,n;l=0,1,\ldots,k_i$. Denote
$\zeta=(U,\zeta_{0k_0},\ldots,\zeta_{00},\ldots,$
$\zeta_{nk_n},\ldots,\zeta_{n0})$. It is easy to show that
$(\YK,\zeta)$ is a generic zero of $\CI$. Let
$\CI_1=\CI\cap\Q[\buk]$. Then $\CI_1$ is a prime ideal with a
generic zero $\zeta$. Since $\Q(\zeta)\subset\Q(\YK,U)$,
$\codim(\CI_1)=|U|+\sum_{i=0}^n(k_i+1)-\trdeg\,\Q(\zeta)/\Q\geq
|U|+|\P^{[\overrightarrow{K}]}|-\trdeg\,\Q(\YK,U)/\Q=|\P^{[\overrightarrow{K}]}|-|\YK|\geq1$.
Thus, $\CI_1\neq \{0\}$. Suppose $f$ is a nonzero polynomial in
$\CI_1$. Clearly, $\ord(f,\bu_{i})\leq k_i$ and $f\in[ \P
]:\mathbbm{m}\cap\Q\{\bu_0,\ldots,\bu_n\} $. By
Lemma~\ref{lm-genericresultant} and Lemma~\ref{lm-order1},
$\ord(\SR,\bu_{i})\leq\ord(f,\bu_i)\leq k_i$. \qed

\begin{lem}\label{le-jac-cons}\cite[Lemma 5.6]{li}
Let $\P$ be a  system  with $J_i\geq0$ for each $i=0,\ldots,n$. Then
$k_i=J_i\,(i=0,\ldots,n)$ satisfy ~\bref{constraint} in the equality
case.
\end{lem}

\begin{cor}\label{th-order-jacb1}
Let $\P$ be a Laurent transformally essential system and $J_i\geq0$
for each $i=0,\ldots,n$. Then $\ord(\SR,\bu_i)\leq
J_i\,(i=0,\ldots,n)$.
\end{cor}

\proof It is a direct consequence of Lemma~\ref{le-order-cons} and
Lemma~\ref{le-jac-cons}. \qed

The above corollary shows that when all the Jacobi numbers are not
less that $0$, then Jacobi numbers are order bounds for the sparse
difference resultant. In the following, we deal with the remaining
case when some $J_i=-\infty$. 
To this end, two more lemmas are needed.

\begin{lem}\cite{cohn2,lando}\label{lem-lando}
Let $A$ be an $m\times n$ matrix whose entries are $0$'s and $1$'s.
Let $\Jac(A)=J<\min\{m,n\}$. Then $A$ contains an $a\times b$ zero
sub-matrix with $a+b=m+n-J$.
\end{lem}

\begin{lem}\label{lm-essord}
Let $\P$ be a Laurent transformally essential system with the
following $(n+1)\times n$ order matrix
\[
\mbox{\bf $A$}=\left(\begin{array}{cc}
A_{11} & \, (-\infty)_{r\times t} \\
A_{21} & \,A_{22}
\end{array}\right),
\]
where $r+t\geq n+1$. Then $r+t=n+1$ and $\Jac(A_{22})\geq0$.
Moreover, when regarded as difference polynomials in
$y_1,\ldots,y_{r-1}$, $\{\P_{0},\ldots,\P_{r-1}\}$ is a Laurent
transformally essential system.
\end{lem}
\proof  The proof is similar to \cite[Lemma 5.9]{li}.\qed

\begin{thm}\label{th-ord-jacobi2}
Let $\P$ be a Laurent transformally essential system and $\SR$ the
sparse difference resultant of $\P$. Then
\[\ord(\SR,\bu_i)=\left\{\begin{array}{lll}
-\infty&& \text{if}\quad\,J_i = -\infty,\\
h_i\leq J_i&& \text{if}\quad\,J_i \geq0.\end{array}\right.\]
\end{thm}
\proof Corollary~\ref{th-order-jacb1} proves the case when
$J_i\geq0$ for each $i$.
 Now suppose there exists at least one $i$ such that $J_i =
-\infty$. Without loss of generality, we assume $J_n = -\infty$ and
let $A_n = (s_{ij})_{0\le i\le n-1; 1\le j\le n}$ be the order
matrix of $\P_{\hat{n}}$. By Lemma~\ref{lem-lando}, we can assume
that $A_n$ is of the following form
\[\mbox{\bf $A_n$}=\left(\begin{array}{cc}
A_{11} & \,(-\infty)_{r\times t} \\
\bar{A}_{21} & \,\bar{A}_{22}
\end{array}\right), \]
where $r+t\geq n+1$. Then the order matrix of $\P$ is equal to
 \[\mbox{\bf $A$}=\left(\begin{array}{cc}
A_{11} & \,(-\infty)_{r\times t} \\
A_{21} & \,A_{22}
\end{array}\right). \]

 Since $\P$ is  Laurent transformally essential, by
Lemma~\ref{lm-essord}, $r+t = n+1$ and $\Jac(A_{22})\ge 0$.
Moreover, considered as difference polynomials in
$y_1,\ldots,y_{r-1}$, $\widetilde{\PS} = \{p_{0},\ldots,p_{r-1}\}$
is Laurent transformally essential and $A_{11}$ is its order matrix.
Let $\widetilde{J}_i=\Jac((A_{11})_{\hat{i}})$. By applying the
above procedure when necessary, we can suppose that
$\widetilde{J}_i\geq0$ for each $i=0,\ldots,r-1$.
Since
$[\P]\cap\Q\{\bu_0,\ldots,\bu_n\}=[\widetilde{\P}]\cap\Q\{\bu_0,\ldots,\bu_{r-1}\}$,
$\SR$ is also the sparse difference resultant of the system
$\widetilde{\P}$ and $\bu_r,\ldots,\bu_{n}$ will not occur in $\SR$.
By Corollary~\ref{th-order-jacb1}, $\ord(\SR,\bu_{i})\leq
\widetilde{J_i}$. Since $J_{i} = \Jac(A_{22})+
\widetilde{J_i}\geq\widetilde{J_i}$ for $0\le i\le r-1$,
$\ord(\SR,\bu_i)\leq J_{i}$ for $0\leq i\leq r-1$ and
$\ord(\SR,\bu_i)=-\infty$ for $i=r,\ldots,n.$ \qed

\begin{exmp}\label{ex-2n} 
Let $n=2$ and
$$\P_0=u_{00}+u_{01}y_1y_1^{(1)},\,
  \P_1=u_{10}+u_{11}y_1,\,
\P_2=u_{10}+u_{11}y_2^{(1)}.$$
The sparse resultant is $\SR = u_{00}u_{11}u_{11}^{(1)} +
u_{01}u_{10}u_{10}^{(1)}$. In this example, the order matrix
of $\P$ is $A=\left(\begin{array}{cc}1&-\infty\\ 0&-\infty \\
-\infty&1
\end{array} \right)$. Thus $J_0=1,J_1=2,J_2=-\infty$. And
$\ord(\SR,\bu_0)=0<J_0, \ord(\SR,\bu_1)=1<J_1,
\ord(\SR,\bu_2)=-\infty.$
\end{exmp}

\begin{cor}
Let $\P$ be super-essential. Then $J_i\ge 0$ for $i=0,\ldots,n$ and
 $\ord(\SR,\bu_i) \le J_i$.
\end{cor}
\proof From the proof of Theorem \ref{th-ord-jacobi2}, if
$J_i=-\infty$ for some $i$, then $\P$ contains a proper
transformally essential subsystem, which contradicts to Theorem
\ref{th-rankessential}. Therefore, $J_i\ge 0$ for $i=0,\ldots,n$.
\qed

We conclude this section by giving two improved order bounds based
on the Jacobi bound given in Theorem~\ref{th-ord-jacobi2}.

For each $j\in\{1,\ldots,n\}$, let
$\underline{o}_j=\min\{k\in\mathbb{N}_0| \,\forall i\, s.t.\,
\deg(\norm(\P_i),y_j^{(k)})>0\}$. In other words, $\underline{o}_j$
is the smallest number such that $y_j^{(\underline{o}_j)}$ occurs in
$\{\norm(\P_0),\ldots,\norm(\P_n)\}$. Let
$B=(s_{ij}-\underline{o}_j)$ be an $(n+1)\times n$ matrix. We call
$\bar{J}_i=\Jac(B_{\hat{i}})$ the {\em modified Jacobi number} of
the system $\P_{\hat{i}}$. Denote
$\underline{\gamma}=\sum_{j=1}^n\underline{o}_j$. Clearly,
$\bar{J}_i=J_i-\underline{\gamma}.$ Then we have the following
result.

\begin{thm} \label{th-jacobi-order3}
Let $\P$ be a Laurent transformally essential system and $\SR$ the
sparse difference resultant of $\P$. Then
\[\ord(\SR,\bu_i)=\left\{\begin{array}{lll}
-\infty&& \text{if}\quad\,J_i = -\infty,\\
h_i\leq J_i-\underline{\gamma}&& \text{if}\quad\,J_i
\geq0.\end{array}\right.\]
\end{thm}
\proof The proof is similar to \cite[Theorem 5.13]{li}. \qed

Now, we assume that $\P$ is a Laurent transformally essential system
which is not super-essential. Let $\SR$ be the sparse difference
resultant of $\P$. We will give a better order bound for $\SR$.
By Theorem \ref{th-rankessential}, $\P$ contains a unique
super-essential sub-system $\P_{\TT}$. Without loss of generality,
suppose $\TT = \{ 0,\ldots, r\}$ with $r< n$.
Let $A_\TT$ be the order matrix of $\P_\TT$ and for $i=0,\ldots,r$,
let $(A_{\TT})_{\hat{i}}$ be the matrix obtained from $A_\TT$ by
deleting the $(i+1)$-th row. Note that $(A_{\TT})_{\hat{i}}$ is an
$r\times n$ matrix.
Then we have the following result.
\begin{thm}\label{th-jb12}
With the above notations, we have
\[\ord(\SR,\bu_i)=\left\{\begin{array}{lll}
h_i\leq \Jac((A_{\TT})_{\hat{i}})&&i=0,\ldots,r,\\
-\infty&&i=r+1,\ldots,n.\\\end{array}\right.\]
\end{thm}
\proof Similarly to the proof of \cite[Theorem 5.16]{li}, it can be
proved.\qed

\begin{exmp} Continue from Example~\ref{ex-2n}.
In this example, $\TT = \{ 0,1\}$. Then
$A_{\TT}=\left(\begin{array}{c}1\\ 0\end{array} \right)$. Thus
$\Jac((A_{\TT})_{\hat{0}})=0,\Jac((A_{\TT})_{\hat{1}})=1$. For this
example, the exact bounds are given:
$\ord(\SR,\bu_0)=0=\Jac((A_{\TT})_{\hat{0}}),
\ord(\SR,\bu_1)=1=\Jac((A_{\TT})_{\hat{1}}),
\ord(\SR,\bu_2)=-\infty.$
\end{exmp}

\subsection{Effective order bound in terms of Jacobi number}
\label{sec-eord}

In this section, we give an improved Jacobi-type bound for the
effective order and order of the sparse difference resultant.

For a difference polynomial $f\in \ff\{y_1,\ldots,y_n\}$ and an
arbitrary variable $y_i$, the {\em least order} of $f$ w.r.t. $y_i$
is $\lord(f,y_i)=\min\{k|\deg(f,y_{i}^{(k)})>0\}$
  and the {\em effective order} of $f$ w.r.t. $y_i$ is $\Eord(f,y_i)=\ord(f,y_i)-\lord(f,y_i)$.
And if $y_{i}$ does not appear in $f$, then set
$\Eord(f,y_i)=-\infty$. Let $\SR$ be the sparse difference resultant
of a Laurent transformally essential system
$\{\P_0,\P_1,\ldots,\P_n\}$ of the form \bref{eq-sparseLaurent}. By
Lemma \ref{lm-order1}, $u_{i0}^{(s)}$ effectively appears in $\SR$
if and only if $u_{ik}^{(s)}$ effectively appears in $\SR$ for each
$k\in\{0,\ldots,l_i\}$. Thus, we can define
$\lord(\SR,\bu_{i})=\lord(\SR,u_{i0})$ and  $\Eord(\SR,\bu_i) =
\ord(\SR,\bu_i)-\lord(\SR,\bu_{i})$ whenever $\bu_{i}$ effectively
appears in $\SR$.

For further discussion, suppose $\P_{\TT}$ is the super-essential
subsystem of $\{\P_0,\P_1,\ldots,\P_n\}$.
Without loss of generality, assume $\TT=\{ 0, 1, \ldots, p\}$.
For each  $i\in\{0,\ldots,p\}$, let
$\underline{s}_i=\min_{j=1}^n\{\lord(\P_i,y_j)|\lord(\P_i,y_j)\neq-\infty\}$
and $\underline{s}=\sum_{i=0}^p\underline{s}_i$.
Let $\widetilde{J}_i = J_i-\underline{s}+\underline{s}_i$. Then,

\begin{thm}\label{th-eord}
The effective order of $\SR$ in $\bu_i$ is bounded by
$\widetilde{J}_i$ for each $0\le i\le p$.
\end{thm}
\proof
Let $m = \max_{i=0}^p \underline{s}_i$. Consider the following
difference system
$$\mathcal{P}_1=\{ \P_0^{(m-\underline{s}_0)}, \ldots, \P_p^{(m-\underline{s}_p)}\}$$
which is also super-essential. Suppose $\SR_1$ is the sparse
difference resultant of $\mathcal{P}_1$. Clearly,
$\SR_1\in\mathcal{I}_\bu=[\P_0,\ldots,\P_p]\cap\Q\{\bu_0,\ldots,\bu_p\}$,
so $\ord(\SR_1,\bu_i)\geq\ord(\SR,\bu_i)$ for each
$i\in\{0,\ldots,p\}$.
Since $y_{j}^{[m-1]}\,(j=1,\ldots,n)$ do not occur in
$\mathcal{P}_1$, by replacing $y_{j}^{(t)}\,(j=1,\ldots,n)$ by
$z_{j}^{(t-m)}$ in $\mathcal{P}_1$, we obtain a new system
$\mathcal{P}_2$. It is clear  that $\SR_1$ is also the sparse
difference resultant of  $\mathcal{P}_2$.
By Theorem \ref{th-jb12}, $\Eord(\SR_1,\bu_i) \le \widetilde{J}_i$
and $\ord(\SR_1,\bu_i) \le \widetilde{J}_i+m-\underline{s}_i$ for
each $i\in\{0,\ldots,p\}$.

Let $h_i=\ord(\SR,\bu_i)$ and $o_i=\lord(\SR,\bu_i)$. We need to
show that $h_i-o_i\leq\widetilde{J}_i$ holds for each
$i\in\{0,\ldots,p\}$. Suppose the contrary, i.e. there exists some
$i_0\in\{0,\ldots,p\}$ such that
$\Eord(\SR,\bu_{i_0})=h_{i_0}-o_{i_0}> \widetilde{J}_{i_0}$.

Suppose $\bar{h}_{i_0}= \ord(\SR_1,\bu_{i_0})$ and $\bar{o}_{i_0} =
\lord(\SR_1,\bu_{i_0})$.
Then, $\bar{h}_{i_0}\geq h_{i_0}$ and $\Eord(\SR_1,\bu_{i_0})$
$=\bar{h}_{i_0}-\bar{o}_{i_0}\le \widetilde{J}_{i_0}<
h_{i_0}-o_{i_0}$.  Clearly, $\sigma^{\bar{h}_{i_0}}u_{i0}$ appears
effectively in both $\sigma^{\bar{h}_{i_0}-h_{i_0}}\SR$ and $\SR_1$.
Let $B_1$ be the Sylvester resultant of
$\sigma^{\bar{h}_{i_0}-h_{i_0}}\SR$ and $\SR_1$ w.r.t.
$\sigma^{\bar{h}_{i_0}}u_{i0}$. We claim that $B_1\neq0$. Suppose
the contrary, then we have $\sigma^{\bar{h}_{i_0} -h_{i_0}}\SR |
\SR_1$, for $\SR$ is irreducible. This is impossible since
$\sigma^{\bar{h}_{i_0}-h_{i_0}+o_{i_0}}u_{i0}$ appears effectively
in $\sigma^{\bar{h}_{i_0}-h_{i_0}}\SR$ while not in $\SR_1$ for
$\bar{h}_{i_0}-h_{i_0}+o_{i_0}< \bar{o}_{i_0}$.

Let $\widetilde{h}_{i_0}=\ord(B_1,u_{i_00})$ and
$\widetilde{o}_{i_0}=\lord(B_1,u_{i_00})$.
Since $B_1$ is the resultant of $\sigma^{\bar{h}_{i_0}-h_{i_0}}\SR$
and $\SR_1$, $\widetilde{h}_{i_0}< \bar{h}_{i_0}$ and
$\widetilde{o}_{i_0}\ge \bar{h}_{i_0}-h_{i_0}+o_{i_0}$.  Then
$\widetilde{h}_{i_0} - \widetilde{o}_{i_0} <\bar{h}_{i_0}
-(\bar{h}_{i_0}-h_{i_0}+o_{i_0}) = h_{i_0}-o_{i_0}$.
Since $B_1\in\mathcal{I}_\bu$, by Lemma~\ref{lm-genericresultant},
$\ord(B_1,u_{i_00})\ge \ord(\SR,u_{i_{0}0})$.
Repeat the above procedure for $B_1$ and
$\sigma^{\widetilde{h}_{i_0}-h_{i_0}}\SR$, we obtain a nonzero
difference polynomial $B_2\in\mathcal{I}_\bu$ and
$\ord(B_2,u_{i_00})< \ord(B_1,u_{i_00})$. Continue the procedure in
this way, one can finally obtain a nonzero $B_l\in\mathcal{I}_\bu$
such that $\ord(B_l,u_{i_00})< \ord(\SR,u_{i_00})$ which contradicts
to Lemma~\ref{lm-genericresultant}.

\qed

By the proof of the above theorem, the order of $\SR_1$ with respect
to $\bu_i$ is bounded by $\widetilde{J}_i+m-\underline{s}_i$. Thus,
we have the following new order bound for $\SR$.

\begin{cor}\label{co-eord} Let $\SR$ and
$\widetilde{J}_i\,( i= 0,   \ldots, p)$ be defined as above. Then
the order of $\SR$ in $\bu_i$ is bounded by $\underline{J_i} =
\widetilde{J}_i+m-\underline{s}_i = J_i-\underline{s}+m$ for each
$0\le i\le p$ where $m = \max_{i=0}^p\underline{s}_i$.
\end{cor}

\begin{exmp}
Let $\P_0 = u_{00}+u_{01}y_1+u_{02}y_2, \P_1 =
u_{10}+u_{11}y_1^{(1)}+u_{12}y_2^{(1)}, \P_2 =
u_{20}+u_{21}y_1^{(1)}+u_{22}y_2^{(1)}$. Then $J_0 = \bar{J}_0 = 2,
J_1 = \bar{J}_1 = 1, J_2 = \bar{J}_2 = 1$, $\widetilde{J}_0 =
\widetilde{J}_1 = \widetilde{J}_2 = 0$. By corollary~\ref{co-eord},
$\underline{J_0}=1, \underline{J_1}=0, \underline{J_2}=0$.  Notice
that $\SR =\left|\begin{array}{ccc}
u_{00}^{(1)}& u_{01}^{(1)} & u_{02}^{(1)}\\
u_{10} & u_{11} & u_{12} \\
u_{20} & u_{21} & u_{22} \end{array} \right|$ and $\widetilde{J}_0 =
\widetilde{J}_1 = \widetilde{J}_2 = 0$, $\underline{J_0}=1,
\underline{J_1}=\underline{J_2}=0$ give the exact effective order
and order of $\SR$ respectively.
\end{exmp}

\section{Sparse difference resultant as algebraic sparse resultant}
\label{sec-sares} In this section, we will show that the sparse
difference resultant is just equal to the algebraic sparse resultant
of certain generic sparse polynomial system, which leads to a
determinant representation for the sparse difference resultant.

\subsection{Preliminary on algebraic sparse resultant}
We first introduce several basic notions and properties on algebraic
sparse resultants which are needed in this paper. For more details
about sparse resultant, please refer to \cite{gelfand,sturmfels}.

Let $\mathcal{B}_0,\ldots,\mathcal{B}_n$ be finite subsets of
$\mathbb{Z}^n$. Assume ${\bf 0}\in\mathcal{B}_i$ and
$|\mathcal{B}_i|\geq 2$ for each $i$. For algebraic indeterminates
$\X=\{x_1,\ldots,x_n\}$ and
$\alpha=(\alpha_1,\ldots,\alpha_n)\in\mathbb{Z}^n$, denote
$\X^{\alpha}=\prod_{i=1}^nx_i^{\alpha_i}$. Let
\begin{equation}\label{eq-algsparse}
\BF_i(x_1,\ldots,x_n)=c_{i0}+\sum_{\alpha\in\mathcal{B}_i\backslash
\{{\bf 0}\}}c_{i\alpha}\X^{\alpha}\,(i=0,\ldots,n)\end{equation} be
generic sparse Laurent polynomials, where $c_{i\alpha}$ are
algebraic indeterminates. We call $\mathcal{B}_i$ the support of
$\BF_i$ and
$\omega_i=\sum_{\alpha\in\mathcal{B}_i}c_{i\alpha}\alpha$ is called
the {\em symbolic support vector} of $\BF_i$. The smallest convex
subset of $\mathbb{R}^n$ containing $\mathcal{B}_i$ is called the
{\em Newton polytope} of $\BF_i$.
For any subset $I\subset\{0,\ldots,n\}$, the matrix $M_{I}$ whose
row vectors are $\omega_i\, (i\in I)$ is called the {\em symbolic
support matrix} of $\{\BF_i:i\in I\}$.
 Denote $\bc_i=(c_{i\alpha})_{\alpha\in\mathcal{B}_i}$ and
$\bc_I=\cup_{i\in I}\bc_i$. 
Similar to the proof of \cite[Theorem 4.17]{li} and use the Jacobi
criterion for algebraic independence, it can be easily shown that
\begin{lem}\label{lm-as1}
For any subset $I\subset\{0,\ldots,n\}$,
$\trdeg\,\mathbb{Q}(\bc_I)(\BF_i:i\in
I)/\mathbb{Q}(\bc_I)=\rk(M_I)$.

\end{lem}
\begin{defn}
Follow the notations introduced above.
\begin{itemize}
\item  A collection of  $\{\BF_i\}_{i\in I}$   is said to be
{\rm weak essential} if $\rk(M_I)=|I|-1$.
\item A collection of $\{\BF_i\}_{i\in I}$  is said to be
{\rm essential} if $\rk(M_I)=|I|-1$ and for each proper subset
$\text{J}$ of $\text{I}$, $\rk(M_J)=|J|$.
\end{itemize}
\end{defn}

Similar to Theorems~\ref{th-cri} and~\ref{th-rankessential}, we have
the following two lemmas.

\begin{lem}\label{th-crialg}
$\{\BF_i\}_{i\in I}$ is weak essential  if and only if $(\BF_i:i\in
I)\cap \Q[\bc_I]$ is of codimension one. In this case,there exists
an irreducible polynomial $\SR\in\Q[\bc_I]$ such that $(\BF_i:i\in
I)\cap \Q[\bc_I] = (\SR)$ and $\SR$ is called the {\em sparse
resultant} of $\{\BF_i:i\in I\}$.
\end{lem}
\begin{lem} \label{th-criess}
$\{\BF_i\}_{i\in\text{I}}$ is essential  if and only if $(\BF_i:i\in
I)\cap \Q[\bc_I]= (\SR)$ and $\bc_i$ appears effectively in $\SR$
for each $i\in I$.

\end{lem}

Suppose an arbitrary total ordering of $\{\BF_0,\ldots,\BF_n\}$ is
given, say $\BF_0 < \BF_1 < \cdots < \BF_n$. Now we define a total
ordering among subsets of $\{\BF_0,\ldots,\BF_n\}$. For any two
subsets $\mathcal {D} = \{ D_0,\ldots, D_s \}$ and $\C =
\{C_0,\ldots, C_t \}$ where $D_0 > \cdots > D_s$ and $C_0 >\cdots>
C_t$, $\mathcal {D}$ is said to be of {\em higher ranking }than
$\C$, denoted by $\D\succ\C$, if 1) there exists an $i\leq
\min(s,t)$ such that $D_0 = C_0, \ldots, D_{i-1} = C_{i-1}$, $D_i >
C_i$ or 2) $s>t$ and $D_i = C_i\,(i=0,\ldots,t)$. Note that if
$\mathcal {D}$ is a proper subset of $\C$, then $\C\succ \mathcal
{D}$.

\begin{lem}\label{lm-sub}
Let $\BF=\{\BF_i:i=0,\ldots,n\}$ be the system given in
\bref{eq-algsparse}. Suppose $\rk(M_\BF) \le n$. Then $\BF$ has an
essential subset with minimal ranking.
\end{lem}
\proof It suffices to show that $\BF$ contains an essential subset,
for the existence of an essential subset with minimal ranking can be
deduced from the fact that ``$\succ$" is a total ordering.

%
Let $\T_i=\BF\backslash\{\BF_0,\ldots,\BF_{i-1}\}\,(i=1,\ldots,n)$
and $\T_0=\BF$.
We claim that at least one of $\T_i$ is weak essential. If
$\rk(M_{\T_0}) = n$, we are done. Otherwise, $\rk(M_{\T_0}) < n$.
It is clear that $\rk(M_{\T_i}) =\rk(M_{\T_{i-1}})$ or
$\rk(M_{\T_i}) =\rk(M_{\T_{i-1}})-1$ for $i=1,\ldots,n-1$, so when
deleting one row from the matrix, the co-rank, i.e.
$|\T_i|-\rk(M_{\T_i})$, will be unchanged or decreased by $1$.
Since $\rk(M_{\T_0}) < n$, the co-rank of $M_{\T_0}$ is larger than
$1$. Since the co-rank of $M_{\T_{n}}$ is $0$, there exists
$k\in\{1,\ldots,n-1\}$ such that the co-rank of $M_{\T_k}$ is $1$.
Then $M_{\T_k}$ is weak essential.
Now, let $\SR$ be the sparse resultant of $\T_k$ and let $\C$ be the
set of $\TT_i\in\T_k$ such that the coefficients of $\BF_i$ occur in
$\SR$ effectively.
Then, $\C$ is an essential subset of $\BF$ by Lemma \ref{th-criess}.
\qed

\begin{lem}\label{lm-vess}
Suppose $\BF_I=\{\BF_i:i\in I\}$ is an essential system. Then there
exist $n-|I|+1$  of the $x_i$ such that by setting these $x_i$ to
$1$, the specialized system
$\widetilde{\BF}_I = \{\widetilde{\BF}_i:\,i\in I \}$ satisfies \\
$(1)$ \,\, $\widetilde{\BF}_I$ is still essential. \\
$(2)$ \,\, $\rk(M_{\widetilde{\BF}_I}) = |I|-1 $ is the number of
variables in $\widetilde{\BF}_I$. \\
$(3)$ \,\, $(\BF_I)\cap \Q[\bc_I] = (\widetilde{\BF}_I)\cap
\Q[\bc_I]$.
\end{lem}
\proof Let $M_I=(m_{ij})_{|I|\times n}$ be the symbolic support
matrix of $\BF_I$. Since $\BF_I$ is essential, $M_I$ contains a
submatrix of rank $|I|-1$. Without loss of generality, we assume the
matrix $M_0=(m_{ij})_{i=1,\ldots,|I|-1;j=1,\ldots,|I|-1}$ is of full
rank.
Then consider the new system $\widetilde{\BF}_I$ obtained by setting
$x_i=1\, (i =|I|,\ldots,n)$ in $\BF_I$.
Since $M_0$ is a submatrix of $M_{\widetilde{\BF}_I}$,
$\widetilde{\BF}_I$ is weak essential.
By Lemma~\ref{th-crialg}, we have $(\BF_I)\cap \Q[\bc_I] = (\SR)$
and $(\widetilde{\BF}_I)\cap \Q[\bc_I]=(\widetilde{\SR})$ where
$\SR, \widetilde{\SR}$ are irreducible polynomials in $\Q[\bc_I]$.
Hence, there exists a monomial $\m\in\Q[x_1,\ldots,x_n]$  such that
$\m\SR = \sum Q_i\BF_i$. Set $x_i=1\, (i =|I|,\ldots,n)$, then we
have $\widetilde{\m}\SR = \sum \widetilde{Q}_i\widetilde{\BF}_{i}$.
Hence $\SR\in(\widetilde\SR)$. Since both $\SR$ and $\widetilde\SR$
are irreducible, $(\widetilde\SR) = (\SR)$ and $(2)$ follows. Thus,
$\bc_i\,(i\in I)$ appears effectively in $\widetilde\SR$, for
$\BF_I$ is essential. By  Lemma~\ref{th-criess}, $\widetilde{\BF}_I$
is essential and (1) is proved. (2) is obvious and the lemma is
proved.\qed

 An essential system $\{\BF_i\}_{i\in I}$  is said to be
{\rm variable-essential} if there are only $|I|-1$ variables
appearing effectively in $\BF_i$. Clearly, if
$\{\BF_i:i=0,\ldots,n\}$ is essential, then it is
variable-essential.

\begin{lem}\label{lm-algess}
Let $\BF=\{\BF_i:i=0,\ldots,n\}$ be an essential system of the form
\bref{eq-algsparse}. Then  we can find an invertible variable
transformation $x_1 = \prod_{j=1}^n z_{j}^{m_{1j}}, \ldots, x_n =
\prod_{j=1}^n z_j^{m_{nj}}$ for $m_{ij}\in \mathbb{Q}$, such that
the image $\mathbb{G}$ of $\BF$ under the above
transformation is a generic sparse Laurent polynomial system satisfying \\
$(1)$ \,\, $\mathbb{G}$ is essential.\\
$(2)$ \,\, $\rm{Span}_\mathbb{Z} (\mathcal {B}) = \mathbb{Z}^n$,
where $\mathcal {B}$ is
the set of the supports of all monomials in $\mathbb{G}$.\\
$(3)$ \,\, $(\BF)\cap \Q[\bc] = ({\mathbb{G}})\cap \Q[\bc]$.\\
\end{lem}
\proof This is a direct consequence of the Smith normal form method
\cite[p. 67]{cohen}. Also see paper \cite{shenly} for an alternative
proof.\qed

We call a variable-essential system $\BF=\{\BF_i:i=0,\ldots,n\}$
{\em strong essential} if $\BF$  also satisfies condition (2) in
Lemma \ref{lm-algess}. Recall that condition (2) is a basic
requirement for studying sparse resultant in historic literatures
and a strong essential system here is just an essential system as
defined in papers~\cite{sturmfels2,dandrea1}. If $\BF$ is strong
essential, a matrix representation for $\SR$ can be derived, that
is, $\SR$ can be represented as the quotient of the determinants of
two matrices as shown in paper \cite{dandrea1}. Moreover, the exact
degree of the sparse resultant $\SR$ can be given in terms of  mixed
volumes \cite{sturmfels2}, famous as the BKK-type degree bound. That
is,

\begin{thm}[\cite{sturmfels2}]\label{le-alg-sparse-deg}
Suppose $\BF=\{\BF_i:i=0,\ldots,n\}$ is a strong essential system of
the form \bref{eq-algsparse}. Then, for each $i\in\{0,1,\ldots,n\}$,
the degree of the sparse resultant in $\bu_i$ is a positive integer,
equal to the {\em mixed volume}
\begin{eqnarray} \quad&\quad&\mathcal
{M}(\CQ_0,\ldots,\CQ_{i-1},\CQ_{i+1},\ldots,\CQ_n)
=\sum_{\text{J}\subset
\{0,\ldots,i-1,i+1,\ldots,n\}}(-1)^{n-|\text{J}|}\vol(\sum_{j\in\text{J}}\CQ_j)
\nonumber
\end{eqnarray} where $\CQ_i$ is the {  Newton polytope} of $\BF_i$,  $\vol(\CQ)$ means the $n$-dimensional volume of
$\CQ\subset\mathbb{R}^n$ and $\CQ_1+\CQ_2$ means the Minkowski sum
of $\CQ_1$ and $\CQ_2$.
\end{thm}

\subsection{Sparse difference resultant as algebraic sparse resultant}
With the above preparation, we now give the main theorem of this
section.

\begin{thm}\label{th-dares}
Let $\SR$ be the  sparse difference resultant of a  Laurent
transformally essential system \bref{eq-sparseLaurent}.
Then we can derive a strong essential generic algebraic sparse
polynomial system $\SC$ from \bref{eq-sparseLaurent}, such that the
sparse resultant of $\SC$ is equal to $\SR$.
%
\end{thm}
\proof 
By Theorem \ref{th-rankessential}, the system
\bref{eq-sparseLaurent} has a unique super-essential subsystem
$\P_{\TT}$. Without loss of generality,  assume $\TT =
\{0,1,\ldots,p\}$. For each $i\in\{0,\ldots, p\}$, let
$k_i=\Jac((A_{\TT})_{\hat{i}})$ as defined in Theorem \ref{th-jb12}
and let
$\overrightarrow{K}=(k_0,k_1,\ldots,k_p)\in\mathbb{N}_0^{p+1}$.
Similar to \bref{eq-pk1}, let \begin{equation}\label{eq-apk}
 \PC=\P^{[\overrightarrow{K}]} = \bigcup_{i=0}^p\norm(\P_i)^{[k_i]}\end{equation}
be the  prolongation of $\P_{\TT}$ with respect to
$\overrightarrow{K}$. Note that $|\PC|=\sum_{i=0}^p k_i + p+1$.
Regarding $\PC$ as a set of algebraic polynomials in  $y_i^{(j)}$
with coefficients $U=\cup_{i=0}^n\bu_{i}^{[k_i]}$, then $\PC$ is a
generic sparse polynomial system.

A total ordering for polynomials in $\PC$ is assigned as follows:
$\sigma^k \P_i < \sigma^{l} \P_j$ if and only if $i < j$ or $i=j$
and $k < l$. A total ordering $\succ$ among subsets of $\PC$ is the
same as the one given in Section 5.1.
By Theorem~\ref{th-jb12}, $\rk(M_\PC)\le \sum_{i=0}^p k_i +
p=|\PC|-1$. By Lemma \ref{lm-sub}, we can construct an essential
subsystem $\PC_1$ of $\PC$ with minimal ranking. Let $\SR_1$ be the
sparse resultant of $\PC_1$, that is, $(\PC_1)\cap\Q[U] = (\SR_1)$.

We claim that $\SR_1=c\SR$ for some $c\in\Q$.
 Since $\P_\TT$ is super
essential, for each $i\in\TT$, $\ord(\SR,\bu_i)\ge0$. By Theorem
\ref{th-jb12}, $\SR\in(\PC)$.
Let $\PC_2$ be the elements of $\PC$ whose coefficients appear
effectively in $\SR$. By Lemma \ref{th-criess}, $\PC_2$ is essential
and $(\PC_2)\cap\Q[U] = (\SR)$.
Let $k_1$ and $k_2$ be the largest integers such that
$\sigma^{k_1}\P_p\in\PC_1$ and $\sigma^{k_2}\P_p\in\PC_2$. Since
$\PC_1$ and $\PC_2$ are essential, $\ord(\SR_1,\bu_p)=k_1$ and
$\ord(\SR,\bu_p)=k_2$.
Since $\PC_2\succ\PC_1$, $k_1\le k_2$. Since
$\SR_1\in(\PC_1)\cap\Q[U]\subset
[\P_\TT]\cap\Q\{\bu_0,\ldots,\bu_p\}$, by
Lemma~\ref{lm-genericresultant}, $k_1 \ge k_2$. Hence, $k_1=k_2$.
Since
$\SR_1\in[\P_\TT]\cap\Q\{\bu_0,\ldots,\bu_p\}=\sat(\SR,\ldots)$ and
$\ord(\SR_1,\bu_p)=\ord(\SR,\bu_p)$, $\SR_1$ is algebraically
reduced to zero by $\SR$. Since both $\SR$ and $\SR_1$ are
irreducible, $\SR= c\SR_1$ where $c\in\Q$.

Apply Lemma \ref{lm-vess} to $\PC_1$, we obtain a variable-essential
 system $\PC_2$ satisfying $(\PC_2)\cap \Q[U] = (\SR)$.
Then apply Lemma \ref{lm-algess} to $\PC_2$, we obtain a strong
essential generic system $\SC$ satisfying $(\SC)\cap \Q[U] = (\SR)$
and the existence of $\SC$ is proved.

We will show that $\SC$ can be given algorithmically. Through the
above procedures, only Lemma \ref{lm-sub} is not constructive.
Since $\PC$ contains an essential subsystem, we can simply check
each subsystems $\SC$ of $\PC$ to see whether $\SC$ is essential and
find the one with minimal ranking. Note that $\SC$ is essential if
and only if $\rk(M_\SC) = |\SC|-1$ and any proper subset $\C$ of
$\SC$ satisfies $\rk(M_\C)= |\C|$. \qed

%

\begin{exmp}
Let $n=3$. Denote $y_{ij}=y_{i}^{(j)}$ and  let
$\P=\{\P_0,\P_1,\P_2,\P_3\}$  where
\begin{eqnarray*}
 \P_0 &=& u_{00}+u_{01}y_{11}^2y_{21}^2y_3+u_{02}y_{1}^2y_2y_3,\\
 \P_1 &=&
 u_{10}+u_{11}y_{12}^4y_{22}^4y_{31}^2+u_{12}y_{11}^2y_{21}y_{31},\\
 \P_2 &=& u_{20}+u_{21}y_{11}^2y_{21}^2y_3+u_{22}y_{1}^2y_2y_3,\\
 \P_3 &=& u_{30}+u_{31}y_{11}y_3.
\end{eqnarray*}
It is easy to show that $\P$ is a Laurent transformally essential
system and $\TT=\{0,1,2\}$. Clearly, $\Jac((A_{\TT})_{\hat{0}}) = 3,
\Jac((A_{\TT})_{\hat{1}}) = 2$ and $ \Jac((A_{\TT})_{\hat{2}})= 3$.
Using the notations in Theorem~\ref{th-dares}, we have $\PC =
\{\P_0^{[3]},\P_1^{[2]},\P_2^{[3]}\}$, and we can compute an
essential subset $\PC_1$ with minimal ranking. Here, we have $\PC_1
= \{ \sigma\P_0, \P_1,\sigma\P_2 \} $.
Using the variable order $y_{11} < y_{12} < y_{21} < y_{22} <
y_{31}$ to obtain the symbolic support matrix of $\PC_1$, the first
$2\times 2$ sub-matrix of $M_{\PC_1}$ is of rank $2$. By the proof
of Lemma~\ref{lm-vess}, we set $y_{21}, y_{22}, y_{31}$ to $1$ to
obtain a variable essential system $\PC_2 =
\{\widetilde{\sigma\P_0}, \widetilde{\P_1},
\widetilde{\sigma\P_2}\}$ where
\begin{eqnarray*}
 \widetilde{\sigma\P_0} &=& u_{00}^{(1)} + u_{01}^{(1)}y_{12}^2 +
u_{02}^{(1)}y_{11}^2,\\
\widetilde{\P_1} &=& u_{10}+u_{11}y_{12}^4 + u_{12}y_{11}^2,\\
\widetilde{\sigma\P_2} &=& u_{20}^{(1)}+u_{21}^{(1)}y_{12}^2 +
u_{22}^{(1)}y_{11}^2.
\end{eqnarray*}
Apply Lemma~\ref{lm-algess} to $\PC_2$, set $z_1 = y_{11}^2, z_2 =
y_{12}^2$, we obtain a strong essential generic system $\PC_3 = \{
Q_0, Q_1 , Q_2 \}$ where
\begin{eqnarray*}
 Q_0 &=& u_{00}^{(1)} + u_{01}^{(1)}z_2 + u_{02}^{(1)}z_1,\\
 Q_1 &=& u_{10}+u_{11}z_2^2 + u_{12}z_1,\\
 Q_2 &=& u_{20}^{(1)}+u_{21}^{(1)}z_2 + u_{22}^{(1)}z_1.
\end{eqnarray*}
The sparse resultant of  system $\PC_3$ is $R =
u_{10}(u_{02}^{(1)}u_{21}^{(1)}-u_{01}^{(1)}u_{22}^{(1)})^2+
u_{11}(u_{00}^{(1)}u_{22}^{(1)}-u_{02}^{(1)}u_{20}^{(1)})^2 +
u_{12}(u_{00}^{(1)}u_{21}^{(1)}-u_{01}^{(1)}u_{20}^{(1)})(u_{02}^{(1)}u_{21}^{(1)}-u_{01}^{(1)}u_{22}^{(1)})$,
which is the sparse difference resultant of $\P$.
\end{exmp}

The following corollary is a direct consequence of the proof of
Theorem \ref{th-dares} and paper \cite{dandrea1}.
\begin{cor}\label{co-smat1}
The  sparse difference resultant $\SR$ of a Laurent transformally
essential system \bref{eq-sparseLaurent} can be represented as the
quotient of two determinants  whose elements are $u_{ij}^{(k)}$ or
their sums for certain $i\in\{0,\ldots,n\},j\in\{0,\ldots,l_i\}$ and
$k\in\{0,\ldots,{J}_i\}$, where ${J}_i$ is the Jacobi number of the
system \bref{eq-sparseLaurent} as defined in Section \ref{sec-ord1}.
\end{cor}

\begin{rem}\label{re-bkksr}
It is desirable to derive a degree bound for $\SR$ from Theorem
\ref{th-dares}. Let $\SC$ be the strong essential set mentioned in
the theorem. Then, the degree of $\SR$ is equal to the mixed volume
of $\SC$ by Theorem \ref{le-alg-sparse-deg}. The problem is how to
express the mixed volume of $\SC$ in terms of certain quantities of
$\P_\TT$ without computing $\SC$.
\end{rem}

\section{A single exponential algorithm to compute the sparse difference resultant}\label{sec-algorithm}
In this section, we give an algorithm to compute the sparse
difference resultant for a  Laurent transformally essential
 system with single exponential complexity.
The idea is to estimate the degree bounds for the resultant and then
to use linear algebra to find the coefficients of the resultant.

\subsection{Degree bound for sparse difference resultant}

In this section,  we give an upper bound for the degree of the
sparse difference resultant, which will be crucial to our algorithm
to compute the sparse resultant. Before proposing the main theorem,
we first give some algebraic results which will be needed in the
proof.

\begin{lem}\label{lm-elimination}\cite[Theorem 6.2]{li} Let $\I$ be a prime ideal in $K[x_1,\ldots,x_n]$
and $\I_k=\I\cap K[x_1,\ldots,x_k]$ for any $1\leq k\leq n$. Then
$\deg(\I_k)\leq \deg(\I)$.
\end{lem}

\begin{lem} \cite[Corollary 2.28]{vogel} \label{le-vogel}
Let $V_1,\ldots,V_r\subset\textbf{P}^n\,(r\geq 2)$ be pure
dimensional projective varieties in $\textbf{P}^n$. Then
$$\prod_{i=1}^r\deg(V_i)\geq \sum_{C}\deg(C)$$
where $C$ runs through all irreducible components of
$V_1\cap\cdots\cap V_r$.
\end{lem}


Now we are ready to give the main theorem of this section.
\begin{thm} \label{th-spardeg}
Let $\P_0,\ldots,\P_n$ be a Laurent transformally essential system
of form  \bref{eq-sparseLaurent} with  $\ord(\norm(\P_i))=s_i$ and
$\deg(\norm(\P_i),\Y)$ $=m_i$.  Suppose
$\norm(\P_{i})=\sum_{k=0}^{t_i}u_{ik}N_{ik}$ and $J_i$ is the Jacobi
number of
$\{\norm(\P_0),\ldots,\norm(\P_{n})\}\backslash\{\norm(\P_{i})\}$.
Denote $m=\max_i\{m_i\}$.
 Let $\SR(\bu_0,\ldots,\bu_n)$ be the sparse difference
resultant of $\P_i\,(i=0,\ldots,n)$.  Suppose $\ord(\SR,\bu_i)=h_i$
for each $i$. Then the following assertions hold:
\begin{description}
\item[1)]
$\deg(\SR)\leq \prod_{i=0}^n
    (m_i+1)^{h_i+1}\leq (m+1)^{\sum_{i=0}^n(J_i+1)}$, where $m=\max_i\{m_i\}$.

\item[2)]
$\SR$ has a representation
\begin{equation}\label{eq-deg5}
\prod_{i=0}^n \prod_{k=0}^{h_i} (N_{i0}^{(k)})^{\deg(\SR)}\cdot
\SR=\sum_{i=0}^n\sum_{k=0}^{h_i}G_{ik} \norm(\P_{i} )^{(k)}
\end{equation}
where
 $G_{ik}\in \Q[\bu_0^{[h_0]},\ldots,\bu_n^{[h_n]},\Y^{[h]}]$
and $h=\max\{h_i+e_i\}$ such that $\deg(G_{ik} \norm(\P_{i})
^{(k)})\leq [m+1+\sum_{i=0}^n(h_i+1)\deg(N_{i0})]\deg(\SR)$.
\end{description}
\end{thm}
\proof  In $\SR$, let $u_{i0}$ be replaced by $\big(
\norm(\P_{i})-\sum_{k=1}^{t_i}u_{ik}N_{ik}\big)/N_{i0}$ for each
$i=0,\ldots,n$ and let $\SR$ be expanded as a difference polynomial
in $\norm(\P_{i})$ and their transforms. Then there exist
$a_{ik}\in\mathbb{N}$ and polynomials $G_{ik}$ such that
$\prod_{i=0}^n\prod_{k=0}^{h_i}\big(
N_{i0}^{(k)}\big)^{a_{ik}}\SR=\sum_{i=0}^n\sum_{k=0}^{h_i}G_{ik}
\norm(\P_{i}) ^{(k)}+T$ with $T\in\Q\{\bu,\Y\}$ free from $u_{i0}$.
Since $T\in\CI=[\norm(\P_0),\ldots,\norm(\P_n)]:\mathbbm{m}$, $T$
vanishes identically, for $\CI\cap\Q\{\bu,\Y\}=\{0\}$ by
Theorem~\ref{th-Mcodim1}. Thus,
$$\prod_{i=0}^n\prod_{k=0}^{h_i}\big(
N_{i0}^{(k)}\big)^{a_{ik}}\SR=\sum_{i=0}^n\sum_{k=0}^{h_i}G_{ik}
\norm(\P_{i}) ^{(k)}.$$

1) Let
$\mathcal{J}=\big(\norm(\P_{0})^{[h_0]},\ldots,\norm(\P_{n})^{[h_n]}\big):\mathbbm{m}^{[h]}$
be an algebraic ideal in $\mathcal{R}=\Q[\Y^{[h]},\bu_0^{[h_0]},$
$\ldots,\bu_n^{[h_n]}]$ where $h=\max_i\{h_i+s_i\}$ and
$\mathbbm{m}^{[h]}$ is the set of  all monomials in $\Y^{[h]}$. Then
$\SR\in\mathcal{J}$ by the above equality.
 Let $\eta=(\eta_1,\ldots,\eta_n)$ be a generic zero of $[0]$ over $\Q\langle\bu\rangle$ and denote $\zeta_i=-\sum_{k=1}^{t_i}u_{ik}\frac{N_{ik}(\eta)}{N_{i0}(\eta)}\,(i=0,\ldots,n)$.
 It is easy to show that
     $\mathcal {J}$ is a   prime ideal in $\mathcal{R}$ with a generic zero
     $(\eta^{[h]};\widetilde{\bu},\zeta_0^{[h_0]},\ldots,\zeta_n^{[h_n]})$
     and $\mathcal
     {J}\cap\Q[\bu_0^{[h_0]},\ldots,\bu_n^{[h_n]}]=(\SR)$, where $\widetilde{\bu}=\cup_i\bu_i^{[h_i]}\backslash
     \{u_{i0}^{[h_i]}\}$.
    Let $H_{ik}$ be the homogeneous polynomial corresponding to
    $ \norm(\P_{i}) ^{(k)}$ with $x_0$ the variable of
    homogeneity. Then $\mathcal {J}^0=((H_{ik})_{1\leq i\leq n;0\leq k\leq h_i}):\widetilde{\mathbbm{m}}$
    is a prime ideal in $\Q[x_0,\Y^{[h]},\bu_0^{[h_0]},\ldots,\bu_n^{[h_n]}]$ 
    where $\widetilde{\mathbbm{m}}$ is the whole set of
      monomials in $\Y^{[h]}$ and $x_0$. And $\deg(\mathcal {J}^0)
      =\deg(\mathcal{J})$.

Since $\V((H_{ik})_{1\leq i\leq n;0\leq k\leq h_i})=\V(\mathcal
{J}^0)\cup\V(H_{ik},x_0)\bigcup
    \cup_{j,l}\V(H_{ik},y_{j}^{(l)})$,  $\V(\mathcal {J}^0)$ is an irreducible component of
$\V((H_{ik})_{1\leq i\leq n;0\leq k\leq h_i})$. By
Lemma~\ref{le-vogel}, $\deg(\mathcal {J}^0)$ $\leq
\prod_{i=0}^n\prod_{k=0}^{h_i}(m_i+1)=\prod_{i=0}^n(m_i+1)^{h_i+1}$.
Thus, $\deg(\mathcal {J})\leq\prod_{i=0}^n(m_i+1)^{h_i+1}$.    Since
$\mathcal {J}\cap\Q[\bu_0^{[h_0]},\ldots,\bu_n^{[h_n]}]=(\SR)$, by
Lemma~\ref{lm-elimination},    $\deg(\SR)\leq
    \deg(\mathcal {J})\leq\prod_{i=0}^n(m_i+1)^{h_i+1}\leq (m+1)^{\sum_{i=0}^n(J_i+1)}$
    follows. The last inequality holds because  $h_i\leq J_i$ by Theorem~\ref{th-jacobi-order3}.

2) To obtain the degree bounds for the above  representation of
$\SR$, that is, to estimate $\deg(G_{ik}\norm(\P_i)^{(k)})$ and
$a_{ik}$,
 we take each monomial $M$ in $\SR$ and
 substitute $u_{i0}$ by $\big(\norm(\P_{i})-\sum_{k=1}^{l_i}u_{ik}N_{ik}\big)/N_{i0}$
 into $M$ and then expand it. To be more precise,  we take
 one  monomial $M(\bu;u_{00},\ldots,$ $u_{n0}) =\bu^{\gamma}\prod_{i=0}^n\prod_{k=0}^{h_i}(u_{i0}^{(k)})^{d_{ik}}$
 with $|\gamma|+\sum_{i=0}^n\sum_{k=0}^{h_i}d_{ik}=
 \deg(\SR)$  for an example, where $\bu^{\gamma}$ represents a difference monomial in $\bu$ and their transforms with exponent vector $\gamma$.
Then
 {\small$$M(\bu;u_{00},\ldots,u_{n0})=
 \bu^{\gamma}\prod_{i=0}^n\prod_{k=0}^{h_i} \Big(\big(\norm(\P_{i})-\sum_{k=1}^{l_i}u_{ik}N_{ik}\big)^{(k)}\Big) ^{d_{ik}}\Big/\prod_{i=0}^n\prod_{k=0}^{h_i}\big(N_{i0}^{(k)}\big)^{d_{ik}}.$$}
When expanded, every term of
$\prod_{i=0}^n\prod_{k=0}^{h_i}\big(N_{i0}^{(k)}\big)^{d_{ik}}M$  is
of degree bounded by
$|\gamma|+\sum_{i=0}^n\sum_{k=0}^{h_i}(m_i+1)d_{ik}\leq
(m+1)\deg(\SR)$ in $\bu_0^{[h_0]},\ldots,\bu_n^{[h_n]}$ and
 $\Y^{[h]}$. Suppose $\SR=\sum_Ma_{M}M$ and $a_{ik}\geq\max_{M}\{d_{ik}\}$.
 Then $$\prod_{i=0}^n\prod_{k=0}^{h_i}\big( N_{i0}^{(k)}\big)^{a_{ik}}\SR=\sum_{i=0}^n\sum_{k=0}^{h_i}G_{ik} \norm(\P_{i}) ^{(k)}$$
 with $\deg(G_{ik}\norm(\P_i)^{(k)})\leq (m+1)\deg(\SR)+\sum_{i=0}^n\sum_{k=0}^{h_i}\deg(N_{i0})a_{ik}$.
Clearly, we can take $a_{ik}=\deg(\SR)$ and then
$\deg(G_{ik}\norm(\P_i)^{(k)})\leq
(m+1+\sum_{i=0}^n(h_i+1)\deg(N_{i0}))\deg(\SR)$.
 Thus,
 \bref{eq-deg5} follows.
 \qed

For a   transformally essential difference polynomial system with
degree $0$ terms, the second part of Theorem \ref{th-spardeg} can be
improved as follows.
\begin{cor} \label{cor-pspardeg}
Let $\P_i=u_{i0}+\sum_{k=1}^{l_i}u_{ik}N_{ik}\,(i=0,\ldots,n)$ be a
transformally essential difference polynomial system
 with $m=\max_i\{\deg( \P_i ,\Y)\}$ and $J_i$ the  Jacobi number of $\{\P_0,\ldots,\P_{n}\}\backslash\{\P_{i}\}$.
 Let $\SR(\bu_0,\ldots,$ $\bu_n)$ be the sparse difference
resultant of $\P_i\,(i=0,\ldots,n)$.  Suppose $\ord(\SR,\bu_i)=h_i$
for each $i$ and $h=\max\{h_i+s_i\}$. Then  $\SR$ has a
representation
$$\SR(\bu_0,\ldots,\bu_n)=\sum_{i=0}^n\sum_{j=0}^{h_i}G_{ij}\P_{i}^{(j)}$$
where $G_{ij}\in \Q[\bu_0^{[h_0]},\ldots,\bu_n^{[h_n]},\Y^{[h]}]$
 such that $\deg(G_{ij}\P_{i}^{(j)})\leq
(m+1)\deg(\SR)\leq(m+1)^{\sum _{i=0}^n(J_i+1)+1}$.
\end{cor}
\proof It is direct consequence of Theorem~\ref{th-spardeg} by
setting $N_{i0}=1$. \qed

The following result gives an effective difference Nullstellensatz
under certain conditions.

\begin{cor}
Let $f_0,\ldots,f_n\in\F\{y_1,\ldots,y_n\}$ have no common solutions
with $\deg(f_i)\leq m$. Let
$\Jac(\{f_0,\ldots,f_{n}\}\backslash\{f_{i}\})=J_i$. If the sparse
difference resultant of $f_0,\ldots,f_n$ is nonzero, then there
exist $H_{ij}\in\F\{y_1,\ldots,y_n\}$ s.t.
$\sum_{i=0}^n\sum_{j=0}^{J_i}H_{ij}f_{i}^{(j)}$ $=1$ and
$\deg(H_{ij}f_{i}^{(j)})\leq (m+1)^{\sum_{i=0}^n(J_i+1)+1}$.
\end{cor}
\proof The hypothesis implies that $\P(f_i)$ form a transformally
essential system. Clearly, $\SR(\bu_0,\ldots,\bu_n)$ has the
property stated in  Corollary~\ref{cor-pspardeg}, where $\bu_i$ are
coefficients of $\P(f_i)$. The result follows directly from
Corollary~\ref{cor-pspardeg} by specializing $\bu_i$ to the
coefficients of $f_i$. \qed

\subsection{A single exponential algorithm to compute sparse difference resultant}

If a polynomial $R$ is the linear combination of some known
polynomials $F_i(i=1,\ldots,s)$, that is $R=\sum_{i=1}^s H_i F_i$,
 and we know the upper bounds of the degrees of $R$ and $H_iF_i$, then a general
idea to estimate the computational complexity of $R$ is  to use
linear algebra to find the coefficients of $R$.

For sparse difference resultant, we already have given its degree
bound and the degrees of the expressions in the linear combination
in Theorem \ref{th-spardeg}.

Now, we give the algorithm {\bf SDResultant} to compute sparse
difference resultants based on the linear algebra techniques.  The
algorithm works adaptively by searching for $\SR$ with an order
vector $(h_0,\ldots,h_n)\in\mathbb{N}_0^{n+1}$ with $h_i \leq J_i$
by Theorem \ref{th-spardeg}. Denote $o=\sum_{i=0}^n h_i$. We start
with $o=0$. And for this $o$, choose one vector $(h_0,\ldots,h_n)$
at a time. For this $(h_0,\ldots,h_n)$, we search for $\SR$ from
degree $d=1$. If we cannot find an $\SR$ with such a degree, then we
repeat the procedure with degree $d+1$ until
$d>\prod_{i=0}^n(m_i+1)^{h_i+1}$. In that case, we choose another
$(h_0,\ldots,h_n)$ with $\sum_{i=0}^nh_i=o$. But if for all
$(h_0,\ldots,h_n)$ with $h_i\leq J_i$ and $\sum_{i=0}^nh_i=o$, $\SR$
cannot be found, then we repeat the procedure with $o+1.$ In this
way, we will find an $\SR$ with the smallest order satisfying
equation \bref{eq-deg5}, which is the sparse resultant.

\begin{algorithm}[ht]\label{alg-dresl}
  \caption{\bf --- SDResultant($\P_0,\ldots,\P_n$)} \smallskip
  \Inp{A generic Laurent transformally essential system $\P_0,\ldots,\P_n$.}\\
  \Outp{The sparse difference resultant $\SR(\bu_0,\ldots,\bu_n)$ of $\P_0,\ldots,\P_n$.}\medskip

  \noindent
  1. For $i=0,\ldots,n$, set $\norm(\P_i)=\sum_{k=0}^{l_i}u_{ik}N_{ik}$ with $\deg(N_{i0})\leq\deg(N_{ik})$.\\
      \SPC Set  $m_i=\deg(\norm(\P_i))$, $m_{i0}=\deg(N_{i0})$, $\bu_i=\coeff(\P_i)$ and $|\bu_i|=l_i+1$.\\
      \SPC Set  $s_{ij}=\ord(\norm(\P_i),y_j)$, $A=(s_{ij})$ and compute $J_i=\Jac(A_{\hat{i}})$.\\
  2. Set $\SR=0$, $o=0$,  $m=\max_i\{m_i\}$.\\
  3. While $\SR=0$ do\\
   \SPC 3.1. For each  $(h_0,\ldots,h_n)\in \mathbb{N}_0^{n+1}$ with
   $\sum_{i=0}^n h_i$$=o$ and $h_i\leq J_i$  do \\
   \SPC\SPC 3.1.1. $U=\cup_{i=0}^n\bu_i^{[h_i]}$,
   $h=\max_i\{h_i+e_i\}$, $d=1$.  \\
   \SPC\SPC 3.1.2. While $\SR=0$ and $d\leq
   \prod_{i=0}^n(m_i+1)^{h_i+1}$ do \\
   \SPC\SPC\SPC 3.1.2.1. Set $\SR_0$ to be a homogeneous GPol of degree $d$ in $U$.\\
   \SPC\SPC\SPC 3.1.2.2. Set $\bc_{0}=\coeff(\SR_0,U)$.\\
   \SPC\SPC\SPC 3.1.2.3. Set $H_{ij}(i=0,\ldots,n;j=0,\ldots,h_i)$ to be GPols of degree \\
  \SPC\SPC\SPC\SPC\SPC\quad$[m+1+\sum_{i=0}^n(h_i+1)m_{i0}]d-m_i-1$
   in $\Y^{[h]},U$.\\
   \SPC\SPC\SPC 3.1.2.4. Set $\bc_{ij}=\coeff(H_{ij}, \Y^{[h]}\cup U)$.
   \\
   \SPC\SPC\SPC 3.1.2.5. Set $\mathcal {P}$ to be the set of coefficients of $\prod_{i=0}^n\prod_{k=0}^{h_i}
(N_{i0}^{(k)})^d\SR_0 -$
\\
  \SPC\SPC\SPC \SPC \SPC\quad $\sum_{i=0}^n\sum_{j=0}^{h_i}H_{ij}(\norm(\P_{i}))^{(j)}$  as a  polynomial in
  $\Y^{[h]},U$.\\
   \SPC\SPC\SPC 3.1.2.6. Solve the linear equation $\mathcal {P}=0$ in
   variables $\bc_{0}$ and
  $\bc_{ij}.$\\
   \SPC\SPC\SPC 3.1.2.7. If $\bc_0$ has a nonzero solution, then
   substitute it into  $\SR_0$ to \\
  \SPC\SPC\SPC \SPC \SPC\quad get $\SR$ and go to  Step 4, else $\SR=0$.\\
  \SPC\SPC\SPC 3.1.2.8. d:=d+1.\\
  \SPC 3.2. o:=o+1.\\
  4. Return $\SR$.\medskip

  \noindent /*/\;  GPol stands for generic algebraic polynomial.\smallskip

  \noindent /*/\; $\coeff(P,V)$ returns the set of coefficients of $P$ as an ordinary  polynomial in
  variables $V$.
\smallskip
\end{algorithm}

\begin{thm}\label{th-cdres}
Let $\P_0,\ldots,\P_n$ be a Laurent transformally essential system
of form \bref{eq-sparseLaurent}. Denote
$\P=\{\norm(\P_0),\ldots,\norm(\P_n)\}$, $J_i=\Jac(\P_{\hat{i}})$,
$J=\max_{i}J_i$ and $m=\max_{i=0}^n \deg(\P_i,\Y)$.
Algorithm {\bf SDResultant} computes the sparse difference resultant
$\SR$ of $\P_0,\ldots,\P_n$ with the following complexities: \vf

 1) In terms of a degree bound $D$ of $\SR$, the algorithm needs at most
$O(D^{O(lJ)}(nJ)^{O(lJ)})$ $\Q$-arithmetic operations, where
$l=\sum_{i=0}^n(l_i+1)$
 is the
size of all $\P_i$.

 \vf

 2) The algorithm needs at most $
O(m^{O(nlJ^2)}(nJ)^{O(lJ)})$ $\Q$-arithmetic operations.
\end{thm}
\proof The algorithm finds a difference polynomial $P$ in
$\Q\{\bu_0,\ldots,\bu_n\}$ satisfying equation \bref{eq-deg5}, which
has the smallest order and the smallest degree in those with the
same order. Existence for such a difference polynomial is guaranteed
by Theorem \ref{th-spardeg}. By the definition of sparse difference
resultant,  $P$ must be $\SR$.

We will estimate the complexity of the algorithm below. Denote $D$
to be the degree bound of $\SR.$ By Theorem \ref{th-spardeg}, $D\leq
(m+1)^{\sum_{i=0}^n(J_i+1)}$.
In each loop of Step 3, the complexity of the algorithm is clearly
dominated by Step 3.1.2, where we need to solve a system of linear
equations $\mathcal {P}=0$ over $\Q$ in $\bc_{0}$ and $\bc_{ij}$.
It is easy to show that $|\bc_{0}|={d+L-1\choose L-1}$ and
$|\bc_{ij}|={d_1-m_i-1+L+n(h+1)\choose L+n(h+1)}$, where
$L=\sum_{i=0}^n (h_i+1)(l_i+1)$ and
$d_1=[m+1+\sum_{i=0}^n(h_i+1)m_{i0}]d$. Then $\mathcal {P}=0$ is a
linear equation system with $N={d+L-1\choose
L-1}+\sum_{i=0}^n(h_i+1){d_1-m_i-1+L+n(h+1)\choose L+n(h+1)}$
variables and $M={d_1+L+n(h+1)\choose L+n(h+1)}$ equations. To solve
it, we need at most $(\max\{M,N\})^{\omega}$ arithmetic operations
over $\Q$, where $\omega$ is the matrix multiplication exponent and
the currently best known $\omega$ is 2.376.

The iteration in Step 3.1.2 may go through $1$ to
$\prod_{i=0}^n(m_i+1)^{h_i+1}\leq(m+1)^{\sum_{i=0}^n(J_i+1)}$, and
the iteration in Step 3.1 at most will repeat
$\prod_{i=0}^n(J_i+1)\leq (n+1)(J+1)$ times, where $J=\max_iJ_i$.
And by Theorem \ref{th-spardeg}, Step 3 may loop from $o=0$ to
$\sum_{i=0}^n(J_i+1)$. The whole algorithm needs at most
\begin{eqnarray}& &\sum_{o=0}^{\sum_{i=0}^n(J_i+1)}\sum_{ h_i\leq
J_i\atop\sum_{i}h_i=o}\sum_{d=1}^{\prod_{i=0}^n(m_i+1)^{h_i+1}}\big(\max\{M,
N\}\big)^{2.376}\nonumber\\ & \leq& O(D^{O(lJ)}(nJ)^{O(lJ)})\leq
O(m^{O(nlJ^2)}(nJ)^{O(lJ)})\nonumber \end{eqnarray}
arithmetic operations over $\Q$. In the above inequalities, we
assume that $(m+1)^{\sum_{i=0}^n(J_i+1)+1 }\geq l(n+1)J$ and
 $l\geq (n+1)^2$, where $l=\sum_{i=0}^n(l_i+1)$.
Our complexity assumes an $O(1)$-complexity cost for all field
operations over $\mathbb{Q}$. Thus, the complexity follows.\qed

\begin{rem}
As we indicated at the end of Section 3.3, if we  first  compute the
super-essential set $\TT$, then the algorithm can be improved by
only considering the Laurent difference polynomials
$\P_i\,(i\in\TT)$ in the linear combination of the sparse resultant.
\end{rem}

\begin{rem}
Algorithm {\bf SDResultant} can be improved by using a better search
strategy. If $d$ is not big enough, instead of checking $d+1$, we
can check $2d$. Repeating this procedure, we may find a $k$ such
that $2^k\leq\deg(\SR)\leq2^{k+1}$. We then bisecting the interval
$[2^k,2^{k+1}]$ again to find the proper degree for $\SR$. This may
lead to a better complexity, which is still single exponential.
\end{rem}

\vskip 10pt For difference polynomials with non-vanishing degree
terms, a better degree bound is given in
Corollary~\ref{cor-pspardeg}. Based on this bound, we can simplify
the Algorithm {\bf SDResultant} to compute the sparse difference
resultant by removing the computation for $\norm(P_i)$ and $N_{i0}$
in the first step where $N_{i0}$ is exactly equal to 1.
\begin{thm}\label{th-0cdres}
Algorithm {\bf SDResultant} computes the sparse difference resultant
for a transformally essential system
$\{\P_i=u_{i0}+\sum_{k=1}^{l_i}u_{ik}N_{ik}\}$ with at most
$O(n^{3.376}J^{O(n)}m^{O(nlJ^2)})$
 $\Q$-arithmetic
operations.
\end{thm}
\proof Follow the proof process of Theorem~\ref{th-cdres}, it can be
shown that the complexity is $O(n^{3.376}J^{O(n)}m^{O(nlJ^2)})$.
 \qed

\section{Difference resultant} \label{differenceresultant}
In this section, we introduce the notion of difference resultant and
prove its basic properties.

\begin{defn}
Let $\mathbbm{m}_{s,r}$ be the set of all difference monomials in
$\Y$  of  order $\leq s$ and degree $\leq r$. Let $\bu=\{u_M\}_{M\in
\mathbbm{m}_{s,r}}$ be a set of difference indeterminates over $\Q$.
Then, $\P=\sum_{M\in \mathbbm{m}_{s,r}}u_M M$ is called a {\em
generic difference polynomial} of order $s$ and degree $r$.
\end{defn}

Throughout this section, a generic difference polynomial is assumed
to be of degree greater than zero. For any vector
$\alpha=(a_1,\ldots,a_m)\in\mathbb{Z}^m$ and $\X=(x_1,\ldots,x_m)$,
denote $x_1^{a_1}x_2^{a_2}\cdots x_m^{a_m}$ by $\X^{\alpha}$. Let
\begin{equation}\label{eq-genericdiffpoly}\P_i=u_{i0}+\sum_{\begin{array}{c}
\alpha \in \mathbb{Z}^{n(s_i+1)}_{\geq 0} \\ 1\leq |\alpha|\leq
 m_i
 \end{array}}u_{i\alpha}(\Y^{[s_i]})^\alpha\,\,(i=0,1,\ldots,n)
 \end{equation}
be $n+1$ generic difference polynomials in $\Y$ of order $s_i$,
degree $m_i$  and coefficients $\bu_i$. Since $\{1,y_1,\ldots,y_n\}$
is contained in the support of each $\P_i$,
$\{\P_0,\P_1,\ldots,\P_n\}$ is a super-essential system and the
sparse difference resultant
$\Res_{\P_0,\ldots,\P_n}(\bu_0,\ldots,\bu_n)$ exists. We define
$\Res_{\P_0,\ldots,\P_n}(\bu_0,\ldots,\bu_n)$ to be the {\em
difference resultant} of $\P_0,\ldots,\P_n.$

Because each generic difference polynomial $\P_i$ contains all
difference monomials of order bounded by $s_i$ and total degree at
most $m_i$, the difference resultant is sometimes called the {\em
dense} difference resultant, in contrary to the sparse difference
resultant.

The difference resultant satisfies all the properties we have proved
for sparse difference resultants in previous sections. Apart from
these, the difference resultant possess other better properties to
be given in the rest of this section.

\subsection{Exact Order and Degree}
In this section, we will give the precise order and degree for the
difference resultant, which is of BKK-style \cite{bkk,cox}.

\begin{thm}\label{th-resultant-deg}
Let $\P_i\,(i=0,\ldots,n)$ be generic difference polynomials of form
\bref {eq-genericdiffpoly} with order $s_i$, degree $m_i$, and
coefficients $\bu_i$, respectively. Let $\SR(\bu_0,\ldots,\bu_n)$ be
the difference resultant of $\P_0,\ldots,\P_n$. Denote
$s=\sum_{i=0}^n s_i$. Then $\SR(\bu_0,\ldots,\bu_n)$ is also the
algebraic sparse resultant of $\P_0^{[s-s_0]},\ldots,\P_n^{[s-s_n]}$
treated as polynomials in $\Y^{[s]}$, and for each
$i\in\{0,1,\ldots,n\}$ and $k=0,\ldots,s-s_i$,
\begin{eqnarray}
 &&\ord(\SR,\bu_i)=s-s_i\label{eq-gord}\\
 &&\deg(\SR,\bu_i^{(k)})=\mathcal {M}\big((\CQ_{jl})_{j\neq i,0\leq
 l\leq s-s_j},\CQ_{i0},\ldots,\CQ_{i,k-1},\CQ_{i,k+1},\ldots,\CQ_{i,s-s_i}\big)\label{eq-gdeg}
\end{eqnarray}
where $\CQ_{jl}$ is the Newton polytope of $\P_j^{(l)}$ as a
polynomial in $\Y^{[s]}$ and
$\bu_i^{(k)}=\{u_{i\alpha}^{(k)},\,u_{i\alpha}\in\bu_i\}$.
\end{thm}
\proof
 Regard
$\P_i^{(k)}~(i=0,\ldots,n;k=0,\ldots,s-s_i)$ as polynomials in the
$n(s+1)$ variables
$\Y^{[s]}=\{y_1,\ldots,y_n,y^{(1)}_1,\ldots,y^{(1)}_n,\ldots,y^{(s)}_1,$
$\ldots,y^{(s)}_n\}$, and we denote its support by
$\mathcal{B}_{ik}$.  Since the coefficients $\bu_i^{(k)}$ of
$\P_i^{(k)}$ can be treated as algebraic indeterminates,
$\P_i^{(k)}$ are generic sparse polynomials with supports
$\mathcal{B}_{ik}$, respectively. Now we claim that
$\overline{\mathcal{B}}$ is strong essential, that is

\vskip5pt C1) $\overline{\mathcal{B}}=\{\mathcal{B}_{ik}:0\leq i\leq
n; 0\leq k \leq
 s-s_i\}$ is an essential set.

\vskip5pt

  C2) $\overline{\mathcal{B}}=\{\mathcal{B}_{ik}:0\leq i\leq n; 0\leq k \leq
 s-s_i\}$ jointly spans the affine lattice $\mathbb{Z}^{n(s+1)}$.

\vskip5pt

Note that $|\overline{\mathcal{B}}|=n(s+1)+1$. To prove C1), it
suffices to show that any $n(s+1)$ distinct $\P_{i}^{(k)}$ are
algebraically independent.
 Without loss of generality, we prove that for a fixed $l\in\{0,\ldots,s-s_0\}$,
 $$S_l=\{(\P_{i}^{(k)})_{1\leq i \leq n;0\leq k\leq s-s_i},\P_{0},\ldots,
   \P_{0}^{(l-1)},\P_{0}^{(l+1)},\ldots,\P_{0}^{(s-s_0)}\}$$
is an algebraically independent set. Clearly,
$\{y_j^{(k)},\ldots,y_j^{(s_i+k)}\big|j=1,\ldots,n\}$ is a subset of
the support of $\P_{i}^{(k)}$. Now we choose a monomial from each
$\P_{i}^{(k)}$ and denote it by $m(\P_{i}^{(k)})$. Let
\begin{eqnarray*} m(\P_{0}^{(k)})=\left\{\begin{array}{ll}
y_1^{(k)}& 0\leq
k\leq l-1 \\ y_1^{(s_0+k)}&l+1\leq k\leq s-s_0 \\
\end{array}\right. \text{and  }\, m(\P_{1}^{(k)})=\left\{\begin{array}{ll}
y_1^{(l+k)}& 0\leq
k\leq s_0 \\ y_2^{(s_1+k)}&s_0+1\leq k\leq s-s_1 \\
\end{array}\right..
\end{eqnarray*}
 For each $i\in\{2,\ldots,n\}$, let
 \begin{eqnarray*} m(\P_{i}^{(k)})=\left\{\begin{array}{lll}
y_i^{(k)}&\quad&0\leq
k\leq \sum_{j=0}^{i-1}s_j \\ y_{i+1}^{(s_i+k)}&\quad&\sum_{j=0}^{i-1}s_j+1\leq k\leq s-s_i \\
\end{array}\right..
\end{eqnarray*}

So $m(S_l)$ is equal to $\{y_j^{[s]}:1\leq j\leq n\}$, which are
algebraically independent over $\mathbb{Q}$. Thus, the $n(s+1)$
members of $S_l$ are algebraically independent over $\Q$. For if
not,  all the $\P_{i}^{(k)}-u_{i0}^{(k)}\,(\P_{i}^{(k)}\in S_l)$ are
algebraically dependent over $\Q(\bv)$ where
$\bv=\cup_{i=0}^{n}\bu_{i}^{[s-s_i]}\backslash\{u_{i0}^{[s-s_i]}\}$.
Now specialize the coefficient of $m(\P_i^{(k)})$ in $\P_i^{(k)}$ to
1, and all the other coefficients of $\P_i^{(k)}-u_{i0}^{(k)}$ to 0,
by the algebraic version of Lemma~\ref{lm-special},
$\{m(\P_i^{(k)}):\,\P_i^{(k)}\in S_l\}$ are algebraically dependent
over $\Q$, which is a contradiction. Thus, claim C1) is proved.
Claim C2) follows from the fact that $1$ and $\Y^{[s]}$ are
contained in the support of $\P_{0}^{[s-s_0]}$.

By C1) and C2), the sparse resultant of $(\P_i^{(k)})_{0\leq i\leq
n; 0\leq k\leq s-s_i}$ exists and we denote it by $G$. Then
$(G)=\big((\P_i^{(k)})_{0\leq i\leq n; 0\leq k\leq
s-s_i}\big)\bigcap$
$\mathbb{Q}[\bu_0^{[s-s_0]},\cdots,\bu_n^{[s-s_n]}]$, and by
Theorem~\ref{le-alg-sparse-deg}, $\deg(G,\bu_{i}^{(k)})=\mathcal
{M}\big((\CQ_{jl})_{j\neq i,0\leq l\leq s-s_j},\CQ_{i0},\ldots,$
$\CQ_{i,k-1},\CQ_{i,k+1},\ldots,\CQ_{i,s-s_i}\big)$, where
$\bu_i^{(k)}=(u_{i0}^{(k)},$ $\ldots,u_{i\alpha}^{(k)},\ldots)$. The
theorem will be proved if we can show that $G=c\cdot \SR$ for some
$c\in\Q.$

Since $G\in[\P_0,\ldots,\P_n]$ and $\ord(G,\bu_i)=s-s_i$,
 by Lemma~\ref{lm-genericresultant}, $\ord(\SR,\bu_i)\leq s-s_i$ for each $i=0,\ldots,n$.
If for some $i$, $\ord(\SR,\bu_i)=h_i<s-s_i$, then
$\SR\in((\P_{j}^{(k)})_{ j \neq i;0\leq k\leq s-s_j},\P_{i},$
$\ldots,
   \P_{i}^{(h_i)})$, a contradiction to C1). Thus, $\ord(\SR,\bu_i)= s-s_i$  and $\SR\in(G)$.
   Since $\SR$ is irreducible, there exists some $c\in\Q$ such that $G=c\cdot \SR$. So $\SR$ is equal to
   the algebraic sparse resultant of $\P_0^{[s-s_0]},\ldots,\P_n^{[s-s_n]}$.
   \qed

As a direct consequence of the above theorem and the determinant
representation for algebraic sparse resultant given by D'Andrea
\cite{dandrea1}, we have the following result.
\begin{cor}
The difference resultant for generic difference polynomials
$\P_i,i=0,\ldots,n$ can be written as the form $\det(M_1)/\det(M_0)$
where $M_1$ and $M_0$ are matrixes whose elements are coefficients
of $\P_i$ and their transforms up to the order $s-s_i$ and $M_0$ is
a minor of $M_1$.
\end{cor}

Based on the matrix representation given in the above corollary, the
efficient algorithms given by Canny,  Emiris, and Pan
\cite{emiris1,emiris2005} can be used to compute the difference
resultant.

\begin{cor}
The  degree of $\SR$ in each coefficient set $\bu_i$ is
$$\deg(\SR,\bu_i) =\sum_{k=0}^{s-s_i}\mathcal {M}\big((\CQ_{jl})_{j\neq i,0\leq
 l\leq s-s_j},\CQ_{i0},\ldots,\CQ_{i,k-1},\CQ_{i,k+1},\ldots,\CQ_{i,s-s_i}\big),$$
 and the total degree of $\SR$ is $$\deg(\SR)=\sum_{i=0}^n\sum_{k=0}^{s-s_i}\mathcal {M}\big((\CQ_{jl})_{j\neq i,0\leq
 l\leq s-s_j},\CQ_{i0},\ldots,\CQ_{i,k-1},\CQ_{i,k+1},\ldots,\CQ_{i,s-s_i}\big).$$
\end{cor}

\begin{rem}
From the proof of Theorem~\ref{th-resultant-deg}, we can see that
for each $i$ and $0\leq k\leq s-s_i$, $\deg(\SR,\bu_i^{(k)})>0$.
Furthermore, by Lemma~\ref{lm-order1}, $\deg(\SR,u_{i0}^{(k)})>0$
and $\deg(\SR,u_{i\alpha}^{(k)})>0$ for each $\alpha$. In
particular, $\deg(\SR,u_{i0})>0$ and $\deg(\SR,u_{i\alpha} )>0$.
\end{rem}

\begin{exmp}
Consider two generic difference polynomials of order one and degree
two in one indeterminate $y$:
\begin{eqnarray*}
 \P_0 &=& u_{00}+u_{01}y+u_{02}y^{(1)}+u_{03}y^2+u_{04}yy^{(1)}+u_{05}(y^{(1)})^2,\\
 \P_1 &=& u_{10}+u_{11}y+u_{12}y^{(1)}+u_{13}y^2+u_{14}yy^{(1)}+u_{15}(y^{(1)})^2.
\end{eqnarray*}
Then the degree bound given by Theorem \ref{th-spardeg} is
$\deg(\SR)\le (2+1)^4=81$.
By Theorem~\ref{th-resultant-deg},
$\deg(\SR,\bu_0)=\mathcal{M}(\CQ_{10},\CQ_{11},\CQ_{00}) +
\mathcal{M}(\CQ_{10},\CQ_{11},\CQ_{01})=8+8=16$  and
 consequently $\deg(\SR)=32$,
 where
 $\CQ_{00}=\CQ_{10}=\conv\{(0,0,0),(2,0,0),(0,2,0)\}$,
 $\CQ_{01}=\CQ_{11}=\conv\{(0,0,0),(0,2,0),(0,0,2)\}$,
%
%
and $\conv(\cdot)$ means taking the convex hull in $\mathbb{R}^3$.
By the proof of Theorem \ref{th-resultant-deg}, $\SR$ is the sparse
resultant of $\P_0, \sigma(\P_0), \P_1, \sigma(\P_1)$.
\end{exmp}

\subsection{Poisson-type product formula}
In this section, we will give a Poisson-type product formula for
difference resultant.

Let $\widetilde{\bu}=\cup_{i=0}^n \bu_i\setminus \{u_{00}\}$ and $
\qq\langle\widetilde{\bu}\rangle$ be the transformally
transcendental extension of $\qq$ in the usual sense. Let
$\qq_{0}=\qq\langle\widetilde{\bu}\rangle(u_{00},\ldots,u_{00}^{(s-s_0-1)})$.
Here, $\qq_0$ is not necessarily a difference overfield of $\qq$,
for the transforms of $u_{00}$ are not defined. In the following, we
will follow Cohn \cite{cohn-manifolds} to obtain algebraic
extensions $\mathcal {G}_i$ of $\qq_0$ and define transforming
operators  to make $\mathcal {G}_i$  difference fields. Consider
$\SR$ as an irreducible algebraic polynomial $r(u_{00}^{(s-s_0)})$
in $\qq_{0}[u_{00}^{(s-s_0)}]$.
In a suitable algebraic extension field of $\qq_{0}$,
$r(u_{00}^{(s-s_0)})=0$ has
$t_0=\deg(r,u_{00}^{(s-s_0)})=\deg(\SR,u_{00}^{(s-s_0)})$ roots
$\gamma_{1},\ldots,\gamma_{t_0}$. Thus
\begin{equation}\label{eq-fac00}
\SR(\bu_{0},\bu_{1},\ldots,\bu_{n})=A\prod^{t_0}_{\tau=1}(u_{00}^{(s-s_0)}-\gamma_{\tau})
\end{equation}
where $A\in\qq_0$.
%
Let $\mathcal
{I}_\bu=[\P_0,\ldots,\P_n]\cap\Q\{\bu_0,\ldots,\bu_n\}$.
%
%
By the definition of the difference resultant, $\mathcal {I}_\bu$ is
an essential reflexive prime difference ideal in the decomposition
of $\{\SR\}$ which is not held by any difference polynomial of order
less than $s-s_0$ in $u_{00}$. Suppose $\SR,\SR_1,\SR_2,\ldots$ is a
basic sequence\footnote{For the rigorous definition of {\em basic
sequence}, please refer to \cite{cohn-manifolds}. Here, we list its
basic properties:
 i) For each $k\geq0$,  $\ord(\SR_k,u_{00})=s-s_0+k$
and  $\SR,\SR_1,\ldots,\SR_k$ is an irreducible algebraic ascending
chain, and ii) $\bigcup_{k\geq0}\asat(\SR,\SR_1,\ldots,\SR_k)$ is a
reflexive prime difference ideal. } of $\SR$ corresponding to
$\mathcal {I}_\bu$. That is, $\mathcal
{I}_\bu=\bigcup_{k\geq0}\asat(\SR,\SR_1,\ldots,\SR_k)$. Regard all
the $\SR_i$ as algebraic polynomials over the coefficient field
$\qq\langle\widetilde{\bu}\rangle$. Denote
$\gamma_{\tau0}=\gamma_\tau.$ Clearly,
$u_{00}^{(s-s_0)}=\gamma_{\tau0}$ is a generic zero of $\asat(\SR)$.
Suppose $\gamma_{\tau i}\,(i\leq k)$ are found in some algebraic
extension field of $\qq_{0}$ such that
$u_{00}^{(s-s_0+i)}=\gamma_{\tau i}\,(0\leq i\leq k)$ is a generic
zero
 of $\asat(\SR,\SR_1,\ldots,\SR_k)$.
 Then let $\gamma_{\tau, k+1}$ be an element such that $u_{00}^{(s-s_0+i)}=\gamma_{\tau i}\,(0\leq i\leq k+1)$ is a generic zero
 of $\asat(\SR,\SR_1,\ldots,\SR_k,\SR_{k+1})$. Clearly, $\gamma_{\tau, k+1}$ is also algebraic over $\Q_0$.
 Let $\mathcal {G}_\tau=\qq\langle\widetilde{\bu}\rangle(u_{00},\ldots,u_{00}^{(s-s_0-1)},\gamma_\tau,\gamma_{\tau1},\ldots)$.
 Clearly, $\mathcal {G}_\tau$ is an algebraic extension of $\qq_0$ and $\mathcal {G}_\tau$ is algebraically isomorphic to the quotient field of $\qq\{\bu_0,\ldots,\bu_n\}/\mathcal{I}_\bu.$
Since the quotient field of
$\qq\{\bu_0,\ldots,\bu_n\}/\mathcal{I}_\bu$ is also a difference
field, we can introduce a transforming operator $\sigma_\tau$ into
$\mathcal {G}_\tau$ to make it a difference field such that the
above isomorphism becomes a difference one. That is,
$\sigma_\tau|_{\qq_0}=\sigma|_{\qq_0}$ and
\[\sigma^{k}_\tau(u_{00})=\left\{ \begin{array}{ll}
u_{00}^{(k)}&0\leq k\leq s-s_0-1\\
\gamma_{\tau,k-s-s_0}&k\geq s-s_0\\
\end{array}
\right.
\]
In this way,  $(\mathcal {G}_\tau,\sigma_\tau)$ is a difference
field.

Let $F$ be a difference polynomial in
$\qq\{\bu_0,\bu_1,\ldots,\bu_n\} = \qq\{\widetilde{\bu},u_{00}\}$.
For convenience,  by the symbol
$F\big|_{u_{00}^{(s-s_0)}=\gamma_{\tau}}$, we mean substituting
$u_{00}^{(s-s_0+k)}$ by $\sigma_\tau^{k}\gamma_{\tau}=\gamma_{\tau
k}\,(k\geq0)$ into $F$.
Similarly, by saying $F$ vanishes at
$u_{00}^{(s-s_0)}=\gamma_{\tau}$, we mean
$F\big|_{u_{00}^{(s-s_0)}=\gamma_{\tau}}=0$. The following lemma is
a direct consequence of the above discussion.

\begin{lem}\label{lm-ftaup}
$F\in \mathcal {I}_\bu$ if and only if $F$ vanishes at
$u_{00}^{(s-s_0)}=\gamma_{\tau}$.
\end{lem}
\proof Since $\mathcal
{I}_\bu=\bigcup_{k\geq0}\asat(\SR,\SR_1,\ldots,\SR_k)$ and
$u_{00}^{(s-s_0+i)}=\gamma_{\tau i}\,(0\leq i\leq k)$ is a generic
zero
 of $\asat(\SR,\SR_1,\ldots,\SR_k)$, the lemma follows.\qed

\begin{rem}\label{rm-ftau}
In order to make $\mathcal {G}_\tau$ a difference field, we need to
introduce a transforming operator $\sigma_\tau$ which is closely
related to $\gamma_\tau$. Since even for a fixed $\tau,$  generic
zeros of $\asat(\SR,\SR_1,\ldots,\SR_k)$ beginning from
$u_{00}^{(s-s_0)}=\gamma_{\tau}$ may not be unique, the definition
of $\sigma_\tau$ also may not be unique, which is  different from
the differential case. In fact, it is a common phenomena in
difference algebra. Here, we just choose one, for they do not
influence the following discussions.
\end{rem}
Now we give the following Poisson type formula for the difference
resultant.
\begin{thm}\label{th-fac1}
Let  $\SR(\bu_{0}, \ldots,\bu_{n})$
 be the difference resultant
of $\P_0,\ldots,\P_n$. Let $\deg(\SR,u_{00}^{(s-s_0)})$ $=t_0$. Then
there exist $\xi_{\tau k}\,(\tau=1,\ldots,t_0;k=1,\ldots,n)$ in
 overfields $(\mathcal {G}_\tau,\sigma_\tau)$ of
$(\qq\langle\widetilde{\bu}\rangle,\sigma)$  such that
\begin{equation} \label{eq-decom1} \SR=A\prod_{\tau=1}^{t_0} \P_0(\xi_{\tau1},\ldots,\xi_{\tau n})^{(s-s_0)},
\end{equation} where $A\in\qq\langle\bu_1,\ldots,\bu_n\rangle[\bu_0^{[s-s_0]}\backslash
u_{00}^{(s-s_0)}]$.
Note that \bref{eq-decom1} is formal and should be understood in the
following precise meaning: $\P_0(\xi_{\tau})^{(s-s_0)}
\stackrel{\triangle}{=}
\sigma^{s-s_0}u_{00}+\sigma_\tau^{s-s_0}(\sum_{\alpha\in\mathcal{B}_0\setminus\{0\}}
u_{0\alpha}(\xi_\tau^{[s-s_0]})^{\alpha})$, where
$\xi_\tau=(\xi_{\tau1},\ldots,\xi_{\tau n})$.
\end{thm}
\proof By Theorem~\ref{th-homo}, there exists $m\in\mathbb{N}$ such
that
$$u_{00}\frac{\partial \SR}{\partial u_{00}}+\sum_{\alpha}u_{0\alpha}\frac{\partial \SR}{\partial u_{0\alpha}}=m\SR.$$
Setting $u_{00}^{(s-s_0)}=\gamma_{\tau}$ in both sides of the above
equation, we have
$$u_{00}\frac{\partial \SR}{\partial u_{00}}\Big|_{u_{00}^{(s-s_0)}=\gamma_{\tau}}+\sum_{\alpha}u_{0\alpha}\frac{\partial \SR}{\partial u_{0\alpha}}\Big|_{u_{00}^{(s-s_0)}=\gamma_{\tau}}=0.$$
Let $\xi_{\tau \alpha}=(\frac{\partial \SR}{\partial
u_{0\alpha}}/\frac{\partial \SR}{\partial
u_{00}})\big|_{u_{00}^{(s-s_0)}=\gamma_\tau}$. Then
$u_{00}=-\sum_{\alpha}u_{0\alpha}\xi_{\tau \alpha}$ with
$u_{00}^{(s-s_0)}=\gamma_{\tau}.$ That is,
$\gamma_{\tau}=-\sigma_\tau^{s-s_0}(\sum_{\alpha}u_{0\alpha}\xi_{\tau
\alpha})=-(\sum_{\alpha}u_{0\alpha}\xi_{\tau \alpha})^{(s-s_0)}.$
Thus, $$\SR=A\prod_{\tau=1}^{t_0}
(u_{00}+\sum_{\alpha}u_{0\alpha}\xi_{\tau \alpha})^{(s-s_0)}.$$
Suppose $\P_0=u_{00}+\sum_{j=1}^nu_{0j0}y_j+T_0.$ Let $\xi_{\tau
j}=(\frac{\partial \SR}{\partial u_{0j0}}/\frac{\partial
\SR}{\partial
u_{00}})\big|_{u_{00}^{(s-s_0)}=\gamma_\tau}\,(j=1,\ldots,n)$ and
$\xi_\tau=(\xi_{\tau1},\ldots,\xi_{\tau n})$. It remains to show
that $\xi_{\tau\alpha}=(\xi_\tau^{[s_0]})^{\alpha}$.

Let
$\zeta_i=-\sum_{\alpha}u_{i\alpha}(\Y^{[s_i]})^{\alpha}\,(i=0,\ldots,n).$
Clearly, $\zeta=(\bu,\zeta_0,\ldots,\zeta_n)$ is a generic zero of
$\mathcal{I}_\bu=[\P_0,\ldots,\P_n]\cap\Q\{\bu_0,\ldots,\bu_n\}$,
where $\bu=\cup_{i=1}^n\bu_i\backslash\{u_{i0}\}$. For each
$(\Y^{[s_0]})^{\alpha}=\prod_{j=1}^n(y_j^{(k)})^{d_{jk}}$, by
equation \bref{eq-partialdiff},
$(\Y^{[s_0]})^{\alpha}=\overline{\frac{\partial \SR}{\partial
u_{0\alpha}}}\big/\overline{\frac{\partial \SR}{\partial
u_{00}}}=\prod\limits_{j=1}^n\prod\limits_{k=0}^{s_0}\Big(\big(\overline{\frac{\partial
\SR}{\partial u_{0j0}}}\big/\overline{\frac{\partial \SR}{\partial
u_{00}}}\big)^{(k)}\Big)^{d_{jk}}$, where $ \overline{\frac{\partial
\SR}{\partial u_{0\alpha}}}=\frac{\partial \SR}{\partial
u_{0\alpha}}\Big|_{u_{i0}=\zeta_i}$. So $\frac{\partial
\SR}{\partial
u_{0\alpha}}\prod\limits_{j=1}^n\prod\limits_{k=0}^{s_0}\Big(\big(
\frac{\partial \SR}{\partial u_{00}}
\big)^{(k)}\Big)^{d_{jk}}-\frac{\partial \SR}{\partial
u_{00}}\prod\limits_{j=1}^n\prod\limits_{k=0}^{s_0}\Big(\big(
\frac{\partial \SR}{\partial u_{0j0}}
\big)^{(k)}\Big)^{d_{jk}}\in\mathcal{I}_\bu.$ By
Lemma~\ref{lm-ftaup},
$\xi_{\tau\alpha}=\prod\limits_{j=1}^n\prod\limits_{k=0}^{s_0} \big(
\xi_{\tau j}^{(k)}\big) ^{d_{jk}}=(\xi_\tau^{[s_0]})^{\alpha}.$
Thus, \bref{eq-decom1} follows. \qed

\begin{thm}\label{th-fac1-genericpoint}
The points $\xi_\tau=(\xi_{\tau1},\ldots,\xi_{\tau
n})\,(\tau=1,\ldots,t_0)$ in \bref{eq-decom1} are  generic zeros of
the difference ideal
$[\P_1,\ldots,\P_n]\subset\Q\langle\bu_1,\ldots,\bu_n\rangle\{\Y\}$.
\end{thm}
\proof Clearly, $\xi_{\tau}$ are $n$-tuples over
$\Q\langle\bu_1,\ldots,\bu_n\rangle$. For each $i=1,\ldots,n,$
rewrite
$\P_i=u_{i0}+\sum\limits_{\alpha}u_{i\alpha}\prod\limits_{j=1}^n\prod\limits_{k=1}^{s_i}
(y_j^{(k)})^{\alpha_{jk}}$. Since
$\zeta_i=-\sum\limits_{\alpha}u_{i\alpha}\prod\limits_{j=1}^n\prod\limits_{k=1}^{s_i}(y_j^{(k)})^{\alpha_{jk}}$
and $y_j=\overline{\frac{\partial \SR}{\partial
u_{0j0}}}\big/\overline{\frac{\partial \SR}{\partial u_{00}}}$,
$\zeta_i+\sum\limits_{\alpha}u_{i\alpha}\prod\limits_{j=1}^n\prod\limits_{k=1}^{s_i}\big((\overline{\frac{\partial
\SR}{\partial u_{0j0}}}\big/\overline{\frac{\partial \SR}{\partial
u_{00}}})^{(k)}\big)^{\alpha_{jk}}=0.$ Let
$a_{jk}=\max_\alpha\alpha_{jk}$. Then
$u_{i0}\prod\limits_{j=1}^n\prod\limits_{k=1}^{s_i}\big((\frac{\partial
\SR}{\partial
u_{00}})^{(k)}\big)^{a_{jk}}+\sum\limits_{\alpha}u_{i\alpha}\prod\limits_{j=1}^n\prod\limits_{k=1}^{s_i}\big((\frac{\partial
\SR}{\partial u_{0j0}})^{(k)}\big)^{\alpha_{jk}}\big((\frac{\partial
\SR}{\partial
u_{00}})^{(k)}\big)^{a_{jk}-\alpha_{jk}}\in\mathcal{I}_{\bu}$. Thus,
by Lemma~\ref{lm-ftaup},
$\P_i(\xi_\tau)=u_{i0}+\sum\limits_{\alpha}u_{i\alpha}\prod\limits_{j=1}^n\prod\limits_{k=1}^{s_i}
(\xi_{\tau j}^{(k)})^{\alpha_{jk}}=0\,(i=1,\ldots,n)$.

On the other hand, suppose
$F\in\Q\langle\bu_1,\ldots,\bu_n\rangle\{\Y\}$ vanishes at
$\xi_\tau$. Without loss of generality, suppose
$F\in\Q\{\bu_1,\ldots,\bu_n,\Y\}$. Clearly, $\P_1,\ldots,\P_n$
constitute an ascending chain in $\Q\{\bu_1,\ldots,\bu_n,\Y\}$ with
$u_{i0}$ as leaders. Let $G$ be the difference remainder of $F$ with
respect to this ascending chain. Then $G$ is free from $u_{i0}$ and
$F\equiv G\,\mod\, [\P_1,\ldots,\P_n]$. Then
$G(\xi_\tau)=G(\widetilde{\bu};\xi_{\tau1},\ldots,\xi_{\tau n})=0$,
where $\widetilde{\bu}=\cup_{i=1}^n\bu_i\backslash\{u_{i0}\}$. So
there exist $a_k\in\mathbb{N}$ such that
$G_1=\prod_{k}\big((\frac{\partial \SR}{\partial
u_{00}})^{(k)}\big)^{a_k}G(\widetilde{\bu};\Y)\in\mathcal {I}_\bu.$
Thus, $G_1$ vanishes at $u_{i0}=\zeta_i\,(i=1,\ldots,n)$ while
$\frac{\partial \SR}{\partial u_{00}}$ does not. It follows that
$G(\widetilde{\bu};\Y)\equiv0$ and $F\in[\P_1,\ldots,\P_n]$. So
$\xi_\tau$ are  generic zeros of
$[\P_1,\ldots,\P_n]\subset\Q\langle\bu_1,\ldots,\bu_n\rangle\{\Y\}$.
\qed

By Theorems~\ref{th-fac1} and \ref{th-fac1-genericpoint}, we can see
that difference resultants have Poisson-type product formula, which
is similar to their algebraic  and differential analogues.

\vf We conclude this section by proving the following theorem, which
explores the relationship between the difference resultant and the
solvability of the given systems.
\begin{thm}\label{th-drcondition}
Let $\SR$ be the difference resultant of $\P_0,\ldots,\P_n$. Suppose
when each $\bu_i$ is specialized to  $\bv_i$, $\P_i$ is specialized
to $\overline{\P}_i.$ If $\overline{\P}_0=\cdots=\overline{\P}_n=0$
has a common difference solution, then $\SR(\bv_0,\ldots,\bv_n)=0$.
Moreover, if $\SR(\bv_0,\ldots,\bv_n)=0$ and
$\frac{\partial\SR}{\partial u_{00}}(\bv_0,\ldots,\bv_n)\neq 0$,
then $\overline{\P}_0=\cdots=\overline{\P}_n=0$ has at most one
solution $(\bar{y}_1,\ldots,\bar{y}_n)$ with each
$\bar{y}_k=\big(\frac{\partial\SR}{\partial
u_{0k}}\big/\frac{\partial\SR}{\partial
u_{00}}\big)(\bv_0,\ldots,\bv_n)$, where
 $u_{0k}$ is the coefficient of $y_k$ in $\P_0$.
\end{thm}
\proof Suppose $\P_i=u_{i0}+T_i\,(i=1,\ldots,n)$ and
$\bu=\cup_{i=0}^n\bu_i\backslash\{u_{i0}\}$. Clearly,
$(\Y;\bu,-T_0(\Y),$ $\ldots,-T_n(\Y))$ is a generic zero of
$[\P_0,\ldots,\P_n]\subset\Q\{\Y;\bu_0,\ldots,\bu_n\}$. Taking the
partial derivative of $\SR(\bu;-T_0(\Y),\ldots,-T_n(\Y))=0$ w.r.t.
$u_{0k}$, we can show that $\frac{\partial \SR}{\partial
u_{00}}y_k-\frac{\partial \SR}{\partial
u_{0k}}\in[\P_0,\ldots,\P_n]\,(k=1,\ldots,n)$. If
$\overline{\P}_0=\cdots=\overline{\P}_n=0$ has a common solution
$\xi$, then $(\xi;\bv_0,\ldots,\bv_n)$ is a common solution of
$[\P_0,\ldots,\P_n]$. Since $\SR\in[\P_0,\ldots,\P_n]$, $\SR$ must
vanish at $(\bv_0,\ldots,\bv_n)$. Now suppose
$\SR(\bv_0,\ldots,\bv_n)=0$ and $\frac{\partial\SR}{\partial
u_{00}}(\bv_0,\ldots,\bv_n)\neq 0$. If
$(\bar{y}_1,\ldots,\bar{y}_n)$ is a common solution of
$\overline{\P}_i=0$, then each $\frac{\partial \SR}{\partial
u_{00}}y_k-\frac{\partial \SR}{\partial u_{0k}}$ vanishes at
$(\bar{y}_1,\ldots,\bar{y}_n;\bv_0,\ldots,\bv_n)$. Thus,
$\bar{y}_k=\big(\frac{\partial\SR}{\partial
u_{0k}}\big/\frac{\partial\SR}{\partial
u_{00}}\big)(\bv_0,\ldots,\bv_n)$, since
$\frac{\partial\SR}{\partial u_{00}}(\bv_0,\ldots,\bv_n)\ne0$.
Hence, the second assertion holds. \qed

\begin{rem}
If Problem~\ref{problem-generator} can be solved positively, then
Theorem~\ref{th-drcondition} can be strengthened as follows: If
$\SR(\bv_0,\ldots,\bv_n)=0$ and $\frac{\partial\SR}{\partial
u_{00}}(\bv_0,\ldots,\bv_n)\neq 0$, then
$\overline{\P}_0=\cdots=\overline{\P}_n=0$  has a unique solution
$(\bar{y}_1,\ldots,\bar{y}_n)$ with each
$\bar{y}_k=\big(\frac{\partial\SR}{\partial
u_{0k}}\big/\frac{\partial\SR}{\partial
u_{00}}\big)(\bv_0,\ldots,\bv_n)$.

\end{rem}

\section{Conclusion and problem}
\label{sec-conc}

In this paper,  we first introduce the concepts of Laurent
difference polynomials and Laurent transformally essential systems
and give a criterion for a difference polynomial system to be
Laurent transformally essential in terms of its supports. Then the
sparse difference resultant for a Laurent transformally essential
system is defined and its basic properties are proved. Furthermore,
order and degree bounds for the sparse difference resultant are
given. Based on these bounds, an algorithm to compute the sparse
difference resultant is proposed, which is single exponential in
terms of the order, the number of variables, and the size of the
Laurent transformally essential system. Besides these, the
difference resultant is introduced and its basic properties are
given, such as its precise order and BKK style degree, determinant
representation, and a Poisson-type product formula.

We now propose several questions for further study apart from
Problem~\ref{problem-generator}.



The degree of the algebraic sparse resultant is equal to the  mixed
volume of certain polytopes generated by the supports of the
polynomials as shown in \cite{Pedersen} or \cite[p.255]{gelfand}.
%
%
And  Theorem~\ref{th-resultant-deg} shows that the degree of
difference resultants is exactly of  such BKK-style. It is desirable
to obtain such a bound for sparse difference resultant. For more
details, see Remark \ref{re-bkksr}.

There exist very efficient algorithms to compute algebraic sparse
resultants\,\cite{emiris0,emiris1,emiris2005,dandrea1}, which are
based on matrix representations for the resultant. How to apply the
principles behind these algorithms to compute sparse difference
resultants is an important problem.

Algebraic resultant and sparse resultant have many interesting
applications \cite{canny1,cox,emiris1999,gelfand}. It is desirable
to develop the corresponding theory for difference polynomial
systems based difference resultant.


\begin{thebibliography}{99}


\bibitem{bkk}
 D.~N. Bernshtein.
 The Number of Roots of a System of Equations.
 {\em Functional Anal. Appl.}, 9(3), 183-185, 1975.


\bibitem{bouziane} D. Bouziane, A. Kandri Rody, and H. Ma$\hat{a}$rouf.
 Unmixed-dimensional Decomposition of a Finitely Generated Perfect Differential Ideal.
 {\em Journal of Symbolic Computation}, 31(6), 631-649, 2001.


%
%

\bibitem{canny1}
 J.~F. Canny.
 Generalized Characteristic Polynomials.
 {\em Journal of Symbolic Computation}, 9, 241-250, 1990.

%
%

\bibitem{cohen}
 H. Cohen.
 {A Course in Computational Algebraic Number Theory}.
 Springer-Verlag, New York, 1993.

 \bibitem{cohn-manifolds}
 R.~M. Cohn.
Manifolds of Difference Polynomials.
 {\em Trans. Amer. Math. Soc.}, 64(1), 1948.
\bibitem{cohn}
R.~M. Cohn.
{\em Difference Algebra}.
Interscience Publishers, New York, 1965.
\bibitem{cohn2}
 R.~M. Cohn.
 Order and Dimension.
 {\em Proc. Amer. Math. Soc.}, 87(1), 1-6, 1983.


\bibitem{cox}
 D. Cox, J. Little, D. O'Shea.
 {\em Using Algebraic Geometry}.
 Springer, 1998.

\bibitem{dandrea1}
 C. D'Andrea.
 Macaulay Style Formulas for Sparse Resultants.
 {\em Trans. of AMS}, 354(7), 2595-2629, 2002.
%
\bibitem{eisenbud}
 D. Eisenbud, F.~O. Schreyer, J. Weyman.
 Resultants and Chow Forms via Exterior Syzygies.
 {\em Journal of Amer. Math. Soc.}, 16(3), 537-579, 2004.


\bibitem{emiris0}
 I.~Z. Emiris.
 On the Complexity of Sparse Elimination.
 {\em J. Complexity}, 12, 134-166, 1996.
\bibitem{emiris1}
  I.~Z. Emiris and J.~F. Canny.
  Efficient Incremental Algorithms for the Sparse Resultant and the
  Mixed Volume.
  {\em  Journal of Symbolic Computation}, 20(2), 117-149, 1995.
%
\bibitem{emiris1999}
  I.~Z. Emiris and B. Mourrain.
  Matrices in Elimination Theory.
  {\em  Journal of Symbolic Computation}, 28(1,2), 3-43, 1999.
%
\bibitem{emiris2005}
  I.~Z. Emiris and V.~Y. Pan.
  Improved algorithms for computing determinants and resultants. {\em  Journal of Complexity}, 21, 43-71, 2005.


\bibitem{gao-chou}
X.~S. Gao and S.~C. Chou. On the Dimension for Arbitrary Ascending
Chains. {\em Chinese Bull. of Sciences,} 38, 396-399, 1993.

\bibitem{gao}
 X.~S. Gao, W. Li, C.~M. Yuan.
 Intersection Theory in Differential Algebraic Geometry:
 Generic Intersections  and the Differential Chow Form.
 {\em Trans. of Amer. Math. Soc.}, 365(9), 4575-4632, 2013.

\bibitem{gao-dcs}
 X.~S. Gao, Y. Luo, C.~M. Yuan.
 A Characteristic Set Method for Ordinary Difference Polynomial Systems.
 {\em Journal of Symbolic Computation}, 44(3), 242-260, 2009.

\bibitem{gelfand}
 I.~M. Gelfand,  M. Kapranov, A. Zelevinsky.
 {\em Discriminants, Resultants and Multidimensional Determinants}.
 Boston, Birkh\"auser, 1994.
%

\bibitem{hodge1}
 W.V.D. Hodge and D. Pedoe.  {\em Methods of Algebraic Geometry, Volume I}.
 Cambridge Univ. Press, 1968.
%
%

\bibitem{Hrushovski1}
 E. Hrushovski.
 The elementary theory of the Frobenius automorphisms.
 Available from http://www.ma.huji.ac.il/\~\,ehud/.

\bibitem{joun1}
 J.~P. Jouanolou.
 Le formalisme du r{\`e}sultant.
 {\em Advances in Mathematics}, 90(2), 117-263, 1991.

\bibitem{Kapranov1}
 M. Kapranov, B. Sturmfels, A Zelevinsky.
 Chow Polytopes and General Resultants.
 {\em Duke Math. J.}, 67, 189-218, 1992.

%
%
%
%

\bibitem{lando}
 B.~A. Lando.
 Jacobi's Bound for the Order of Systems of First Order Differential Equations.
 {\em Trans. Amer. Math. Soc.} 152, 119-135, 1970.

\bibitem{levin}
 A. Levin.
 {\em Difference Algebra}.
 Springer, 2008.

\bibitem{li-pdcf}
 W. Li and X.S. Gao.
 Differential Chow Form for Projective Differential Variety.
 {\em Journal of Algebra}, 370: 344-360, 2012.

\bibitem{li1}
 W. Li, X.~S. Gao, C.~M. Yuan.
 Sparse Differential Resultant.
 {\em Proc. ISSAC 2011},  225-232, ACM Press, New York, 2011.

\bibitem{li}
 W. Li, C.~M. Yuan, X.~S. Gao.
 Sparse Differential Resultant for Laurent Differential Polynomials.
 {\em ArXiv:1111.1084v3,}  70 pages, 2012.

\bibitem{li-2013}
 W. Li, C.~M. Yuan, X.~S. Gao.
 Sparse Difference Resultant,
 {\em Proc. ISSAC 2013}, Boston, ACM Press, 2013.

%
%
\bibitem{Pedersen}
 P. Pedersen and B. Sturmfels.
 Product Formulas for Resultants and Chow Forms.
 {\em Mathematische Zeitschrift}, 214(1), 377-396, 1993.

%
%
%
%
%
%
%

 \bibitem{sonia-arxiv}
 S.~L. Rueda.
 Linear Sparse Differential Resultant Formulas.
 {\em Linear Algebra and its Applications}.
 438(11), 4296-4321, 2013.

 \bibitem{sonia-raf}
 S.~L. Rueda and J.~R. Sendra.
 Linear complete differential resultants and the implicitization of
 linear DPPEs.
 {\em Journal of Symbolic Computation}, 45(3), 324-341, 2010.

 \bibitem{shenly}
 L. Shen, E. Chionh, X.~S. Gao, J. Li.
 Proper Reparametrization for inherently improper unirational varieties.
 {\em Journal of Systems Science and Complexity}, 24(2), 367-380, 2011.

\bibitem{sturmfels}
 B. Sturmfels.
 Sparse Elimination Theory.
 In {\em Computational Algebraic Geometry and Commutative Algebra}, Eisenbud, D., Robbiano, L. eds.
 264-298, Cambridge University Press, 1993.
%
\bibitem{sturmfels2}
 B. Sturmfels.
 On The Newton Polytope of the Resultant.
 {\em Journal of Algebraic Combinatorics}, 3, 207-236, 1994.

\bibitem{vogel}
 W. Vogel. {\em Lectures on Results on Bezout's Theorem}.
Springer-Verlag,  Berlin-Heidelberg-New York-Tokyo, 1984.


\bibitem{wibmer}
 M. Wibmer.
 {\em Lecture Notes on Algebraic Difference Equations}.
 Preprint, February, 2013.

\bibitem{wu}
 W.~T. Wu.
 {\em Mathematics Machenization}.
 Science Press/Kluwer, Beijing, 2003.


\bibitem{dres-matrix}
 Z. Y. Zhang, C. M. Yuan,  X. S. Gao.
 Matrix Formula of Differential Resultant for First Order Generic
 Ordinary Differential Polynomials.
 arXiv:1204.3773, 2012.

\end{thebibliography}
\end{document}